\documentclass[a4paper]{article}%

\usepackage{amsmath}%
\usepackage{amsfonts}%
\usepackage{amssymb}%
\usepackage{graphicx}

\usepackage{mathrsfs}
\usepackage{xspace}
\usepackage{mathtools, amssymb}

\usepackage[affil-it]{authblk}
\usepackage{geometry}
\usepackage{amsmath,amssymb,graphicx}
\usepackage{tabularx}
\newcolumntype{C}[1]{>{\centering\arraybackslash}p{#1}}
\usepackage{hyperref}

\usepackage{xcolor}

\usepackage{amsmath,scalerel}
\DeclareMathOperator*{\Bigcdot}{\scalerel*{\cdot}{\bigodot}}

\newcommand{\be}{\begin{equation}}
\newcommand{\ee}{\end{equation}}
\newcommand{\bea}{\begin{eqnarray}}
\newcommand{\eea}{\end{eqnarray}}

\usepackage{mathrsfs}

\usepackage{mathtools}

\usepackage{tikz}
\usetikzlibrary{positioning,shapes,fit,arrows}

\definecolor{myblue}{RGB}{56,94,141}
\definecolor{myred}{RGB}{30,94,100}

\begin{document}
\title{Dimensional aspects of Lovelock-Lanczos gravity}

\author{Aimeric Coll\'{e}aux\thanks{Electronic address: \texttt{aimeric.colleaux@alumni.unitn.it \& aimeric.colleaux@gmail.com}} } 

\affil{%
Department of Physics, Trento University, Via Sommarive 14-38123 Trento, Italy\\}

\maketitle

\textbf{Abstract} :
\\

There has recently been an increasing interest in regularizations of Lovelock-Lanczos gravity (LLG) in four dimensions, in which dimensional poles and possibly counter-terms are introduced to compensate the vanishing of the Lovelock field equations in critical and lower dimensions.  In this paper, we review and extend some of these results. We first find a class of LLG theories whose perturbative expansion around a given (A)dS vacuum can be regularized up to arbitrary order, the simplest one being close to Lovelock gravities with a unique vacuum. If well-defined, these models might be interpreted as effective field theories of gravitons in four dimensions, or might be combined with other regularization approaches. Among those, we establish the general procedure to obtain $4$D covariant and background independent regularizations from metric transformations. 
In the conformal (and critical) case, we generalize previous results obtained for the Gauss-Bonnet theory to the full Lovelock series. Similarly, the regularization of Gauss-Bonnet gravity from the breaking of $4$D-covariance down to $3$D is generalized to arbitrary curvature order, seemingly resulting in new Minimally Modified gravities propagating solely the two degrees of freedom of the graviton. 
 Finally, we present general results regarding the minisuperspace regularization of specific sectors of LLG. Non-perturbative (in curvature) regularized theories admitting non-singular black holes as well as non-singular past-dS$_4$ and cyclic closed cosmologies are found. We conclude with the non-uniqueness of these background regularizations by finding inequivalent regularizations of the Bianchi I sector of Lovelock-Lanczos gravity in four dimensions.

\allowdisplaybreaks[1]

\maketitle

\tableofcontents

\section{Introduction}

In the metric formalism, the Lovelock-Lanczos (LL) gravitational action, depending solely on the metric field, invariant under $d$-dimensional diffeomorphisms and leading to second order field equations is given by :
\begin{eqnarray}
I\left[ g_{\mu\nu} \right]= \frac{1}{2 \kappa_0} \int d^d x \sqrt{-g} \mathcal{L}\left[ g_{\mu\nu} \right]\,, \label{LLGAction}
\end{eqnarray}
where its Lagrangian density can be expressed in terms of the Lovelock-Lanczos scalars $\mathcal{L}_p$ as
\begin{eqnarray}
\mathcal{L} := \sum_{p=0}^{t} \alpha_p \mathcal{L}_p \; , \;\;\; \mathcal{L}_p := \frac{1}{2^p}  \delta^{\mu_1 \nu_1 \dots \mu_p \nu_p}_{\sigma_1 \rho_1 \dots \sigma_p \rho_p} \prod  _{r=1}^p R_{\mu_r \nu_r}^{\sigma_r \rho_r}\,, \label{LagrangianDensity}
\end{eqnarray}
with $R^{\mu \nu}_{\sigma \rho} :=g^{\gamma\mu} R_{\sigma \rho\gamma}^{\phantom{\sigma \rho\gamma}\nu}$ and $\delta^{\mu_1  \dots \mu_s}_{\nu_1  \dots \nu_s}$ is the generalized Kronecker delta (GKD) defined by\footnote{In this paper, all the (anti)symmetrizations are taken without normalization, for example, $X_{[ab]}:=X_{ab}-X_{ba}$.}
\begin{eqnarray}
\delta^{\mu_1  \dots \mu_s}_{\nu_1  \dots \nu_s}:= \delta^{\mu_1}_{[\nu_1} \dots  \delta^{\mu_s}_{\nu_s]} \,,
\end{eqnarray}
while the $\alpha_p$ and $\kappa_0$ are coupling constants ensuring the proper dimensionality of the action and $t$, a positive integer, is a free parameter of the theory. In the presence of matter, described by a stress-energy tensor $T$, the components of the field equations of this theory are given by
\begin{equation}
\mathcal{G}_{\mu \nu} \big[g_{\mu\nu} \big] := \sum_{p=0}^t \alpha_p \, \mathcal{G}^{(p)}_{\mu\nu} = \kappa_0 \, T_{\mu\nu}, 
\label{NonPerturbativeFieldEquations}
\end{equation}
where each Lovelock-Lanczos scalar of curvature order $p$ contributes as
\begin{equation}
\mathcal{G}_{\phantom{(p)} \beta}^{(p) \alpha}\big[g_{\mu\nu} \big]  :=\left[ \frac{g^{\alpha\gamma}}{\sqrt{-g}}\frac{ \delta}{\delta g^{\gamma\beta}} \right] \left(\sqrt{-g} \mathcal{L}_p \right)=  -\frac{1}{2^{p+1}} \delta^{\alpha \mu_1 \nu_1 \dots \mu_p \nu_p}_{\beta \sigma_1 \rho_1 \dots \sigma_p \rho_p}\prod  _{r=1}^p R_{\mu_r \nu_r}^{\sigma_r \rho_r} \,,  \label{NonPerturbativeFieldEquations2}
\end{equation}
which is the reason why the field equations are second order. Stated at the level of the field equations, the Lovelock theorem \cite{Lovelock:1971vz,Lovelock:1972vz} ensures that the unique covariant $d$-dimensional symmetric tensor containing a number of $2p$ derivatives of the metric (and no additional fields) while being second order (i.e. containing at most terms like $\left( \partial \partial g\right)^p$) and divergence-free is the tensor $\mathcal{G}_{\phantom{(p)}}^{(p)}$ whose components are given in the previous equation. Furthermore, due to the GKD, only one Riemann tensor of each products can contain second time derivatives of the metric.

Similarly, the order-$p$ tensors $\mathcal{G}_{\phantom{(p)}}^{(p)}$ are by construction identically (i.e. \textit{algebraically}) vanishing in dimensions $d\leq2p$, as their GKD have $(2p+1)$ antisymmetrized indices. This is the reason why General Relativity (GR) with a cosmological constant is the only purely metric theory of gravity having non-vanishing covariant second order field equations in four dimensions, because at zeroth order in curvature, $\mathcal{L}_0=1$ and $\mathcal{G}_{\phantom{(p)} \beta}^{(0) \alpha}= -\delta^\alpha_\beta /2$ while at first order, $\mathcal{L}_1=R$ and $\mathcal{G}_{\phantom{(p)} \beta}^{(1) \alpha} =G_\beta^\alpha$ which is the Einstein tensor.
\\

Nonetheless, there has recently been a renewed interest in the so-called ``regularization" of Lovelock-Lanczos gravities in four dimensions. Initially, it was noted by Tomozawa in \cite{Tomozawa:2011gp} that when evaluated on a static spherically symmetric ansatz, the field equation tensor associated with the Gauss-Bonnet scalar $\mathcal{L}_2 = R^2-4 R_{\mu\nu}R^{\mu\nu}+R^{\mu \nu}_{\sigma \rho}R_{\mu \nu}^{\sigma \rho}$ can be written as $\mathcal{G}_{\phantom{(2)}}^{(2)} = (d-4) \mathscr{G}^{(2)}$, for a well-defined static spherically symmetric tensor $\mathscr{G}^{(2)}$, in any dimension $d>2$. It means that the effect of such symmetric reduction of the field equation is that the \textit{algebraic} vanishing of this tensor is being replaced by an \textit{analytic} factor of the dimension. It was therefore suggested that the limit as $d\to 4$ of the tensor $\mathcal{G}_{\phantom{(2)}}^{(2)} / (d-4)$ is well-defined \textit{after being evaluated on such metric field}. This result was later extended to the Friedmann-Lema\^{i}tre–Robertson-Walker (FLRW) cosmological ansatz in \cite{Cognola:2013fva}, where it was also shown that such model yields a logarithmic correction to the gravitational entropy, similarly to some quantum gravity results, see e.g. \cite{Carlip} and \cite{PerezNoui}. Such peculiar property indicates a kind of dimensional universality of these sectors of Gauss-Bonnet gravity and might also be related to the conformal anomaly, as the same minisuperspace field equations appear in that context, see \cite{conformal}.

It is only recently that some improvement on that topic appeared, as it was shown in \cite{Glavan:2019inb} that after being perturbed linearly around (A)dS$_d$ vacua, the vanishing of this tensor once again becomes analytic due to a common factor $(d-4)$ in front of all its components. Around the same time, we showed in our thesis \cite{Aim} that this feature of the Gauss-Bonnet scalar in spherical symmetry and cosmology is shared by all the LLG series (for any curvature order $p$), so that within this class of metric fields, non-perturbative four-dimensional LLG can be considered. In particular, following \cite{Maeda1, Maeda2}, we showed that many cosmological and black hole solutions without singularities can be found in this way, as we will also see in this paper. This is, we think, one of the main interest of these models, as the presence of singularities within General Relativity is one of its main self-inconsistency as a physical theory. Finally, together with our co-authors, we showed in \cite{AMS} that the result at linear perturbative order is also preserved for higher (than Gauss-Bonnet) Lovelock-Lanczos scalars and also found some new non-perturbative regular black holes from this approach.  During the later stages of this work, a paper with similar results regarding singularity avoidance appeared \cite{nonsing}. 
\\

In order to understand this regularization procedure, it is important to see why it is working for specific classes of metric fields $\bar{g}$. There are two necessary ingredients to extract the \textit{analytic} factors of the dimension from the \textit{algebraic} vanishing of the LL field equations. First, the dimensional factors arise from the following properties of the GKD :
\begin{eqnarray}
\delta_{\nu_1 \dots \nu_s \nu_{s+1} \dots \nu_q}^{\mu_1 \dots \mu_s \mu_{s+1} \dots \mu_q} \delta^{\nu_{s+1} \dots  \nu_q}_{\mu_{s+1} \dots  \mu_q} = (q-s)! \frac{(d-s)!}{(d-q)!} \delta^{\mu_1 \dots \mu_s}_{\nu_1 \dots \nu_s} \;,   \quad \delta^{\mu_1 \dots \mu_s \mu_{s+1} \dots \mu_q}_{\nu_1 \dots \nu_s \mu_{s+1} \dots \mu_q} = \frac{(d-s)!}{(d-q)!} \delta^{\mu_1 \dots \mu_s}_{\nu_1 \dots \nu_s}\,. \label{GdeltaProp}
\end{eqnarray}
Using this relation, one finds that the only quantities that are regularizable in critical dimension $d \to 2p$ and in a background independent way are the traces of the Lovelock tensors :
\begin{eqnarray}
\mathcal{G}_{\phantom{(p)} \sigma}^{(p) \sigma} = - \frac{1}{2} \left( d-2p \right) \mathcal{L}_p \label{RegTrace}\,.
\end{eqnarray}
 Secondly, the $d$-dimensional metric ansatz $\bar{g}^{(d)}$ should yield the necessary (but not sufficient) condition that, when evaluated on such metric, \textit{some} components of the curvature can be written as $R_{ij}^{kl}(x) =  \phi(x) \delta_{ij}^{kl} +\psi_i^k(x) \delta_{j}^{l} + K_{ij}^{kl}(x)$, where $\phi, \psi, K$ only involve derivatives of the metric, i.e. that the tensorial and functional structures of these components of the Riemann tensor must split, for $i,j,k,l$ belonging to some subset of $\left\{ 0, \dots, d-1 \right\}$. It is then possible to use the properties \eqref{GdeltaProp} of the GKD to extract some dimensional factors, as we will see throughout this paper.

 Furthermore, to the question whether it is possible or not to regularize the Gauss-Bonnet theory or more generally LLG in critical or lower dimensions for general metric fields and end up with new covariant $D$-dimensional theories leading to second order field equations \footnote{I.e. to end up with a new rank-two, divergence-free, $D$-dimensional (covariant) tensor depending solely on a \textit{general metric} at most through terms like $\left( \partial \partial g\right)^p$.} : the answer is negative, as it must be clear from Eq\eqref{GdeltaProp} applied to a Weyl decomposition of Eq(\ref{NonPerturbativeFieldEquations2}), from the Lovelock theorem stated above, from the discussions of \cite{TekinGB,Yetanother,HornGB3} or from simple counter-examples like the Bianchi I sector of Gauss-Bonnet gravity \cite{DeruelleBianchi}. Therefore, this approach of finding ansatz from which overall dimensional factors can be extracted from the field equations, that we will call ``background regularization", can only yield to new minisuperspace models.
 \\

However, it was recently realized that the results of \cite{Tomozawa:2011gp, Cognola:2013fva, Glavan:2019inb} can be recovered from a well-defined $4$D covariant Horndeski scalar-tensor theory arising from two different kinds of regularization of the Gauss-Bonnet theory. The first one, found in \cite{HornGB1,HornGB2}, is based on the Kaluza-Klein reduction of Lovelock-Lanczos gravity (see \cite{KKLLG, Charmousis}), in which one evaluates the LLG action on a warped $(d=D+n)$-dimensional metric $g^{(d)}$ containing a metric $g^{(D)}$ and a warped ``radius" scalar field $\phi$, both depending solely on the coordinates of the $D$-dimensional sub-manifold. The $n$-dimensional space was considered maximally symmetric in \cite{HornGB1,HornGB2}. After integrating out the angular coordinates, the action ends up reducing to a $D$-dimensional Horndeski scalar-tensor theory. Discarding the boundary term and dividing by $(d-D)$, the limit $d\to D$ is then well-defined. However, as noted in \cite{ConfGB2}, this procedure is highly non-unique due to the large freedom in parametrizing the metric components associated with the higher dimensions.

Surprisingly, providing that the maximally symmetric space is flat, this procedure agrees with the Mann-Ross conformal regularization, initially applied to $2$D General Relativity in \cite{GRD=2} and extended to the Gauss-Bonnet case in \cite{ConfGB1,ConfGB2}, see also \cite{ConGB12}. As we will see in the present paper, this regularization starts by applying a conformal transformation to LLG, from which dimensional factors arise. Provided that suitable counter-terms be added to such theory, it is then possible to introduce dimensional poles and take a well-defined limit resulting in a unique $4$D Horndeski theory, whose observational constraints have been discussed in \cite{ClifClif}. It is noticeable that this theory has also been found in \cite{RenGroupAnomCCRGB} in the context of renormalization group flows and trace anomalies, see also \cite{Riegert}.

That being said, this theory was shown to suffer from strong coupling around flat space in \cite{HornGB2} and \cite{Amplitudes}, meaning that its Lovelock branch should not be dynamically reachable from more general solutions of the theory, see also \cite{StrongCouplingQTG} for a similar issue in the context of Quasi-Topological gravities \cite{Quasi-Top1,EinCubic,Quasi-Top2,GQTEffAct1,GQTEffAct2,1703.01631}, which share many features with the Lovelock branch of the conformal critical regularization of Gauss-Bonnet gravity.

For this reason, a four dimensional metric theory propagating solely the two degrees of freedom of the graviton but invariant under $3$D-diffeomorphisms only was extracted from a regularization of the Gauss-Bonnet theory in the recent paper \cite{3Dcov1}, see also \cite{3Dcov2,3Dcov3} for its application to gravitational waves and inflation. The idea in this case was to consider the peace of the Gauss-Bonnet Hamiltonian that do not contain the Weyl parts of the $3$D Riemann curvature and quadratic product of extrinsic curvature tensors. By definition, that peace is proportional to $(d-4)$ and can thus be regularized. In some sense, such theory is even closer to ($d$-dimensional) GR than LLG because this former is the only LLG theory without Weyl parts. These metric theories with the same degrees of freedom as GR belong more generally to the class of Minimally Modified gravity, see for example \cite{MMG1,MMG2,MMG3,MMG4,MMG5}. 
\\

A remarkable feature of all these ``regularized" models is that, when evaluated on FLRW or spherical symmetry, the resulting field equations admit a branch which agrees with the background regularization of Gauss-Bonnet gravity (for some specific configurations of the extra scalar field in the conformal case), meaning that these theories really preserves \textit{some of the properties and solutions} that are usually restricted to higher dimensions. Remark also that another $3$D-covariant model able to reach the spherically symmetric sector of Gauss-Bonnet gravity while preserving the two degrees of freedom of General Relativity was discovered some years ago in the context of Ho\v{r}ava-Lifshitz gravity \cite{Horava}. Together with the mentioned relations with conformal anomaly, logarithmic corrections to the entropy and renormalization group flows, this might suggest possible relations between these regularized theories and quantum gravity.

Given this already substantial zoology of regularizations of LLG, our aim is to review and extend some of these results to arbitrary curvature order $p$ : at perturbative level around (A)dS vacua, via metric transformations, in particular the conformal one, via the breaking of $4$D-diffeomorphisms down to $3$D ones, and at the background level. 
\\

This paper is organized as follows. In Sec.1, the dimensional dependence of the perturbative expansion of LLG around AdS$_d$ vacua is understood. From this, we find a class of fine-tuned Lovelock-Lanczos gravity for which the regularization is well-defined up to an arbitrary (perturbative) order $m$ around one of these vacua.

More precisely, we show that it is possible to cancel all the terms that remain irreducibly identically vanishing from second order up to $m^{\text{th}}$ order in perturbations, providing that the coupling constants $\alpha_p$, for $2 \leq p \leq m$, are chosen in a specific form, the simplest one being close to that of LL-Born-Infeld, LL-Chern-Simons and more generally to LLG with a unique vacuum (see for example \cite{ZanelliBICS, Zanelli,Lovelock-Born-Infeld1,Lovelock-Born-Infeld2,Lovelock-Chern-Simons}). If well-defined, these theories might be interpreted as effective field theories (trajectories in theory space) of gravitons on (A)dS$_4$, as their coupling constants must be running with the perturbative order $m$ above which the regularization breaks down. By construction, they have the same perturbative expansion as General Relativity (up to order $m$), modulo a shift of its coupling constants and possibly different vacua. 
\\

In Sec.2, the regularization of LLG from a general metric transformation is considered and the resulting $4$D covariant theory is obtained. Some subtleties of this approach are addressed. For instance, we make a distinction between the critical ($d\to 2p$) regularizations and lower ones, such as the limit $d\to4$ of the cubic Lovelock theory. In the conformal case, the resulting models belong to Horndeski theory, but only the first ones are shift-invariant.  The (critical) conformal regularization obtained in the literature for the Gauss-Bonnet case is generalized to the whole Lovelock series and the four dimensional metric and scalar field equations are obtained.
\\

In Sec.3, we first review the ADM formulation of LLG via its covariant form, found by  Horndeski in \cite{HorndeskiVector}, from which we give a direct derivation of the Hamiltonian formalism of LLG, originally found in \cite{TeitZan}. These formulations are sufficient to investigate possible regularizations of LLG given that they split the LLG action into a boundary term and an identically vanishing part in dimensions $d\leq 2p$. After a brief discussion regarding their possible generalizations to the case of null boundaries, which in principle could also be regularized, we then focus on the generalizations of the $3$D-covariant regularization of the Gauss-Bonnet theory \cite{3Dcov1} for arbitrary curvature order $p$. It is shown that the regularizations of the Hamiltonian and ADM Lagrangian are inequivalent, although both seems to yield $3$D-covariant metric theories propagating solely the two degrees of freedom of the graviton. 
\\

Finally, the Sec.4 is devoted to the study of the background regularizations. We first review the existing results (mainly taken from \cite{Maeda}) about the Regularization of LLG on $2$D-covariant dynamical spherically symmetric spacetimes (DSS) (which includes curved FLRW and static spherically symmetric metric fields), in particular the so-called Unified first law of thermodynamics, the black hole solutions and the associated logarithmic correction to their entropy, as well as the singularities at their center. It is shown that the regularization of the field equations is well-defined at background level, but breaks down at respectively first and second order perturbations around static spherically symmetric and (curved) FLRW geometries, as expected from Lovelock theorem. Two well-defined regularized Lagrangian are found, generalizing the findings of \cite{ActionRLLG}.

Simple $4$D regularized models are presented, involving non-perturbative curvature corrections to the EH action (i.e. $t\to \infty$ in Eq\eqref{NonPerturbativeFieldEquations}). One of these yields a cosmological generalization of the well-known non-singular Hayward black hole \cite{Hayward_1} (see \cite{Maeda1, Maeda2} where it was originally found within this approach), as well as a non-singular flat cosmology, behaving as dS$_4$ in the infinite past and transitioning smoothly to the GR regime. For closed FLRW cosmology, the solution can either describe non-singular cyclic or bouncing universes. It is argued that, although not necessary, this kind of models admitting regular solutions are quite common within this approach.

Finally, we investigate the non-uniqueness of the background regularization. We remark that the Bianchi I sector of LLG cannot be regularized \textit{as it is}, what was also noted in \cite{BianchiIGB} in the Gauss-Bonnet case, and a solution to this issue is provided, based on a ``periodization" of the $(d-1)$-dimensional spatial metric into $3$D (or less) repeated building blocks. We explicitly show that different choices can be made, resulting in different well-defined $4$D limits of Bianchi I LLG. Only the less symmetric choices are in agreement with the Kantowski-Sachs sector arising from dynamical spherically symmetric reduction, while the more symmetric one has a weaker dimensional universality, existing only in $d=1+3n$ dimensions. In the later stages of this work, we became aware of a similar result \cite{BianchiIGB2} regarding the Bianchi I sector of Gauss-Bonnet gravity.

\section{Perturbative Lovelock-Lanczos Gravity}

Our starting point to assess the possibility of regularizing LLG in four dimensions will be to understand its dimensional dependence when it is expanded into perturbations around its maximally symmetric (A)dS vacua with metric $\bar{g}$. In this case, the components of the background Riemann tensor reduce to :
\begin{equation}
R^{\mu \nu}_{\sigma \rho}\left[  \bar{g} \right]  = \Lambda \delta^{\mu \nu}_{\sigma \rho} \, .
\label{RiemannAdS}
\end{equation}{}
Considering a general perturbative expansion of the metric field around these vacua
\begin{eqnarray}
g_{\mu\nu}=\bar{g}_{\mu\nu}  + h_{\mu\nu} + h_{\mu\alpha}h^{\alpha}_\nu + \dots + \left[ h^n \right]_{\mu\nu} + \dots  
\end{eqnarray}
allows to write the full field equations Eq\eqref{NonPerturbativeFieldEquations} of LLG as
\begin{eqnarray}
\mathcal{G}^{\mu}_{\nu} \big[ g_{\mu\nu}  \big] = \sum_{p=0}^t \sum_{n=0}^\infty  \alpha_p \, \mathcal{G}^{(p)\mu}_{(n)\nu}\left[ h^n  \right]= \sum_{n=0}^\infty \sum_{p=0}^t   \alpha_p \, \mathcal{G}^{(p)\mu}_{(n)\nu}\left[  h^n \right]  \label{FieldEqExp} \,,
\end{eqnarray}
where by definition,
\begin{eqnarray}
 \mathcal{G}^{(p)\mu}_{(n)\nu} :=  -\frac{1}{2^{p+1}} \delta^{\mu \mu_1 \nu_1 \dots \mu_p \nu_p}_{\nu \sigma_1 \rho_1 \dots \sigma_p \rho_p}\left[ \prod  _{r=1}^p   R_{\mu_r \nu_r}^{\sigma_r \rho_r}  \right]^{[n]}   \label{GpnDef} \,,
\end{eqnarray} 
where the notation $\left[ K \right]^{[n]}$ corresponds to the $n^{\text{th}}$ order term in the perturbative expansion of a tensor $K$. In order to have a grasp on the tensorial structure of $\mathcal{G}^{(p)}_{(n)} $, note that we have for some integer $q$,
\begin{eqnarray}
\begin{split}
 \delta^{\mu \nu \alpha_1 \dots \alpha_q}_{\sigma \rho \beta_1 \dots \beta_q} \left[R_{\mu \nu}^{\sigma \rho}\right]^{[1]} =&-2  \delta^{\mu \nu \alpha_1 \dots \alpha_q}_{\sigma \rho \beta_1 \dots \beta_q} \left( \nabla_{\mu} \nabla^{\sigma} + \delta_{\mu}^{\sigma} \Lambda \right)h_{\nu}^{\rho}\,, \\
  \delta^{\mu \nu \alpha_1 \dots \alpha_q}_{\sigma \rho \beta_1 \dots \beta_q} \left[R_{\mu \nu}^{\sigma \rho}\right]^{[2]} =&\, \delta^{\mu \nu \alpha_1 \dots \alpha_q}_{\sigma \rho \beta_1 \dots \beta_q} \Big( 2 \Lambda  h_\mu^\gamma h_\gamma^\sigma \delta_\nu^\rho + \nabla^\gamma h_\nu^\rho \left( \nabla_\mu h_\gamma^\sigma +\nabla^\sigma h_{\gamma\mu} \right)+ 2 h^{\gamma \sigma} \left( \nabla_\mu  \nabla_\gamma h_\nu^\rho + \nabla_\nu  \nabla^\rho h_{\gamma \mu} \right)\\
  &\phantom{ \delta^{\mu \nu \alpha_1 \dots \alpha_q}_{\sigma \rho \beta_1 \dots \beta_q} \Big(}+ \frac{1}{2} \left( \nabla_\gamma h^\sigma_\nu \nabla^\gamma h^\rho_\mu + \nabla_\mu h_\gamma^\rho \left( \nabla_\nu h^{\gamma\sigma} + 2  \nabla^\sigma h^{\gamma}_\nu \right) + \nabla^\sigma h_{\gamma \nu} \nabla^\rho h^\gamma_\mu \right)  \Big) \,,
 \end{split}
\end{eqnarray}
with the complexity quickly rising with the expansion order. Let us first isolate the zero and first order perturbations from Eq\eqref{FieldEqExp}, corresponding respectively to the equations determining the vacua and linear dynamics of the theory. As $\mathcal{G}_{\phantom{(0)} \nu}^{(0) \mu} =-\frac{1}{2} \delta^\mu_\nu$, this term contributes only at zero-order : $\mathcal{G}_{(n)\nu}^{(0) \mu} =0$ for $n>0$. Therefore, dropping the indices for simplicity gives
\begin{eqnarray}
\mathcal{G}  =\sum_{p=0}^t \alpha_p  \mathcal{G}^{(p)}_{(0)} +
\sum_{n=1}^\infty   \sum_{p=1}^t   \alpha_p \, \mathcal{G}^{(p)}_{(n)} \,, \label{GFull}
\end{eqnarray}
where the first two terms are well-known (see e.g. \cite{PerturbLLG} and \cite{AMS}) and given by
\begin{eqnarray}
\sum_{p=0}^t \alpha_p  \mathcal{G}^{(p)}_{(0)}=\left( \sum_{p=0}^t \alpha_p \frac{(d-1)!}{(d-2p-1)!} \Lambda^p \right)  \mathcal{G}^{(0)}_{(0)} \; , \quad \sum_{p=1}^t  \alpha_p  \mathcal{G}^{(p)}_{(1)}    = \left( \sum_{p=0}^t p \alpha_p \frac{(d-3)!}{(d-2p-1)!} \Lambda^{p-1} \right)  \mathcal{G}^{(1)}_{(1)} \,.
\end{eqnarray}
That is, all the higher order (in curvature) Lovelock-Lanczos terms collapse to $ \mathcal{G}^{(0)}_{(0)}$ and $ \mathcal{G}^{(1)}_{(1)}$ at zero and first order in perturbation, meaning that these terms constitute one-dimensional basis at these orders. As we will see, it is a general feature of Lovelock-Lanczos gravity that at a given perturbative order $n$, all the terms of curvature-order $p>n$ can be decomposed into the $p\leq n$ ones. 

\subsection{Dimensionality and regularizations}

In order to do so, we have to understand the general dimensional structure of $\mathcal{G}^{(p)\mu}_{(n)\nu}$. Dropping the indices and representing the free indices of the GKD as $\Bigcdot$, we rewrite Eq\eqref{GpnDef} in the following way :
\begin{eqnarray}
 \mathcal{G}^{(p)}_{(n)} :=  -\frac{1}{2^{p+1}} \delta^{\Bigcdot 1 \dots p}_{\Bigcdot 1 \dots p}\left[R_1 \dots R_p \right]^{[n]} =-\frac{1}{2^{p+1}} \delta^{\Bigcdot 1 \dots p}_{\Bigcdot 1 \dots p} \sum_{k=1}^p \binom{p}{k} R^{[0]} \dots R^{[0]} \Big\{ R_1 \dots R_k \Big\}^{[n]}\,, \label{Gpn}
\end{eqnarray}
where we extracted all the zeroth order of the Riemann tensor, so that $\Big\{ R_1 \dots  R_k \Big\}^{[n]}$ is defined by :
\begin{eqnarray}
\Big\{ R_1 \dots  R_k \Big\}^{[n]} := \sum_{\sum_{i=1}^k j_i =n \, ; j_i \neq 0} R^{[j_1]} \dots R^{[j_k]} \,, \label{accolades}
\end{eqnarray}
 which is identically vanishing for $k>n$. Expanding the $(p-k)$ terms $R^{[0]}$ with the generalized Kronecker delta, using Eq\eqref{RiemannAdS} and Eq\eqref{GdeltaProp} we obtain the following decomposition :
\begin{eqnarray}
\mathcal{G}^{(p)}_{(n)} = \sum_{k=1}^p u_{(p,k)} \mathcal{K}^{(k)}_{(n)} \; , \quad \; u_{(p,k)} := \binom{p}{k}  \Lambda^{p-k} \frac{(d-2k-1)!}{(d-2p-1)!} \; , \quad \; \mathcal{K}^{(k)}_{(n)} :=  -\frac{1}{2^{k+1}} \delta^{\Bigcdot 1 \dots k}_{\Bigcdot 1 \dots k}\Big\{ R_1 \dots  R_k \Big\}^{[n]}  \,. \label{KBasis}
\end{eqnarray}
It means that the $n^{\text{th}}$ perturbative order of the $p^{\text{th}}$ Lovelock-Lanczos term can be fully decomposed into a finite dimensional basis $\big\{ \mathcal{K}^{(k)}_{(n)} \big\}_{1 \leq k \leq n}$ with coefficients given by a kind of (truncated in the first column due to $\mathcal{G}_{(n)\nu}^{(0) \mu} =0$ for $n>0$) lower-triangular Pascal matrix $u_{(p,k)}$. Thus, $\mathcal{G}^{(p)}_{(n)}$ and $\mathcal{K}^{(k)}_{(n)}$ are related by a kind of binomial transform whose application (see Appendix\ref{BCid}\ref{BinInversion}) allows to consider directly the basis $\big\{ \mathcal{G}^{(k)}_{(n)} \big\}_{1 \leq k \leq n}$ on which the terms $\mathcal{G}^{(p)}_{(n)} $ for $p > n$ decompose as : 
\begin{eqnarray}
\mathcal{G}^{(p)}_{(n)} = \sum_{k=1}^{n} \gamma_{(p,n,k)} \Lambda^{p-k} \frac{(d-2k-1)!}{(d-2p-1)!}  \mathcal{G}^{(k)}_{(n)} \,, \;\;\;  \gamma_{(p,n,k)}:= (-1)^{n+k} \binom{p}{k}\binom{p-k-1}{p-n-1} \,. \label{DecompPerturbationLLG}
\end{eqnarray}
For example,
\begin{eqnarray}
\begin{split}
&\mathcal{G}^{(3)}_{(2)}= -3  (d-3)\dots (d-6) \Lambda^2  \mathcal{G}^{(1)}_{(2)} + 3 (d-5)(d-6) \Lambda \mathcal{G}^{(2)}_{(2)} \,,\\
&\mathcal{G}^{(4)}_{(2)}= -8 (d-3)\dots (d-8) \Lambda^3 \mathcal{G}^{(1)}_{(2)} + 6 (d-5)\dots(d-8) \Lambda^2 \mathcal{G}^{(2)}_{(2)} \,. \label{RelationsLLPertu}
\end{split}
\end{eqnarray}
It is from these kinds of identities that one can hope to fine-tune the coupling constants of LLG in order to suppress the terms for which the factor $(d-4)$ does not come out. As we see, a necessary condition to do so is to have $\Lambda \neq 0$, i.e. that vacuum cannot be a Minkowski geometry. A simple reason explaining why it is working this way is that for any couple of positive integers $0 \leq i < j$, we have
\begin{eqnarray}
\left[  W^{\mu \nu}_{\sigma \rho} \right]^{[0]}  =0 \;\; \implies \;\;\; \left[ \prod_{q=1}^j W^{\mu_q \nu_q}_{\sigma_q \rho_q}  \right]^{[i]} = 0\,,
\end{eqnarray} 
where $W$ is the Weyl tensor. Thus, if the Lovelock tensors are decomposed in a Weyl basis (see Eq\eqref{WeylDecompLovelock}), the above Weyl products, which do not contain Kronecker deltas able to produce dimensional factors when contracted with the GKD, will disappear from the perturbative expansion around (A)dS vacua. 
\\

Back to the full pertubative expansion Eq\eqref{GFull}, we first set,
 \begin{eqnarray}
 \alpha_p =  \frac{(d-2p-1)!}{(d-1)!}\; \tilde{\alpha}_p \,,  \label{Regularization}
 \end{eqnarray}
so that 
  \begin{eqnarray}
 \begin{split}
 \mathcal{G} - \left( \sum_{p=0}^t \tilde{\alpha}_p  \Lambda^p \right)  \mathcal{G}^{(0)}_{(0)}  &= \sum_{n=1}^\infty \sum_{k=1}^n \left( \sum_{p=k}^{t} \tilde{\alpha}_p \binom{p}{k}  \Lambda^{p-k}  \right) \mathcal{K}^{(k)}_{(n)} \frac{(d-2k-1)!}{(d-1)!}  \\
 &=   \sum_{n=1}^\infty \sum_{k=1}^n   \left( \tilde{\alpha}_k +\sum_{p=n+1}^{t}   \tilde{\alpha}_p \, \gamma_{(p,n,k)} \Lambda^{p-k}  \right)  \mathcal{G}^{(k)}_{(n)}\frac{(d-2k-1)!}{(d-1)!} \,. \label{RLLGFieldEq}
 \end{split}
 \end{eqnarray}
Meaning that the effect of this redefinition is to factorize the dimensional factors, making them dependant on the label $k$ of the basis, instead of on the curvature order $p$. Now recall that both $\mathcal{K}^{(k)}_{(n)}$ and $\mathcal{G}^{(k)}_{(n)}$ are of the form 
\begin{eqnarray*}
\mathcal{H}^{(k)\mu}_{(n)\nu} = -\frac{1}{2^{k+1}}  \delta^{\mu \mu_1 \nu_1 \dots \mu_k \nu_k}_{\nu \sigma_1 \rho_1 \dots \sigma_k \rho_k} \mathcal{R}^{(k)\sigma_1 \rho_1 \dots \sigma_k \rho_k}_{(n)\mu_1 \nu_1 \dots \mu_k \nu_k}\,,
\end{eqnarray*}
so that they identically vanish for $d\leq2k$, by definition of the GKD. Thus, the following quantities
\begin{eqnarray}
\tilde{\mathcal{H}}^{(k)\mu}_{(n)\nu} := \frac{(d-2k-1)!}{(d-1)!} \mathcal{H}^{(k)\mu}_{(n)\nu} = -\frac{1}{2^{k+1}}  \frac{\delta^{\mu \mu_1 \nu_1 \dots \mu_k \nu_k}_{\nu \sigma_1 \rho_1 \dots \sigma_k \rho_k}}{(d-1)(d-2)\dots(d-(2k-1))(d-2k)} \mathcal{R}^{(k)\sigma_1 \rho_1 \dots \sigma_k \rho_k}_{(n)\mu_1 \nu_1 \dots \mu_k \nu_k} \,, \label{tilde}
\end{eqnarray}
yields an indeterminate form $0 \times \frac{\mathcal{R}}{0}$ as one takes naively any of the limits $d\to 1, 2, \dots 2k$.  This is a necessary condition to start considering the limit $d\to 4$.

Up to now, four dimensional regularizations of perturbative Lovelock-Lanczos gravities have restricted to the first order case \cite{Glavan:2019inb, AMS}. From Eq\eqref{RLLGFieldEq}, it amounts to write :
\begin{eqnarray*}
\mathcal{G} = \left( \sum_{p=0}^t \tilde{\alpha}_p  \Lambda^p \right)  \mathcal{G}^{(0)}_{(0)} +  \left( \sum_{p=0}^t p \tilde{\alpha}_p \Lambda^{p-1} \right) \frac{\mathcal{G}^{(1)}_{(1)}}{(d-1)(d-2)} + O\left( h^2 \right) \,,
\end{eqnarray*}
which is well-defined in any dimensions $d>2$, if the higher order perturbations are discarded before setting the value of the dimension. Without a definite way to deal with these higher orders, such procedure is of course problematic and more care is needed when they are taken into account. From the full perturbative expansion of LLG Eq\eqref{RLLGFieldEq}, one sees that there are two ways to do so.

The first one uses the relations like Eq\eqref{RelationsLLPertu}. For example, \textit{after} being evaluated at $n^{\text{th}}$-order in perturbation around the (A)dS$_d$ vacuum $\Lambda$, i.e. on a metric  $g_{\mu\nu} = \bar{g}_{\mu\nu} + h_{\mu\nu} + \dots + \left[ h^{n} \right]_{\mu\nu}$, the following combinations of Lovelock tensors are well-defined in the limit $d\to 4$, for respectively $n=1,2, n=1,2,3$ and $n=1,2,3,4$ : 
\begin{eqnarray}
\begin{split}
&\frac{1}{d-4} \left(-3 \Lambda \mathcal{G}^{(2)} +\frac{\mathcal{G}^{(3)}}{(d-5)(d-6)}\right)\\
&\frac{1}{d-4} \left(6 \Lambda^2 \mathcal{G}^{(2)}-\frac{4 \Lambda \mathcal{G}^{(3)}}{(d-5)(d-6)} +\frac{\mathcal{G}^{(4)}}{(d-5)\dots(d-8)} \right)\\
&\frac{1}{d-4} \left(-10 \Lambda^3 \mathcal{G}^{(2)}+\frac{10\Lambda^2 \mathcal{G}^{(3)}}{(d-5)(d-6)}-\frac{5 \Lambda \mathcal{G}^{(4)}}{(d-5)\dots(d-8)} + \frac{\mathcal{G}^{(5)}}{(d-5)\dots(d-10)} \right)  , \,\, \text{etc} \dots \label{ExampleReg}
\end{split}
\end{eqnarray}
If this regularization procedure is to be interpreted as usual dimensional regularization in a (perturbative and background dependent) quantum theory of gravity, it might be interesting to put these results in perspective with \cite{OneLoop}\footnote{This reference was provided to us and others by Bayram Tekin, that we thank for it.}. It was noticed there (quite a long time ago) that the dimensional regularization of the Gauss-Bonnet scalar is ill-defined around flat space-time. However, we see that these specific combinations of LLG tensors seem to be well-defined in dimensional regularization around (A)dS$_4$, up to a finite order in perturbation theory if the combinations of LLG terms contain a finite number of terms, or up to infinite order if non-perturbative LLG theories are considered. Thus, if the LL terms are to be interpreted as loop corrections preserving the second order character of the field equations, these theories are such that any new loop correction just collapses to General Relativity-like perturbative terms.

The second one is directly to make sense of the quantities $\tilde{\mathcal{H}}$ for these values of $d$, i.e. to find a way to extract these analytic dimensional factors, so that, if we want a $(d\to D)$-regularization, we need : 
\begin{eqnarray*}
\delta^{\mu \mu_1 \nu_1 \dots \mu_k \nu_k}_{\nu \sigma_1 \rho_1 \dots \sigma_k \rho_k}\mathcal{R}^{(k)\sigma_1 \rho_1 \dots \sigma_k \rho_k}_{(n)\mu_1 \nu_1 \dots \mu_k \nu_k}= (d-D) \mathcal{Z}^{(k)\mu}_{(n)\nu}\,, 
\end{eqnarray*}
where $\mathcal{Z}$ would be a well-defined tensor without dimensional poles.  If this is possible, 
\begin{eqnarray*}
\tilde{\mathcal{H}}^{(k)\mu}_{(n)\nu}  =  -\frac{1}{2^{k+1}}  \frac{(d-D)}{(d-1)(d-2) \dots (d-D) \dots (d-(2k-1))(d-2k)}  \mathcal{Z}^{(k)\mu}_{(n)\nu}  
\end{eqnarray*}
would be well-defined in the limit $d\to D$. As mentioned in the introduction, this is what has first been done in the two-dimensional case and in a background independent way in \cite{GRD=2} by Mann and Ross, by using a conformal rescaling of the two-dimensional metric of the Einstein-Hilbert action. We will further discuss this possibility in the next section of this paper, so we stop here on that option.

\subsection{Effective regularization by fine-tuning}

As exemplified in Eq\eqref{ExampleReg}, it is possible to use the relations given by Eq\eqref{DecompPerturbationLLG} to fine-tuned the coupling constants $\tilde{\alpha}_p$ of LLG in order to cancel the dimensional poles $1/(d-4)$ up to an arbitrary order $m \leq t-1$ in perturbation around a given vacuum $\Lambda$. As we are interested in the four-dimensional case, it will be convenient to rewrite the perturbative LLG tensor of Eq\eqref{RLLGFieldEq} in the following way : 
 \begin{eqnarray}
 \begin{split}
 &\mathcal{G} - \left( \sum_{p=0}^t \tilde{\alpha}_p  \Lambda^p \right)  \mathcal{G}^{(0)}_{(0)} -  \sum_{n=1}^\infty  \left( \tilde{\alpha}_1 +\sum_{p=n+1}^{t}   \tilde{\alpha}_p \, \gamma_{(p,n,1)} \Lambda^{p-1}  \right)  \frac{\mathcal{G}^{(1)}_{(n)}}{(d-1)(d-2)} \\
 =&    \sum_{n=2}^m \sum_{k=2}^n   \left( \tilde{\alpha}_k +\sum_{p=n+1}^{t}   \tilde{\alpha}_p \, \gamma_{(p,n,k)} \Lambda^{p-k}  \right)  \mathcal{G}^{(k)}_{(n)} \frac{(d-2k-1)!}{(d-1)!}\\
 +&   \sum_{n=m+1}^\infty \sum_{k=2}^n   \left( \tilde{\alpha}_k +\sum_{p=n+1}^{t}   \tilde{\alpha}_p \, \gamma_{(p,n,k)} \Lambda^{p-k}  \right)  \mathcal{G}^{(k)}_{(n)} \frac{(d-2k-1)!}{(d-1)!} \,.\label{RLLGFieldEq2}
 \end{split}
 \end{eqnarray}
Let's now consider the class of Lovelock-Lanczos theories such that for any integer $2 \leq k \leq m$, 
 \begin{eqnarray}
 \begin{split}
 \tilde{\alpha}_k = - \sum_{p=m+1}^t \tilde{\alpha}_p \gamma_{(p,m,k)} \Lambda^{p-k} \iff  \alpha_k =  \sum_{p=m+1}^t \alpha_p  \, \Lambda^{p-k} (-1)^{m+k+1} \binom{p}{k}\binom{p-k-1}{p-m-1}  \,\frac{(d-2k-1)!}{(d-2p-1)!} \label{FineTuning} 
 \end{split}
 \end{eqnarray}
while the coupling constants for $m<k\leq t$ remain arbitrary for the moment. Notice that the constant $\Lambda$ appearing in these coupling constants has to be chosen to be one of the vacua of the theory, i.e. it is a function of $\tilde{\alpha}_0, \tilde{\alpha}_1$ and the remaining coupling constants $\tilde{\alpha}_k$ for $m<k\leq t$.

The effect of this choice of coupling constants is that the first term on the RHS of Eq\eqref{RLLGFieldEq2} vanishes. As this term contains all the ``non-regularizable" perturbations from order $O\left( h^2 \right)$ up to $O\left( h^m \right)$, it reduces the RHS of that equation to be $O\left( h^{m+1} \right)$, for the \textit{arbitrary integer} $m \leq t-1$. Indeed, for $2 \leq k \leq n \leq m \leq t-1$ we obtain,
\begin{eqnarray}
  \tilde{\alpha}_k +\sum_{p=n+1}^{t}   \tilde{\alpha}_p \, \gamma_{(p,n,k)} \Lambda^{p-k} = \sum_{p=m+1}^t \tilde{\alpha}_p \Lambda^{p-k} \left[ \gamma_{(p,n,k)}-\gamma_{(p,m,k)}- \sum_{q=n+1}^m \gamma_{(p,m,q)}\gamma_{(q,n,k)} \right]=0 \,, \label{EqGamma}
\end{eqnarray}
because the term inside the brackets is vanishing, as shown in the Appendix\ref{BCid}\ref{gammarelation}.

With this choice of couplings, the non-perturbative and background independent field equations \eqref{NonPerturbativeFieldEquations} becomes :
\begin{eqnarray}
\begin{split}
\mathcal{G}^\mu_\nu =&- \frac{1}{2}\tilde{\alpha}_0 \delta^{\mu}_\nu + \frac{\tilde{\alpha}_1}{(d-1)(d-2)} G^\mu_\nu \\
&+ \sum_{k=m+1}^t  \tilde{\alpha}_k \left\{ \mathcal{G}^{(k)\mu}_\nu - \sum_{q=2}^m \mathcal{G}^{(q)\mu}_\nu \Lambda^{k-q}  \gamma_{(k,m,q)} \frac{(d-2q-1)!}{(d-2k-1)!} \right\}  \frac{(d-2k-1)!}{(d-1)!}  = \kappa_0 \, T^\mu_\nu \,. \label{RegularizedPerturbativeLLG}
\end{split}
\end{eqnarray}
 Using the relation \eqref{EqGamma} in the LHS of Eq\eqref{RLLGFieldEq2} as well, the perturbative expansion of the field equations of this class of LLG theories around the vacuum $\Lambda$ becomes
\begin{eqnarray}
 \begin{split}
 &\mathcal{G}^{\mu}_{\nu} \Big[ \bar{g}_{\mu\nu} + h_{\mu\nu} + h_{\mu\alpha}h^{\alpha}_\nu + \dots + \left[ h^m \right]_{\mu\nu} +\dots  \Big] \\
 =& -\frac{1}{2}  \left\{ \tilde{\alpha}_0 + \tilde{\alpha}_1 \Lambda + \sum_{q=m+1}^t \tilde{\alpha}_q \Lambda^q  \binom{q-2}{m-1} \left(q - \frac{q-1}{m} \right)\left(-1\right)^{m+1} \right\} \, \delta^\mu_\nu \\
 &+\frac{1}{(d-1)(d-2)}    \left\{ \tilde{\alpha}_1 + \sum_{q=m+1}^t \tilde{\alpha}_q \, \Lambda^{q-1} \binom{q-2}{m-1} q  \left(-1\right)^{m+1}   \right\}  \,  \sum_{n=1}^m G^\mu_\nu\left[ h^n \right] + O\left( h^{m+1} \right)\\
 =& \kappa_0 \, T^\mu_\nu \,, \label{DUDUDUDU} 
 \end{split}
 \end{eqnarray}
 where $G^\mu_\nu\left[ h^n \right]$ is the $n^{\text{th}}$ order expansion of the Einstein tensor and all the dimensional poles $1/(d-4)$ are contained into the term
 \begin{eqnarray}
  O\left( h^{m+1} \right) :=  \sum_{n=m+1}^\infty \sum_{k=1}^n   \left( \tilde{\alpha}_k +\sum_{p=n+1}^{t}   \tilde{\alpha}_p \, \gamma_{(p,n,k)} \Lambda^{p-k}  \right) \mathcal{G}^{(k)}_{(n)} \frac{(d-2k-1)!}{(d-1)!} \,.
 \end{eqnarray}
 Therefore, up to the arbitrary order $m$, the tensorial structure of the perturbative expansion of this LLG theory around its vacuum $\Lambda$ is identical to the one of General Relativity, while the coupling constants $\tilde{\alpha}_0$ and $\tilde{\alpha}_1$ are shifted with respect to their GR values.

 Moreover, as the order $(m+1)$ at which the dimensional regularization $d\to 4$ breaks down can be chosen to be arbitrary large, this class of RLLG might constitute a suitable effective field theory (i.e. a trajectory in theory space) of gravitons, propagating in (A)dS$_4$.

 To be more precise, we see from Eq\eqref{RegularizedPerturbativeLLG} that the coupling constants (and so $\Lambda$) must be running with the perturbative order $m$ at which the four-dimensional limit breaks down. Thus,  for a given set of free coupling constants $\tilde{\alpha}_p$ for $p>m$, one could interpret that equation as describing a trajectory of theories $\mathcal{G}\left(t,m\right)$, valid up to order $m$ in perturbation and incorporating an arbitrary number $t$ of ``loop corrections" preserving the second order structure of the field equations. Requiring such regularized theory to be well-defined for any perturbative order $m$ implies to consider non-perturbative curvature corrections $t\to \infty$. However, without a clear way to choose the remaining coupling constants, this might yield to convergence issues.

 Furthermore, it is important to note that, if such procedure is well-defined, the four-dimensional limit should be taken \textit{only after} having evaluated Eq\eqref{RegularizedPerturbativeLLG} at $m^{\text{th}}$ order around its (A)dS$_d$ vacuum $\Lambda$, i.e. at the level of Eq\eqref{DUDUDUDU}. Otherwise, it would yield to indeterminate forms as we said previously. In particular, the perturbative expansions of the corresponding higher dimensional theories $\mathcal{G}\left(t,m\right)$ around all the other vacua are not regularizable, so that it would be desirable to find trajectories of theories admitting a single series of vacua $\Lambda_m$. We will soon discuss this point further.

  \subsection{Born-Infeld-Chern-Simons-like minimal theory}

The simplest trajectories of theories that might be well-defined in four dimensions up to arbitrary perturbative order $m$ is obtained by considering a finite number $(t = m+1)$ of Lovelock-Lanczos invariants which, following the previous possible interpretation, would grow as the theory is probed at higher and higher energies, i.e. as $m \to \infty$. For $2 \leq k \leq m$, the coupling constants of that theory reads,
 \begin{eqnarray}
\tilde{\alpha}_k=  \tilde{\alpha}_{m+1}  \binom{m+1}{k} \left(- \Lambda \right)^{m+1-k}  \,,\label{BICScouplings}
\end{eqnarray}
 where $\tilde{\alpha}_{m+1}$ is the only additional scale with respect to General Relativity.

 \subsubsection{Higher-dimensional theories}
 
Interestingly,  if we would instead consider such couplings for $0 \leq k \leq m$, we would get the class of Lovelock-Lanczos theory with a unique vacuum found in \cite{Zanelli}. In even dimensions and for $m+1=d/2$, this yields to Lovelock-Born-Infeld gravity, while in odd dimensions and $m+1=(d-1)/2$, it would give Lovelock-Chern-Simons gravity, see \cite{ZanelliBICS}.

A direct application of Eqs\eqref{RLLGFieldEq2}\eqref{EqGamma} then proves that the perturbative expansion of this class of higher-dimensional theories is very much degenerate around their (A)dS vacua because the first order at which non-vanishing field equations appear is at $h^{m+1}$, where $m+1$ also corresponds to the highest curvature order appearing in the theory. This might indicate that gravitons are strongly coupled around the (A)dS vacua of these theories.

For example, the quadratic theory given by Eq($19$) in \cite{Zanelli} does not have first order field equations around its unique AdS vacuum : its perturbative expansion starts at order $h^2$. Similarly, considering $t=m+2=4$ and the more general theories Eq\eqref{FineTuning} for $0 \leq k \leq m$ yields to the following higher dimensional LLG : 
\begin{eqnarray}
\begin{split}
I = \frac{1}{2 \kappa_0} \int d^d x \sqrt{-g} \bigg(& - \Lambda^3 \left( \tilde{\alpha}_3 + 3 \Lambda \tilde{\alpha}_4 \right) + \frac{\Lambda^2 \left( 3 \tilde{\alpha}_3 + 8 \Lambda \tilde{\alpha}_4 \right)}{(d-1)(d-2)} R -  \frac{3 \Lambda \left( \tilde{\alpha}_3 + 2 \Lambda \tilde{\alpha}_4 \right)}{(d-1)(d-2)(d-3)(d-4)} \mathcal{L}_2 \\
&+ \frac{\tilde{\alpha}_3}{(d-1)\dots(d-6)} \mathcal{L}_3 + \frac{\tilde{\alpha}_4}{(d-1)\dots (d-8)} \mathcal{L}_4\bigg)\,.
\end{split}
\end{eqnarray}
At background level around an (A)dS geometry given by $R^{\mu \nu}_{\sigma \rho}  = \lambda \delta^{\mu \nu}_{\sigma \rho}$, we obtain the polynomial $\left(\Lambda-\lambda\right)^3 \left(\tilde{\alpha}_3+ \tilde{\alpha}_4 \left( \lambda+3 \Lambda \right)  \right) =0$, so that for $\tilde{\alpha}_4 \neq 0$, the theory admits two vacua given by $\lambda =\{ \Lambda, - \left(\tilde{\alpha}_3+3 \Lambda \tilde{\alpha}_4 \right) / \tilde{\alpha}_4 \}$ and a unique vacuum otherwise. When perturbed at respectively first and second order, its field equation tensor $\mathcal{G}^\mu_\nu$ becomes 
\begin{eqnarray}
\begin{split}
\left[ \mathcal{G}^\mu_\nu \right]^{[1]} &= \left(\Lambda-\lambda\right)^2 \left( 3 \tilde{\alpha}_3 + 4 \tilde{\alpha}_4\left( \lambda + 2 \Lambda \right) \right) \frac{G^\mu_\nu \left[h\right]}{(d-1)(d-2)}\\
\left[ \mathcal{G}^\mu_\nu \right]^{[2]} &=\left(\Lambda-\lambda\right) \left(   \frac{\left( 3 \left( \lambda+\Lambda\right) \tilde{\alpha}_3 + 2 \left(\lambda^2 +\Lambda\lambda + \lambda^2 \right) \tilde{\alpha}_4 \right) G^\mu_\nu \left[h^2\right]}{(d-1)(d-2)} - \frac{3 \left( \tilde{\alpha}_3+ 2 \left(\Lambda+\lambda \right) \tilde{\alpha}_4 \right) \mathcal{G}^{(2)}_{(2)}}{(d-1)(d-2)(d-3)(d-4)} \right)\,.
\end{split}
\end{eqnarray}
Therefore, around the vacuum $\lambda= \Lambda$, the dynamics starts at $O\left( h^3 \right)$, as expected for a theory with $m=2$.
\\

As we see, the effect of the more general couplings \eqref{FineTuning} is to add new vacua to the theory, with the perturbations around the $\Lambda$-one being as degenerate as they are in the unique vacuum theories.  This proliferation of vacua for $t>m+1$ is also to be expected in the case where one imposes the previous couplings for $2 \leq k \leq m$. For a given trajectory $\mathcal{G}\left(t,m\right)$, we want to minimize the number of non-regularizable series of vacua $\lambda_m$, so that the choice Eq\eqref{BICScouplings} is suited for that purpose.

 \subsubsection{Regularized four-dimensional theories}

We now set without loss of generality $\tilde{\alpha}_0= -  \Lambda_0$ and $\tilde{\alpha}_1= 1$. Due to the inverse factor of $(d-1)(d-2)$ in front of the Einstein-Hilbert term (which preserves the symmetry in the dimensionality of each LL terms), the gravitational constant is given by $\tilde{\kappa}= (d-1)(d-2) \kappa_0$. To make the dimension of the coupling $\tilde{\alpha}_{m+1}$ obvious, we set :
\begin{eqnarray}
 \tilde{\alpha}_{m+1} = \frac{\varepsilon_m}{m+1} \left(-  \mathfrak{A} \right)^{m} \,,
\end{eqnarray}
where $\mathfrak{A}$ has dimension of an area and $\varepsilon_m$ dimensionless. From Eqs\eqref{RLLGFieldEq2} and \eqref{BICScouplings}, the (A)dS vacua of these theories defined by $R^{\mu \nu}_{\sigma \rho}  = \lambda \delta^{\mu \nu}_{\sigma \rho}$ are determined by solving :
\begin{eqnarray}
-\Lambda_0 + \lambda +  \frac{\varepsilon_m}{m+1}  \, \left(  \Lambda \mathfrak{A} \right)^{m}  \left( 1- \left( 1+m\right) \frac{\lambda}{\Lambda} - \left( 1- \frac{\lambda}{\Lambda}\right)^{m+1}\right)\Lambda=0 \,, \label{VacgenBI}
\end{eqnarray}
while the ``regularizable vacuum" is obtained by solving the polynomial equation in $\Lambda$, 
\begin{eqnarray}
 \Lambda_0 = \Lambda \left(1 -\frac{m \varepsilon_m}{m+1}  \left( \mathfrak{A} \Lambda \right)^m \right) \,, \label{SolBI}
\end{eqnarray}
which provides its dependence on the free parameters of the theory $\Lambda\left( \Lambda_0,\mathfrak{A} ; m\right)$. If we wish to interpret the present model as a trajectory in theory space, we need to consider a trajectory of vacua $\Lambda_m$ which does not change sign as $m$ increases, exists for all $m$ and, hopefully, which converges to a finite $\Lambda_\infty$ as the number of Lovelock correction becomes non-perturbative $m\to \infty$.

Alternatively, it is simpler to solve Eq\eqref{SolBI} in terms of $\Lambda_0$, in which case Eq\eqref{VacgenBI} reduces to : 
\begin{eqnarray}
\left( \lambda-\Lambda \right)\left( 1 -  \frac{\varepsilon_m}{m+1} \left(\mathfrak{A} \Lambda \right)^m \left(1+m- \left( 1 - \frac{\lambda}{\Lambda} \right)^m\right) \right)=0\,. \label{soll0}
\end{eqnarray}
For $\lambda=\Lambda$, the effective gravitational constant given by the coefficient in front of the perturbed Einstein tensor in Eq\eqref{DUDUDUDU} becomes : 
\begin{eqnarray}
\kappa_{\text{eff}} = \tilde{\kappa} \left( 1 - \varepsilon_m \left( \mathfrak{A} \Lambda \right)^m \right)^{-1} \,.
\end{eqnarray}
After some inspection, one can draw some general conclusions regarding the solutions of Eq\eqref{soll0}. First, let us define the following relevant quantity, $\epsilon_m := \left(\mathfrak{A} \Lambda \right)^{-m}$, so that $\varepsilon_m = \epsilon_m$ corresponds to the unique vacuum theories with maximal multiplicity, i.e. with Eq\eqref{soll0} reducing to $- \Lambda \left( 1 - \lambda/\Lambda \right)^{m+1}/(m+1)= 0$. However, we saw previously that these theories do not have GR-like perturbations and so are not suitable for our purpose. For arbitrary $\varepsilon_m$, the second factor of Eq\eqref{soll0} admits one real solution for $m$ odd. In order not to have a real solution for $m$ even, we need 
\begin{eqnarray}
0<\varepsilon_m<\epsilon_m \,, \label{UniqVacaa}
\end{eqnarray}
so that the only real vacuum of the trajectory, which exists for all $m$, is the ``regularizable" one $\Lambda$. Moreover, assuming for simplicity that $\varepsilon_{m\to\infty} \to c$, for a non-vanishing positive constant $c$, means that if $\mathfrak{A} \Lambda < 1$, the limit $m\to\infty$ simply reduces to perturbative General Relativity because $\Lambda_0\to \Lambda$ and $\kappa_{\text{eff}} \to \tilde{\kappa}$. However, if $\mathfrak{A}=1/\Lambda$ and $\varepsilon_m$ satisfies Eq\eqref{UniqVacaa}, we obtain $\kappa_{\text{eff}} \to \tilde{\kappa} / (1-c)$ and $\Lambda_0 \to \Lambda(1-c)$, which does constitute a non-trivial kind of theory given that the vacuum $\Lambda$ is different from the one of GR. In this case, if only the limit $m\to\infty$ of the trajectory is to be considered, we need $c\neq 1$ for the same reason as previously. But if each theory $\mathcal{G}\left(m\right)$ is considered to be a valid description of processes at perturbative order $m$, choosing $c=1$ would mean that the theory freezes at high energy $m\to \infty$.
\\

To conclude this section about these quite exotic regularized theories,  remark that such construction cannot work with perturbations around other backgrounds, as we will show in the fifth section \ref{Sec.BR}\ref{Sec.DSS}\ref{ssSec.PerturbationDSS}, meaning that it either describes gravitons around exotic (A)dS$_4$ vacua (if well-defined), or might be combined with other regularization techniques.

\section{~$\,$Covariant Regularizations of Lovelock-Lanczos Gravity via metric transformations} \label{sec.covReg}

Let us now turn to more conventional regularizations of $d$-dimensional Lovelock-Lanczos gravity resulting in $(D<d)$-dimensional theories containing additional fields. The method that will follow generalises the Mann-Ross conformal regularization of $2$D Einstein gravity \cite{GRD=2}, as well as the four dimensional conformal regularization of the Gauss-Bonnet theory introduced in \cite{ConfGB1,ConfGB2}. In these papers, the resulting theories belong to the Horndeski class of scalar-tensor theories.

As we already said previously, it is from the use of the property \eqref{GdeltaProp} of the GKD that dimensional factors can be extracted from the Lovelock action or field equations, thus it is not a peculiarity of the conformal transformations and we can start considering a general metric transformations 
\begin{eqnarray}
\bar{g}_{\mu\nu} \left( g_{\mu\nu},R^{\mu\nu}_{\sigma\rho}, \nabla, \phi_i  \right)\,, \label{Transfo}
\end{eqnarray}
where $\phi_i$ is an arbitrary set of fields and $R^{\mu\nu}_{\sigma\rho}$, $\nabla$ are respectively the Riemann tensor and the covariant derivative associated with $g_{\mu\nu}$. We assume this transformation to be such that the components of the Riemann tensor associated with the metric $\bar{g}_{\mu\nu}$ decompose as 
\begin{eqnarray}
\bar{R}^{\mu\nu}_{\sigma\rho} = e^{-\Phi}\left(  \mathcal{R}^{\mu\nu}_{\sigma\rho} + \frac{1}{4} \delta_{[\sigma}^{[\mu} \Psi_{\rho]}^{\nu]} + \frac{\Theta}{2} \delta_{[\sigma}^{\mu}\delta_{\rho]}^{\nu} \right) \,, \label{RiemannTransfo}
\end{eqnarray}
where $\Phi\left(g_{\mu\nu},R^{\mu\nu}_{\sigma\rho}, \nabla, \phi_i  \right)$,$\Theta\left( g_{\mu\nu},R^{\mu\nu}_{\sigma\rho}, \nabla, \phi_i  \right)$, $\Psi_\mu^\nu \left( g_{\mu\nu},R^{\mu\nu}_{\sigma\rho}, \nabla, \phi_i  \right)$ and $\mathcal{R}^{\mu\nu}_{\sigma\rho}  = R^{\mu\nu}_{\sigma\rho} + \mathcal{Q}^{\mu\nu}_{\sigma\rho}$, where the tensor $\mathcal{Q}\left(g_{\mu\nu},R^{\mu\nu}_{\sigma\rho}, \nabla, \phi_i  \right)$ has the same symmetries as the Riemann tensor. Once again, the antisymmetry is taken without normalization. This seems to be the most general transformation from which analytic dimensional factors can be extracted in a covariant way from the Lovelock-Lanczos action or field equations, i.e. from which one can hope to regularize Lovelock-Lanczos gravity in a $(d\to D)$-covariant way. 
\\

Let's denote by a $\Bigcdot$ an arbitrary number $q$ of free indices. Then using the property Eq\eqref{GdeltaProp} of the GKD, we obtain the identity : 
\begin{eqnarray}
\begin{split}
&\delta^{\Bigcdot \mu_1 \nu_1 \dots \mu_p \nu_p}_{\Bigcdot \sigma_1 \rho_1 \dots \sigma_p \rho_p} \prod  _{r=1}^p \bar{R}_{\mu_r \nu_r}^{\sigma_r \rho_r} - e^{-p \, \Phi} \delta^{\Bigcdot \mu_1 \nu_1 \dots \mu_p \nu_p}_{\Bigcdot \sigma_1 \rho_1 \dots \sigma_p \rho_p} \prod  _{r=1}^p \mathcal{R}_{\mu_r \nu_r}^{\sigma_r \rho_r}  \\
=& e^{-p \, \Phi}  \, \sum_{k=1}^p \sum_{l=0}^k \Bigg\{ \frac{( d- 2p-q+k+l )!}{( d- 2p-q )!} \binom{p}{k}\binom{k}{l} \Theta^l \delta^{\Bigcdot \mu_1 \nu_1 \dots \mu_{p-k} \nu_{p-k} \alpha_1 \dots \alpha_{k-l}}_{\Bigcdot \sigma_1 \rho_1 \dots \sigma_{p-k} \rho_{p-k}\beta_1 \dots \beta_{k-l}} \prod  _{r=1}^{p-k} \mathcal{R}_{\mu_r \nu_r}^{\sigma_r \rho_r}  \prod  _{r=1}^{k-l} \Psi^{\beta_r}_{\alpha_r} \Bigg\}\,, \label{CovReg}
\end{split}
\end{eqnarray}
where we extracted the $k=1$ term because it is the only one which does not contain any dimensional factor. Remark that if respectively $\Theta =0$ or $\Psi =0$, we need to enforce these conditions in the previous equation by setting $l=0$ or $l=k$,  i.e. to replace the associated sum by Kronecker deltas $ \sum_{l=0}^k\; \longrightarrow \delta^l_0$ or $\sum_{l=0}^k\; \longrightarrow \delta^l_k$.

\subsection{Lagrangian regularizations}

In order to regularize the Lagrangian of LLG via a metric transformation, the procedure requires to consider a series expansion around $d=D$, i.e. to know \textit{all} the \textit{analytic} dimensional dependence of the theory. However, if the transformation Eq\eqref{Transfo} gives rise to an explicit dependence of $\Phi, \Psi, \Theta$ or $\mathcal{Q}$ on, say, the square of the Weyl tensor $C^{\mu\nu}_{\sigma\rho}C_{\mu\nu}^{\sigma\rho}$, then these functions would acquire some new analytic dimensional dependence. Although the regularizations could also be carried out in this case, we will restrict for simplicity to transformations of the form 
\begin{eqnarray}
\bar{g}_{\mu\nu} \left( g_{\mu\nu}, \nabla, \phi_i  \right)\,, \label{Transfo2}
\end{eqnarray}
for which  $\Phi, \Psi, \Theta$ and $\mathcal{Q}$ \textit{do not} contain any analytic dimensional factors.

Therefore, as we extracted \textit{all} the analytic dimensional factors from the Lovelock-Lanczos scalar associated with the metric $\bar{g}_{\mu\nu}$ (Eq\eqref{CovReg} with $q=0$), the series expansion around $d=D$ will not affect the following term :
\begin{eqnarray}
\Xi^{\Bigcdot}_{\Bigcdot(p,k,l)}\left(g_{\mu\nu},R^{\mu\nu}_{\sigma\rho}, \nabla, \phi_i  \right):= \binom{p}{k}\binom{k}{l} \Theta^l \delta^{\Bigcdot \mu_1 \nu_1 \dots \mu_{p-k} \nu_{p-k} \alpha_1 \dots \alpha_{k-l}}_{\Bigcdot \sigma_1 \rho_1 \dots \sigma_{p-k} \rho_{p-k}\beta_1 \dots \beta_{k-l}} \prod  _{r=1}^{p-k} \mathcal{R}_{\mu_r \nu_r}^{\sigma_r \rho_r}  \prod  _{r=1}^{k-l} \Psi^{\beta_r}_{\alpha_r} \,.\label{DefXi}
\end{eqnarray}
At this point we need to make a last assumption on the transformation Eq\eqref{Transfo2}, in order to fix how the determinants of both metric fields are related, because it is the only remaining piece that can contain a dimensional factor. Having in mind a general disformal transformation, we suppose that it decomposes in the following way 
\begin{eqnarray}
\sqrt{-\bar{g}} = e^{\frac{d \psi}{2}} \Omega  \sqrt{-g}\,,
\end{eqnarray}
where $\Omega\left(g_{\mu\nu}, \nabla, \phi_i  \right)$ and $\psi\left(g_{\mu\nu}, \nabla, \phi_i  \right)$. Setting $q=0$ and normalizing Eq\eqref{CovReg} to consider directly the Lovelock-Lanczos Lagrangians Eq\eqref{LagrangianDensity} gives, 
\begin{eqnarray}
\sqrt{-\bar{g}} \bar{\mathcal{L}}_p
 =\sqrt{-g}   e^{\frac{d \psi -2p \Phi}{2}} \frac{\Omega}{2^p} \sum_{k=0}^p \sum_{l=0}^k \, \Xi_{(p,k,l)}  \, \prod_{j=1}^{k+l}  (d-2p+j) \,,
\end{eqnarray}
where we have used products in order to avoid dealing with negative factorials when the regularized dimension $D$ is such that $D< 2p$, i.e. lower than critical.

In  any dimensional regularization it is assumed that the dimension $d$ is a continuous parameter, so that it becomes possible to expand the previous expression around $d=D$, for some real number $D$,
\begin{eqnarray}
\frac{\sqrt{-\bar{g}} }{\sqrt{-g} } \bar{\mathcal{L}}_p = e^{\frac{D \psi -2p \Phi}{2}} \Omega \Bigg(  \frac{1}{2^p} \sum_{k=0}^p \sum_{l=0}^k\, \Xi_{(p,k,l)}   \, \prod_{j=1}^{k+l}  (D-2p+j) + \mathfrak{L}_p (d-D)+ O\left((d-D)^2\right)  \Bigg)\,, \label{SeriesExpansion}
\end{eqnarray}
where 
\begin{eqnarray}
\mathfrak{L}_p := \frac{1}{2^{p+1}}  \sum_{k=0}^p \sum_{l=0}^k  \Xi_{(p,k,l)}   \left( \psi \prod_{j=1}^{k+l}  (D-2p+j)  + 2  \sum_{j=1}^{k+l} \prod_{r=1}^{j-1} (D-2p+r) \prod_{r=j+1}^{k+l} (D-2p+r) \right)\,. \label{CovariantRegularizedLLG}
\end{eqnarray}
Therefore, the following theory, defined as a dimensional limit involving Riemann polynomials non-minimally coupled with the set of fields $\phi_i$ is well-defined :  
\begin{eqnarray}
L_p := \sqrt{-g} e^{\frac{D \psi -2p \Phi}{2}} \Omega \; \mathfrak{L}_p  
= \lim_{d\to D} \left\{ \sqrt{-\bar{g}} \bar{\mathcal{L}}_p  - \sqrt{-g} e^{\frac{D \psi -2p \Phi}{2}} \frac{\Omega}{2^p}   \sum_{k=0}^p \sum_{l=0}^k\, \Xi_{(p,k,l)}   \, \prod_{j=1}^{k+l}  (D-2p+j)   \right\} \frac{1}{d-D} \label{LimitCovariantRegularizedLLG}
\end{eqnarray}
for any regularized dimension $D$ and any Lovelock order $p$. We see that the theory in the RHS from which we take the limit is not a pure Lovelock theory, but requires a certain amount of counter-terms in order to extract the dimensional factor and take a well-defined limit. Some of these terms can be divergences, but are nonetheless essential.

It is important to understand that the theories $L_p$ can be studied in any dimension $\mathcal{D}$, possibly different from $D$. For example, there are different ways to study the regularization of cubic LLG in four dimensions: either one directly regularizes that theory for $D=4$ or uses the $D=5$ or $D=6$ regularized theories in four dimensions.  Thus, we can consider the regularized dimension $D$ as a free parameter and suppose that we are working in $\mathcal{D}$ dimensions. Then the theory $L_p$ can be further reduced by noting that the GKD appearing in the set of scalars $\Xi_{(p,k,l)}$ contains $2p-k-l$ indices, meaning that $\Xi_{(p,k,l)}=0$, identically, for $2p-\mathcal{D} > k+l$. As we have $k+l \geq 0$, this condition does not reduce further the theory for  $\mathcal{D}\geq 2p$. 
\\

The case $\Theta \neq 0$ is of particular interest as it allows to consider arbitrary high order $p$ theories $L_p$, for any regularized dimension $D$, i.e. \textit{non-perturbative Regularized Lovelock-Lanczos gravities} containing arbitrary high powers of $\Theta$ :
\begin{eqnarray}
\mathfrak{L}_p := \frac{1}{2^{p+1}} \sum_{k= p - \lfloor \frac{\mathcal{D}}{2} \rfloor}^p \sum_{l=2p - \mathcal{D} - k}^k \zeta_{(k+l;p)}^{(D)}  \binom{p}{k}\binom{k}{l} \Theta^l \delta^{\mu_1 \nu_1 \dots \mu_{p-k} \nu_{p-k} \alpha_1 \dots \alpha_{k-l}}_{\sigma_1 \rho_1 \dots \sigma_{p-k} \rho_{p-k}\beta_1 \dots \beta_{k-l}} \prod  _{r=1}^{p-k} \mathcal{R}_{\mu_r \nu_r}^{\sigma_r \rho_r}  \prod  _{r=1}^{k-l} \Psi^{\beta_r}_{\alpha_r} \,, \label{THEORY1}
\end{eqnarray}
where $\lfloor \frac{\mathcal{D}}{2} \rfloor$ is the floor function giving the greatest integer less or equal to $\mathcal{D}/2$. If $ p - \lfloor \frac{\mathcal{D}}{2} \rfloor<0$ or $2p - \mathcal{D} - k<0$, the convention $\binom{k}{l}=0$ for $l<0$ ensures that the sum have the proper bounds. We have also defined the function,
\begin{eqnarray}
\zeta_{(k+l;p)}^{(D)}\left[\psi\right] := \psi \prod_{j=1}^{k+l}  (D-2p+j)  + 2  \sum_{j=1}^{k+l} \prod_{r=1}^{j-1} (D-2p+r) \prod_{r=j+1}^{k+l} (D-2p+r) \,.
\end{eqnarray}
If $D$ is an integer, it can as well be evaluated as a limit to either $D$ for a given $p$, or to $p$ for a given $D$, by using the Gamma function $\Gamma(z) = \int_0^\infty t^{z-1} e^{-t} dt$, which has simple poles at the negative integers and the harmonic numbers $H_n = \sum_{m=1}^n 1/m$, as follows :
\begin{eqnarray*}
\zeta_{(k+l;p)}^{(D)}= \lim_{q\to p} \frac{\Gamma\left(D-2q+1+k+l\right)}{\Gamma\left(D-2q+1\right)} \Big( \psi +2 \left( H_{D-2q+k+l}- H_{D-2q} \right) \Big)\,.
\end{eqnarray*}

On the other hand, if $\Theta=0$, which is enforced from Eq\eqref{CovariantRegularizedLLG} by setting $l=0$, the condition to have non-vanishing $\Xi$ becomes : $2p-\mathcal{D} \leq k \leq p$, i.e. $p \leq \mathcal{D}$. Thus, the regularization yields non-trivial results up to the quartic Lovelock-Lanczos theory in four dimensions.  In this case the theory reduces to : 
\begin{eqnarray}
\mathfrak{L}_p := \frac{1}{2^{p+1}} \sum_{k=2p-\mathcal{D}}^p  \zeta_{(k;p)}^{(D)} \;  \binom{p}{k}\delta^{\mu_1 \nu_1 \dots \mu_{p-k} \nu_{p-k} \alpha_1 \dots \alpha_{k}}_{\sigma_1 \rho_1 \dots \sigma_{p-k} \rho_{p-k}\beta_1 \dots \beta_{k}} \prod  _{r=1}^{p-k} \mathcal{R}_{\mu_r \nu_r}^{\sigma_r \rho_r}  \prod  _{r=1}^{k} \Psi^{\beta_r}_{\alpha_r} \,. \label{THEORY2}
\end{eqnarray}
This is the most minimal way to regularize higher-order Lovelock-Lanczos gravities in a $D$-dimensional covariant way, as only one Kronecker delta appears in the decomposition of the Riemann tensor Eq\eqref{RiemannTransfo} in this case. As we will see, conformal transformations yield $\Theta\neq0$, but these kinds of minimal theories might still be possible to reach for more general ones. 
\\

The transformation \eqref{Transfo2} being quite general, it might be useful to explicitly write the theories $L_p$ in four dimensions $\mathcal{D}=4$, in order to have a grasp on the common form of these theories, irrespectively of the specific metric transformation and regularized dimension $D$.  For conformal ones, it is still manageable to deal with the general case as we will see, but for disformal or even more complicated ones, such formula might be of some help.  We will note $\Psi^\mu_\mu = \Psi$, $\mathcal{R} := \delta_{\alpha\mu}^{\beta\nu} \mathcal{R}^{\alpha\mu}_{\beta\nu}/2$, $\mathfrak{G} :=\delta^{\mu\nu\alpha\beta}_{\sigma\rho\delta\lambda} \mathcal{R}_{\mu\nu}^{\sigma\rho}\mathcal{R}^{\delta\lambda}_{\alpha\beta} /4$ and 
\begin{eqnarray}
\sigma_{(i)} := \zeta_{(2p-i;p)}^{(D)}\,,
\end{eqnarray}
where the dimension and the order are now implicit. In four dimensions $\mathcal{D}=4$, the Eq\eqref{THEORY1} reduces to
 \begin{eqnarray}
\begin{split}
\mathfrak{L}_p=   &\Bigg(\sigma_{(0)} \Theta^4  + p \left( 2 \sigma_{(2)} \mathcal{R} + \sigma_{(1)} \Psi \right) \Theta^3 + \frac{p(p-1)}{2} \left(4 \sigma_{(4)}\mathfrak{G} + 2 \sigma_{(3)} \delta^{\mu\nu\alpha}_{\sigma\rho\beta} \mathcal{R}^{\sigma\rho}_{\mu\nu} \Psi^\beta_\alpha  +\sigma_{(2)} \delta_{\alpha\mu}^{\beta\nu} \Psi^\alpha_\beta \Psi^\mu_\nu \right) \Theta^2  \\
&+ \frac{p(p-1)(p-2)}{6} \left( 3 \sigma_{(4)}  \delta^{\mu\nu\alpha\beta}_{\sigma\rho\delta\lambda} \mathcal{R}^{\sigma\rho}_{\mu\nu}  \Psi^\delta_\alpha \Psi^\lambda_\beta + \sigma_{(3)} \delta^{\mu\nu\alpha}_{\sigma\rho\beta}  \Psi^\sigma_\mu \Psi^\rho_\nu \Psi^\beta_\alpha    \right) \Theta  \\
&+\frac{p(p-1)(p-2)(p-3)}{24} \sigma_{(4)} \delta^{\mu\nu\alpha\beta}_{\sigma\rho\delta\lambda}\Psi^\sigma_\mu \Psi^\rho_\nu   \Psi^\delta_\alpha \Psi^\lambda_\beta    \Bigg)\Theta^{p-4}  \frac{1}{2^{p+1}} \,, \label{4DReg}
\end{split}
\end{eqnarray}
while the results for lower dimensions $\mathcal{D}=3,2$ are reached by setting respectively $\delta^{\mu\nu\alpha\beta}_{\sigma\rho\delta\lambda}=0$ and $\delta^{\mu\nu\alpha}_{\sigma\rho\beta}=0$.
 
\subsubsection{Critical order theories} \label{CriticalRegul}

Among the different regularized dimensions $D$, the critical one $D=2p$ is of particular interest because it is the dimension for which the Lovelock scalars becomes topological and thus share a common structure. For instance, the factor $(d-2p)$ is already shared by all the terms but one in the field equation tensor associated with the theory $\sqrt{-\bar{g}} \bar{\mathcal{L}}_p$, as it can be seen in Eq\eqref{CovReg} with $q=1$. On the other hand, more counter-terms are needed below the critical order $d \to D < 2p$ in order to extract a factor $(d-D)$ from the Lovelock tensor. At the level of the action, it means that non-trivial counter terms are required, so that in the non-critical case, the regularization is not that of a pure Lovelock theory.

Moreover, from the recent work of \cite{3DGB} on the conformal Mann-Ross regularization, it seems that the Lovelock branch admitting the regularized Gauss-Bonnet black hole (whose existence is one of the main interest of this approach) can be reached in $3$D from the $4$D (critical) regularization of the Gauss-Bonnet theory. This further indicates the peculiar status of the critical order regularization. Setting $D=2p$ in Eq\eqref{LimitCovariantRegularizedLLG}, we get :
\begin{eqnarray}
\begin{split}
L^\text{critical}_p :=&\lim_{d\to 2p} \left( \sqrt{-\bar{g}} \bar{\mathcal{L}}_p  - \sqrt{-g} e^{p( \psi -  \Phi)} \frac{\Omega}{2^p}   \sum_{k=0}^p \sum_{l=0}^k\, \Xi_{(p,k,l)}   \, (k+l)!  \right) \frac{1}{d-2p} 
\\
=& \sqrt{-g} e^{p\left( \psi - \Phi \right)}   \frac{\Omega}{2^{p+1}}  \sum_{k=0}^p \sum_{l=0}^k  \Xi_{(p,k,l)}  (k+l)! \left(\psi +2 H_{k+l} \right) =:  \sqrt{-g} e^{p\left( \psi - \Phi \right)}  \Omega \mathfrak{L}_p  \,, \label{CriticalRegulari}
\end{split}
\end{eqnarray}
while the Eq\eqref{4DReg} is still valid, with $\sigma_{(i)} = (2p-i)! \left( \psi + 2 H_{2p-i} \right)$.

As Lovelock-Lanczos gravities are topological in critical dimensions, the counter-term appearing in the limit in the previous equation must be a total derivative (whatever the transformation of the metric is) for $d=2p$. Thus, we know that the term multiplied by $e^{p\left( \psi - \Phi \right)} \Omega \psi$ in $L^\text{critical}_p$ is a total derivative. 

\subsubsection{Regularized theories from higher orders in dimensional expansion}

Finally, for the sake of completeness, note that other regularized theories can be reached from higher orders in the dimensional series expansion. Considering the critical regularization $d\to 2p$ for simplicity, we see that at second order, $(d-2p)^2$, Eq\eqref{SeriesExpansion} becomes : 
\begin{eqnarray}
\frac{\sqrt{-\bar{g}} }{\sqrt{-g} } \bar{\mathcal{L}}_p = e^{p\left( \psi - \Phi \right)} \Omega \Bigg(  \frac{1}{2^p} \sum_{k=0}^p \sum_{l=0}^k\, \Xi_{(p,k,l)}   \, (k+l)!  + \mathfrak{L}_p (d-D) + \mathfrak{L}_p^{(2)} (d-D)^2+ O\left((d-D)^3\right)  \Bigg) \,, \label{SecondOrderCCRLLG}
\end{eqnarray}
where 
\begin{eqnarray}
\mathfrak{L}_p^{(2)} := \frac{2^{-(p+3)}}{3} e^{p\left(\psi - \Phi\right)} \sum_{k=0}^p \sum_{l=0}^k\, \Xi_{(p,k,l)}   \,  (k+l)! F_{(k+l)}\left[ \psi \right] \,,
\end{eqnarray}
and 
\begin{eqnarray}
F_{(k+l)}\left[ \psi \right] := \left( -2 \pi^2 + 3 \psi^2 + 12 H_{k+l} \left( \psi + H_{k+l} \right) + 12 \partial^2_z  \log\left( \Gamma\left(1+z\right)\right)|_{z=k+l} \right)\,,
\end{eqnarray}
which involves the trigamma function. Therefore the following regularized theory is well-defined : 
\begin{eqnarray}
\begin{split}
L^{(2)}_p &:= \sqrt{-g} e^{p( \psi - \Phi)} \Omega \; \mathfrak{L}^{(2)}_p  
\\ &= \lim_{d\to 2p} \left( \sqrt{-\bar{g}} \bar{\mathcal{L}}_p  - \sqrt{-g} e^{p\left( \psi - \Phi \right)} \frac{\Omega}{2^p}   \sum_{k=0}^p \sum_{l=0}^k\, \Xi_{(p,k,l)}   \, (k+l)!  - (d-2p) L_p   \right) \frac{1}{(d-2p)^2}  \,.
 \end{split}
\end{eqnarray}
 In particular, if the counter-terms at zeroth order are total derivatives or yields second order field equations, it implies that the theory $L_p$ will also have second order metric field equations, which in turns implies that the second order theory $L^{(2)}_p$ does also and so on. Thus, this process defines a series of theories $L^{(s)}_p$ coming from the dimensional structure of the given Lovelock invariant for $d\to D$. We will not discuss further this class of theories here, but it could interesting to investigate if they share some properties with higher dimensional Lovelock theories as well.

\subsection{Conformal regularization}

The last step to directly compare our results with \cite{GRD=2} and \cite{ConfGB1,ConfGB2} is to restrict the general transformation Eq\eqref{Transfo2} to be conformal, i.e. 
\begin{eqnarray}
\bar{g}_{\mu\nu}  = e^{\phi} g_{\mu\nu} \,, \label{ConformalTrans}
\end{eqnarray}
such that in particular $\psi = \phi$, $\Omega=1$ and so $\sqrt{-\bar{g}} = \sqrt{-g} e^{\frac{d}{2} \phi}$. In this case we can use the Eq(2.12) of \cite{ConfGB2} to obtain
\begin{eqnarray}
\Phi :=\phi  \,\, , \;\;\; \Theta :=- \frac{1}{2}  \nabla_\mu \phi \nabla^\mu \phi  \,\, , \;\;\; \Psi^\mu_\nu :=  \left(  \nabla^\mu \phi \nabla_\nu \phi - 2 \nabla^\mu \nabla_\nu \phi \right)  \,\, , \;\;\; \mathcal{Q}^{\mu\nu}_{\sigma\rho}=0 \,.
\end{eqnarray}
It is therefore possible to expand further the quantity $\Xi$, defined by Eq\eqref{DefXi}, appearing in both the theory to be regularized (RHS of \eqref{LimitCovariantRegularizedLLG}) as well as in the regularized one \eqref{CovariantRegularizedLLG} as, 
\begin{eqnarray}
\begin{split}
\Xi^{\Bigcdot}_{\Bigcdot(p,k,l)}\left(g_{\mu\nu},R^{\mu\nu}_{\sigma\rho}, \nabla, \phi_i  \right)& = \sum_{j=0}^{k-l} \binom{p}{k}\binom{k}{l}\binom{k-l}{j} (-1)^{k-j} 2^{k-2l-j} \left(\nabla_\mu \phi \nabla^\mu \phi\right)^l \\
&\times \delta^{\Bigcdot \mu_1 \nu_1 \dots \mu_{p-k} \nu_{p-k} \alpha_1 \dots \alpha_{j} \lambda_1 \dots \lambda_{k-l-j}}_{\Bigcdot \sigma_1 \rho_1 \dots \sigma_{p-k} \rho_{p-k}\beta_1 \dots \beta_{j} \gamma_1 \dots \gamma_{k-l-j} } \prod  _{r=1}^{p-k} R_{\mu_r \nu_r}^{\sigma_r \rho_r}  \prod  _{r=1}^{j} \nabla^{\beta_r} \phi \nabla_{\alpha_r} \phi  \prod  _{r=1}^{k-l-j} \nabla^{\gamma_r}  \nabla_{\lambda_r} \phi \\
& = \binom{p}{k}\binom{k}{l} (-1)^{k-1} 2^{k-2l-1} \left(\nabla_\mu \phi \nabla^\mu \phi\right)^l  \left( (k-l) \nabla^\beta \phi \nabla_\alpha \phi -2 \nabla^\beta \nabla_\alpha \phi \right) \\
&\times  \delta^{\Bigcdot \mu_1 \nu_1 \dots \mu_{p-k} \nu_{p-k}  \lambda_1 \dots \lambda_{k-l-1}\alpha}_{\Bigcdot \sigma_1 \rho_1 \dots \sigma_{p-k} \rho_{p-k} \gamma_1 \dots \gamma_{k-l-1}\beta } \prod  _{r=1}^{p-k} R_{\mu_r \nu_r}^{\sigma_r \rho_r}  \prod  _{r=1}^{k-l-1} \nabla^{\gamma_r}  \nabla_{\lambda_r} \phi \,,  \label{JJJJJ}
\end{split}
\end{eqnarray}
where in the second line we used the antisymmetry of the GKD. 

\subsubsection{Non-critical conformal regularizations}

Before restricting to the critical regularization $d\to D=2p$ for the reasons that we mentioned previously, let's illustrate briefly the other classes of regularization in the conformal case. Although not shift-symmetric as in the critical case, they belong nonetheless to Horndeski theory, as it can be seen by a direct application of Eq\eqref{LimitCovariantRegularizedLLG} : for $D=2p-1$, we obtain 
\begin{eqnarray}
L_p := \sqrt{-g} e^{-\frac{\phi}{2}}\mathfrak{L}_p  
= \lim_{d\to 2p-1}  \frac{\sqrt{-\bar{g}} \bar{\mathcal{L}}_p  - \sqrt{-g} e^{-\frac{\phi}{2}}   \mathcal{L}_p}{d-2p+1} \, , \;\; \mathfrak{L}_p := \frac{1}{2^{p}}  \left( \frac{\phi}{2} \mathcal{L}_p +  \sum_{k=1}^p\sum_{l=0}^k \Xi_{(p,k,l)} (k+l-1)! \right) \label{NonCrit1}
\end{eqnarray}
As the theory whose limit gives $L_p$ belongs to Horndeski scalar-tensor theories, so does the limit. As we see, this is almost the regularization of a pure Lovelock theory as no counter-terms (not even boudary terms) are required to take the limit. However, for smaller regularized dimensions, Horndeski-type counter-terms appears in \eqref{LimitCovariantRegularizedLLG}. For example, for $D=2p-2$,
\begin{eqnarray}
L_p :=  \lim_{d\to 2(p-1)}  \frac{\sqrt{-\bar{g}} \bar{\mathcal{L}}_p  -  \sqrt{-g} e^{-\phi}   \left(  \mathcal{L}_p -p\mathcal{G}_{\phantom{(p-1)} \beta}^{(p-1) \alpha}   \nabla^\beta \nabla_\alpha \phi  \right) + p \nabla^\beta \left( \sqrt{-g} e^{-\phi}   \mathcal{G}_{\phantom{(p-1)} \beta}^{(p-1) \alpha} \nabla_\alpha \phi \right)  }{d-2(p-1)} \label{NonCrit2}
\end{eqnarray}
and the same pattern seems to continue for lower regularized dimensions.

\subsubsection{Critical regularization of the field equations \& Lagrangians}

We now focus on the (conformal) critical regularization of Lovelock gravity $D=2p$ and work in arbitrary dimension $\mathcal{D}\leq D$. After some manipulations that can be found in Appendix\ref{ConfCubic}, we find the identity :
\begin{eqnarray}
\sum_{k=1}^{p} \sum_{l=0}^{k} \Xi_{(p,k,l)} (k+l)! =4   \sum_{k=0}^{p-1} \sum_{l=0}^{k} (p-k) \alpha_{(k,l)} \mathcal{D}V_{(k,l)} \,,  \label{XIDV} 
\end{eqnarray}
where the sequence of constants $\alpha_{(k,l)}$ is defined by 
\begin{eqnarray}
\alpha_{(k,l)} :=  \binom{p}{k}\binom{k}{l} (-1)^{k-1} 2^{k-2l-1} (k+l)! \,,
\end{eqnarray} 
and $\mathcal{D}V_{(k,l)}$ is a (dimensionally independent) divergence given by : 
\begin{eqnarray}
\mathcal{D}V_{(k,l)} :=  \delta^{ \mu_1 \nu_1 \dots \mu_{p-k-1} \nu_{p-k-1}  \lambda_1 \dots \lambda_{k-l+1}}_{ \sigma_1 \rho_1 \dots \sigma_{p-k-1} \rho_{p-k-1} \gamma_1 \dots \gamma_{k-l+1} }   \nabla_{\lambda_1} \left( \nabla^{\gamma_1}\phi  \left(\nabla_\alpha \phi \nabla^\alpha \phi\right)^l \prod_{r=1}^{p-k-1} R_{\mu_r \nu_r}^{\sigma_r \rho_r}  \prod_{r=2}^{k-l+1} \nabla^{\gamma_r}  \nabla_{\lambda_r} \phi \right)\,.
\end{eqnarray}
Note that the same divergence appears in the context of the ``Kaluza-Klein dimensional reduction" of Lovelock-Lanczos gravity in \cite{KKLLG}. This is not surprising as it was shown in \cite{ConfGB1,ConfGB2} and \cite{HornGB1,HornGB2} that both approaches yields the same $4$D regularized Gauss-Bonnet theory when the compactification space is flat.

Using this relation, the definition of the critical regularization Eq\eqref{CriticalRegulari} in the conformal case becomes: 
\begin{eqnarray}
L^\text{critical}_p :=\lim_{d\to 2p} \left\{ \sqrt{-\bar{g}} \bar{\mathcal{L}}_p  - \sqrt{-g} \mathcal{L}_p-  \sqrt{-g} \frac{1}{2^{p-2}}    \sum_{k=0}^{p-1} \sum_{l=0}^{k} (p-k) \alpha_{(k,l)} \mathcal{D}V_{(k,l)}   \right\} \frac{1}{d-2p} =:\sqrt{-g} \mathfrak{L}_p \,.
\end{eqnarray}
Therefore, without any knowledge on $\mathfrak{L}_p$, we can already know from this result what are the metric field equations associated with $L^\text{critical}_p$. Indeed, the field equation tensor associated with the not-yet-regularized theory reads
\begin{eqnarray}
\begin{split}
G_{\phantom{(p)} \beta}^{(p) \alpha}\big[\bar{g}_{\mu\nu} \big] :&= \left[ \bar{g}^{\alpha\gamma} \frac{ \delta}{\delta \bar{g}^{\gamma\beta}} \right] \left(\sqrt{-\bar{g}} \bar{\mathcal{L}}_p  - \sqrt{-g} \mathcal{L}_p-  \sqrt{-g} \frac{1}{2^{p-2}}    \sum_{k=0}^{p-1} \sum_{l=0}^{k} (p-k) \alpha_{(k,l)} \mathcal{D}V_{(k,l)}  \right) \\
&= \sqrt{-\bar{g}} \bar{\mathcal{G}}_{\phantom{(p)} \beta}^{(p) \alpha} - \sqrt{-g}\mathcal{G}_{\phantom{(p)} \beta}^{(p) \alpha} \\
&=  \sqrt{-g} \left( \mathcal{G}_{\phantom{(p)} \beta}^{(p) \alpha} \left[ e^{\frac{(d-2p)\phi}{2}} -1 \right] -\frac{d-2p}{2^{p+1}} e^{\frac{(d-2p)\phi}{2}}  \sum_{k=1}^{p} \sum_{l=0}^{k}  \frac{( d- 2p-1+k+l )!}{( d- 2p )!}\; \Xi^{\alpha}_{\beta(p,k,l)} \right)\\
&= \sqrt{-g}  \left(\frac{\phi}{2} \mathcal{G}_{\phantom{(p)} \beta}^{(p) \alpha}  -\frac{1}{2^{p+1}}  \sum_{k=1}^{p} \sum_{l=0}^{k}  (k+l-1)!  \; \Xi^{\alpha}_{\beta(p,k,l)} \right)(d-2p) + O\left( \left(d-2p\right)^2 \right) \,, \label{ulula}
\end{split}
\end{eqnarray}
where we applied the chain rule in the second line and Eq\eqref{CovReg} (for $q=1$) in the third one. As we are considering the critical regularization $d\to 2p$ (and possibly consider the theory in dimension $\mathcal{D}<d$), the Lovelock tensor $\mathcal{G}_{\phantom{(p)} \beta}^{(p) \alpha}$ is identically vanishing, so that the metric field equation tensor of the regularized theory can finally be written as :  
\begin{eqnarray}
\begin{split}
\mathfrak{G}_{\phantom{(p)} \nu}^{(p) \mu} &:=\left[ \frac{g^{\mu\gamma}}{\sqrt{-g}} \frac{ \delta}{\delta g^{\gamma\nu}} \right] \left(\sqrt{-g} \mathfrak{L}_p \right) \\
&=- \frac{1}{2^{p+1}} \sum_{k=1}^{p} \sum_{l=0}^{k} \frac{\alpha_{(k,l)}}{k+l} \left(\nabla_\zeta \phi \nabla^\zeta \phi\right)^l  \left( (k-l) \nabla^\beta \phi \nabla_\alpha \phi -2 \nabla^\beta \nabla_\alpha \phi \right) \\
&\phantom{=- \frac{1}{2^{p+1}} \sum_{k=1}^{p} \sum_{l=0}^{k} \frac{\alpha_{(k,l)}}{k+l}} \times  \delta^{\mu \mu_1 \nu_1 \dots \mu_{p-k} \nu_{p-k}  \lambda_1 \dots \lambda_{k-l-1}\alpha}_{\nu \sigma_1 \rho_1 \dots \sigma_{p-k} \rho_{p-k} \gamma_1 \dots \gamma_{k-l-1}\beta } \prod  _{r=1}^{p-k} R_{\mu_r \nu_r}^{\sigma_r \rho_r}  \prod  _{r=1}^{k-l-1} \nabla^{\gamma_r}  \nabla_{\lambda_r} \phi \,. \label{CCRMetricFieldEq}
\end{split}
\end{eqnarray}

As for the scalar field equation, we can use the fact that Lovelock gravity is the unique gravitational theory whose Palatini formalism always admit a branch of solution with a Levi-Civita connection, i.e. equivalent to the metric theory \cite{PalatiniMetric}. Rewriting the conformal Lovelock Lagrangian as
\begin{eqnarray}
\sqrt{-\bar{g}} \bar{\mathcal{L}}_p = \sqrt{-g} e^{\frac{(d-2p)\phi}{2}} \delta^{ \mu_1 \nu_1 \dots \mu_p \nu_p}_{ \sigma_1 \rho_1 \dots \sigma_p \rho_p} \prod  _{r=1}^p g^{\sigma_r  \gamma_r} \bar{R}_{\mu_r \nu_r \gamma_r}^{\phantom{\mu_r \nu_r \gamma_r} \rho_r}\,,
\end{eqnarray}
we can consider $\bar{g}\left( \phi, g\right)$ and the connection $\bar{\Gamma}$ as independent variables. The field equations associated with $\bar{\Gamma}$ will ensures it to be the Levi-Civita connection associated with the metric $\bar{g}$, i.e. $\bar{\Gamma}\left( \Gamma\left[g\right], \nabla\phi\right)$. In this way, both $\phi$ and $g_{\mu\nu}$ can be considered as Lagrange multipliers in the above expression : the field equation associated with $g_{\mu\nu}$ yields Eq\eqref{ulula} while the one associated with $\phi$ is given by :  
\begin{eqnarray}
\frac{\delta}{\delta \phi} \left(\sqrt{-\bar{g}} \bar{\mathcal{L}}_p  - \sqrt{-g} \mathcal{L}_p-  \sqrt{-g} \frac{1}{2^{p-2}}    \sum_{k=0}^{p-1} \sum_{l=0}^{k} (p-k) \alpha_{(k,l)} \mathcal{D}V_{(k,l)}  \right) =\frac{\delta}{\delta \phi} \sqrt{-\bar{g}} \bar{\mathcal{L}}_p =  \frac{(d-2p)}{2} \sqrt{-\bar{g}} \bar{\mathcal{L}}_p  ,
\end{eqnarray}
so that the scalar field equation of the critical regularized theory reads 
\begin{eqnarray}
\mathscr{E}_\phi := \left[\frac{1}{\sqrt{-g}} \frac{\delta}{\delta \phi}  \right]\left(\sqrt{-g} \mathfrak{L}_p \right) = \frac{1}{2} \frac{\sqrt{-\bar{g}}}{\sqrt{-g}} \bar{\mathcal{L}}_p  = \frac{1}{2} \left( \mathcal{L}_p +\frac{1}{2^{p-2}}    \sum_{k=0}^{p-1} \sum_{l=0}^{k} (p-k) \alpha_{(k,l)} \mathcal{D}V_{(k,l)}  \right)=0 \,. \label{EOMphi}
\end{eqnarray}
This generalizes the findings of \cite{GRD=2,ConfGB1,ConfGB2} in which it was noted that the scalar field equations associated with the regularized General Relativity and Gauss-Bonnet theory are equivalent to the vanishing of the associated conformal Lovelock scalars.

Moreover, if the dimension in which one considers the theory is critical as well (i.e. $\mathcal{D}= D=2p$), then, using the property Eq\eqref{GdeltaProp} of the GKD and Eq\eqref{XIDV}, the trace of the metric field equation tensor Eq\eqref{CCRMetricFieldEq} reduces to a divergence : 
\begin{eqnarray}
\mathfrak{G}_{\phantom{(p)} \mu}^{(p) \mu} = -\frac{1}{2^{p-1}} \sum_{k=0}^{p-1} \sum_{l=0}^{k} (p-k) \alpha_{(k,l)} \mathcal{D}V_{(k,l)}  \,,
\end{eqnarray}
corresponding to a conservation equation in the vacuum. It follows that, still in critical dimension, the field equations of this theory satisfies the following off-shell relation : 
\begin{eqnarray}
\mathfrak{G}_{\phantom{(p)} \mu}^{(p) \mu}  + \mathscr{E}_\phi = \frac{1}{2}  \mathcal{L}_p \,,
\end{eqnarray}
meaning that this theory truly constitutes a regularization of Lovelock-Lanczos gravity as (minus) the LHS yields the regularization of the trace of LLG given by Eq\eqref{RegTrace}.
\\

Back to a general dimension $\mathcal{D}$, we show in Appendix\ref{ConfCubic}) that the Lagrangian density of this theory is equivalent up to dimensionally independent boundary terms (represented by the symbol $\equiv$) to 
\begin{eqnarray}
\begin{split}
\mathfrak{L}_p &:= \frac{1}{2} \phi \mathcal{L}_p + \frac{1}{2^{p+1}}  \sum_{k=1}^p \sum_{l=0}^k  \Xi_{(p,k,l)}  (k+l)! \left(\phi +2 H_{k+l} \right)
\\
&\equiv \frac{1}{2} \phi \mathcal{L}_p + \frac{p}{4} \left( \nabla_\mu \phi \nabla^\mu \phi \right) \mathcal{L}_{(p-1)} + \frac{1}{2^{p-1}} \sum_{k=1}^{p-1} \sum_{l=0}^{k+1}  \frac{\alpha_{(k+1,l)}}{(k+l)(k+l+1)} \Omega_{(k,l)}  \,, \label{LagrangianDensityCCRLLG}
\end{split}
\end{eqnarray}
where
\begin{eqnarray}
\Omega_{(k,l)} :=  \delta^{ \mu_1 \nu_1 \dots \mu_{p-k-1} \nu_{p-k-1}  \lambda_1 \dots \lambda_{k-l+1}}_{ \sigma_1 \rho_1 \dots \sigma_{p-k-1} \rho_{p-k-1} \gamma_1 \dots \gamma_{k-l+1} } \left( \nabla_\alpha \phi \nabla^\alpha \phi \right)^l \prod  _{r=1}^{p-k-1} R_{\mu_r \nu_r}^{\sigma_r \rho_r}  \prod  _{r=1}^{k-l+1} \nabla^{\gamma_r}  \nabla_{\lambda_r} \phi  \,.
\end{eqnarray}
Therefore, all these theories are shift-symmetric (see for example \cite{Steer}) because it is well-known that the critical Lovelock scalars $\mathcal{L}_p$ are total derivatives (see the Horndeski formula Eq\eqref{HorndeskiFormula} found in \cite{HorndeskiVector}).

\subsubsection{Four dimensional critical conformal regularization of LLG}

Let's conclude this section by applying the previous critical regularization in four dimensions $\mathcal{D}=4$, so that the sums appearing in the Lagrangian \eqref{LagrangianDensityCCRLLG} and field equations \eqref{CCRMetricFieldEq}\eqref{EOMphi} can be reduced as previously according to the number of indices appearing in their GKD. Denoting the kinetic term of the scalar field as $X=-\frac{1}{2}\nabla_\zeta \phi \nabla^\zeta \phi$, we obtain for the Gauss-Bonnet regularization $p=2$,
\begin{eqnarray}
\mathfrak{L}_2 = \frac{1}{2} \phi \mathcal{L}_2 -  \delta_{\mu\nu}^{\alpha\beta} \phi^{;\mu}_{;\alpha} \phi^{;\nu}_{;\beta} +  \left(\Box \phi -R \right) X  -\frac{X^2}{2}  \,, \label{GB4DRCCR}
\end{eqnarray}
while for higher orders $p>2$ we have :
\begin{eqnarray}
\begin{split}
\mathfrak{L}_p = - \frac{1}{4}\Bigg\{& 2(p-1)(p-2)(2p-3)(2p-5)X^4 -2 p (p-2)(2p-5) \left(  (2p-3) \Box \phi - R \right) X^3 
\\
&+ p(p-1)\left( (2p-5) \left( 2(p-2) \delta_{\mu\nu}^{\alpha\beta} \phi^{;\mu}_{;\alpha} \phi^{;\nu}_{;\beta}-  \delta_{\mu\nu\sigma}^{\alpha\beta\rho} R^{\mu\nu}_{\alpha\beta} \phi^{;\sigma}_{;\rho}   \right) +  \mathcal{L}_2 \right) X^2 
\\
&- \frac{1}{3} p(p-1)(p-2) \left( 2 (2p-5)  \delta_{\mu\nu\sigma}^{\alpha\beta\rho} \phi^{;\mu}_{;\alpha} \phi^{;\nu}_{;\beta} \phi^{;\sigma}_{;\rho}  -3  \delta_{\mu\nu\sigma\gamma}^{\alpha\beta\rho\lambda} R^{\mu\nu}_{\alpha\beta} \phi^{;\sigma}_{;\rho}  \phi^{;\gamma}_{;\lambda}  \right) X 
\\
&+ \frac{1}{3} p(p-1)(p-2)(p-3) \delta_{\mu\nu\sigma\gamma}^{\alpha\beta\rho\lambda}    \phi^{;\mu}_{;\alpha} \phi^{;\nu}_{;\beta} \phi^{;\sigma}_{;\rho}   \phi^{;\gamma}_{;\lambda} \Bigg\}   (p-3)! (2p-7)!!  \, X^{p-4} \,.
\end{split}
\end{eqnarray}
As we are now working in four dimensions, it is possible to further reduce this theory noting that the following quantity is a boundary term in four dimensions :
\begin{eqnarray*}
\mathcal{L}_6^{\text{NH}} \left[ G\left(X\right) \right]=  \delta_{\mu\nu\sigma\gamma}^{\alpha\beta\rho\lambda} \left( \frac{3 \, G(X)}{4}  R^{\mu\nu}_{\alpha\beta}R^{\sigma\gamma}_{\rho\lambda}  +3 G_X(X) R^{\mu\nu}_{\alpha\beta} \phi^{;\sigma}_{;\rho}  \phi^{;\gamma}_{;\lambda} +  G_{XX}(X) \phi^{;\mu}_{;\alpha} \phi^{;\nu}_{;\beta} \phi^{;\sigma}_{;\rho}   \phi^{;\gamma}_{;\lambda} \right) \,,
\end{eqnarray*}
as it can be seen for example in Eq(98) and Eq(109) of  \cite{Unified}. In our case, $G(X) :=- (1/12) p (p-1) (p-3)! (2p-7)!! X^{p-2}$. Using that $\delta_{\mu\nu\sigma}^{\alpha\beta\rho} R^{\mu\nu}_{\alpha\beta} \phi^{;\sigma}_{;\rho} = -4 G^\mu_\nu  \phi^{;\nu}_{;\mu}$, we get 
\begin{eqnarray}
\begin{split}
\mathfrak{L}_p \equiv - \frac{1}{2}  &(p-3)! (2p-5)!! \Bigg\{ (p-1)(p-2)(2p-3)X^3 - p (p-2) \left(  (2p-3) \Box \phi - R \right) X^2
\\
&+ p(p-1) \left( (p-2) \delta_{\mu\nu}^{\alpha\beta} \phi^{;\mu}_{;\alpha} \phi^{;\nu}_{;\beta}+2 G^\mu_\nu  \phi^{;\nu}_{;\mu}  \right) X - \frac{1}{3} p(p-1)(p-2)  \delta_{\mu\nu\sigma}^{\alpha\beta\rho} \phi^{;\mu}_{;\alpha} \phi^{;\nu}_{;\beta} \phi^{;\sigma}_{;\rho}    \Bigg\}    \, X^{p-3} ,\label{Lag4DCCRLLG}
\end{split}
\end{eqnarray}
which is just a Horndeski theory, defined by the four Lagrangian terms :
\begin{eqnarray}
\begin{split}
\mathcal{L}^{\text{H}}_2\left[ G_2 \right] &:= G_2\left(X\right) \; ,\;\;\;
\mathcal{L}^{\text{H}}_3\left[ G_3 \right] := -G_3\left(X\right) \Box \phi  \;,\;\;\;
\mathcal{L}^{\text{H}}_4\left[ G_4 \right] &:=G_4\left(X\right) R + G_{4,X} \delta_{\mu\nu}^{\alpha\beta} \phi^{;\mu}_{;\alpha} \phi^{;\nu}_{;\beta}  \\
\mathcal{L}^{\text{H}}_5\left[ G_5 \right] &:= G_5\left(X\right) G^\mu_\nu  \phi^{;\nu}_{;\mu} -\frac{1}{6} G_{5,X}  \delta_{\mu\nu\sigma}^{\alpha\beta\rho} \phi^{;\mu}_{;\alpha} \phi^{;\nu}_{;\beta} \phi^{;\sigma}_{;\rho}   \,.
\end{split}
\end{eqnarray}
In our case, 
\begin{eqnarray}
\begin{split}
G_2(X) &= -\frac{1}{2^p} (2p-2)! X^p\; , \quad G_3(X) = -\frac{p}{2^{p-1}} (2p-3)! X^{p-1} \; , \\ 
G_4(X) &= -\frac{p}{2^{p-1}} (2p-4)! X^{p-1} \; , \quad G_5(X) = -\frac{p(p-1)}{2^{p-3}} (2p-5)! X^{p-2}   \,.
\end{split}
\end{eqnarray}
Remark that the functions $G_{2,3,4}$ match the result of Eq\eqref{GB4DRCCR} for $p=2$, while the last function behaves as 
\begin{eqnarray}
G_5(X) =\left( -\frac{2}{p-2} -3 + 4 \gamma + \log\left(4\right) \right) - 2 \log\left(X\right) + O\left(p-2\right) \,,
\end{eqnarray}
where $\gamma$ is the Euler-Mascheroni constant. When $G_5(X)$ is constant, the corresponding term $\mathcal{L}^{\text{H}}_5$ becomes a total derivative via the Bianchi identity, so that, up to a constant and divergent boundary term, we have $G_5(X)=- 2 \log\left(X\right) $ for $p=2$, what indeed matches the result of \cite{Ginflation} (see Eq(A.6) to Eq(A.10)) for the decomposition  of $\phi \mathcal{L}_2/2$ on the Horndeski basis, meaning that Eq\eqref{Lag4DCCRLLG} can also be used for $p=2$. Of course it also agrees with the integration by part of the Gauss-Bonnet-Horndeski divergence (see Eq\eqref{HorndeskiFormula}), 
\begin{eqnarray}
\mathcal{L}_2 = -2  \nabla_{\mu} \left(  \delta^{\mu\nu \mu_1 \nu_1}_{\sigma\rho \sigma_1 \rho_1}  \left(  \frac{u^{\sigma} \nabla_{\nu} u^{\rho} }{u^{\gamma}u_{\gamma}}  \right) \left( R_{\mu_{1} \nu_{1}}^{\sigma_{1} \rho_{1}} +\frac{4}{3}  \frac{\nabla_{\mu_1} u^{\sigma_1} \nabla_{\nu_1} u^{\rho_1} }{u^{\gamma}u_{\gamma}}  \right)  \right)\,,  \label{GaussBonnetHorndeski} 
\end{eqnarray}
for the choice $u^\mu= \nabla^\mu \phi$. Similarly, taking into account the number of indices of the GKD appearing in \eqref{CCRMetricFieldEq}, we obtain the metric field equation tensor :
\begin{eqnarray}
\begin{split}
\mathfrak{G}_{\phantom{(p)} \nu}^{(p) \mu} =   \bigg\{& -2(p-1)(2p-1)(2p-3) \delta^\mu_\nu X^3 + 2 p(2p-3) \left( 2 G^\mu_\nu + (p-1)\delta^{\mu\rho}_{\nu\sigma} \left( 2 \phi^{;\sigma}_{;\rho} -\phi^{;\sigma} \phi_{;\rho} \right) \right)X^2  \\
&- p(p-1) \left( \delta_{\nu\alpha\sigma\gamma}^{\mu\beta\rho\lambda} R^{\alpha\sigma}_{\beta\rho}  \left( \phi^{;\gamma} \phi_{;\lambda} - 2  \phi^{;\gamma}_{;\lambda} \right) +2 (2p-3)\delta_{\nu\alpha\sigma}^{\mu\beta\rho} \phi^{;\alpha}_{;\beta}  \left(  \phi^{;\sigma}_{;\rho}-  \phi^{;\sigma} \phi_{;\rho}  \right)  \right) X \\
&+\frac{2 p(p-1)(p-2)}{3}  \delta_{\nu\alpha\sigma\gamma}^{\mu\beta\rho\lambda} \phi^{;\alpha}_{;\beta}\phi^{;\sigma}_{;\rho} \left( 2 \phi^{;\gamma}_{;\lambda} - 3 \phi^{;\gamma} \phi_{;\lambda}  \right)  \bigg\}  \frac{(2p-4)!}{2^{p+1}}\; X^{p-3}\,, \label{EOMgCCR}
\end{split}
\end{eqnarray}
and from Eq\eqref{EOMphi}, the scalar field equation : 
\begin{eqnarray}
\mathscr{E}^{(p)}_\phi := - \nabla_\mu \mathcal{V}_{(p)}^{\mu}\,,
\end{eqnarray} 
where
\begin{eqnarray}
\begin{split}
\mathcal{V}_{(p)}^{\mu} =    \bigg\{& 2(2p-3) \delta^{\mu}_{\nu} X^3 - 2   \left( (2p-3) \delta^{\mu\alpha}_{\nu\beta}\phi^{;\beta}_{;\alpha} +2 G^\mu_\nu  \right)X^2 
+  \phi^{;\beta}_{;\alpha} \left( 2(p-2) \delta^{\mu\alpha\sigma}_{\nu\beta\rho}  \phi^{;\rho}_{;\sigma} -\delta^{\sigma\gamma\mu\alpha}_{\rho\lambda\nu\beta}  R^{\rho\lambda}_{\sigma\gamma} \right) X \\
&- \frac{2(p-3)}{3}\delta^{\mu\alpha\sigma\gamma}_{\nu\beta\rho\lambda}\phi^{;\beta}_{;\alpha}  \phi^{;\rho}_{;\sigma} \phi^{;\lambda}_{;\gamma}\bigg\}  \frac{p(p-1)(2p-4)!}{2^{p}} \;  \phi^{;\nu} \, X^{p-4}\,.  \label{4DscalarEOMCCR}
\end{split}
\end{eqnarray}
This result agrees with the usual scalar field equations derived from shift-invariant scalar-tensor theories, for which we have generically \cite{Shiftscalar}
\begin{eqnarray*}
\mathcal{V}_{(p)}^{\mu} := G_{2X} J^\mu_2 + G_{3X} J^\mu_3 + G_{4X} J^\mu_{(4,1)} + G_{4XX} J^\mu_{(4,2)} + G_{5X} J^\mu_{(5,1)} + G_{5XX} J^\mu_{(5,2)}\,,
\end{eqnarray*}
with
\begin{eqnarray*}
\begin{split}
J^\mu_2 &= - \phi^{;\mu} \, , \quad J^\mu_3 = \delta^{\mu\alpha}_{\nu\beta} \phi^{;\nu} \phi^{;\beta}_{;\alpha}\, , \quad J^\mu_{(4,1)} = 2 G^\mu_\nu \phi^{;\nu} \, , \quad J^\mu_{(4,2)} = -  \delta^{\mu\alpha\sigma}_{\nu\beta\rho} \phi^{;\nu}  \phi^{;\beta}_{;\alpha} \phi^{;\rho}_{;\sigma} \, , \\
J^\mu_{(5,1)} &=\frac{1}{4} \delta^{\mu\alpha\sigma\gamma}_{\nu\beta\rho\lambda} \phi^{;\nu} R_{\alpha\sigma}^{\beta\rho} \phi^{;\lambda}_{;\gamma}\, , \quad  J^\mu_{(5,2)}= \frac{1}{6}  \delta^{\mu\alpha\sigma\gamma}_{\nu\beta\rho\lambda} \phi^{;\nu}  \phi^{;\beta}_{;\alpha} \phi^{;\rho}_{;\sigma}   \phi^{;\lambda}_{;\gamma} \,.
\end{split}
\end{eqnarray*}
As for the case $p=2$, the Eq\eqref{4DscalarEOMCCR} agrees with the scalar field equations of the Gauss-Bonnet regularization because $\mathcal{L}_2$ directly appears in its divergence form \eqref{GaussBonnetHorndeski}. For $p=1$, GR can be recovered from Eqs\eqref{Lag4DCCRLLG}\eqref{EOMgCCR}\eqref{4DscalarEOMCCR} via the following limits : $\lim_{p\to1} \left( (p-1)\mathfrak{L}_p \right)=\left( R + \Box \phi \right)/2$, $\lim_{p\to1} \left( (p-1)\mathfrak{G}_{\phantom{(p)} \nu}^{(p) \mu}  \right)= G^\mu_\nu /2$ and $\lim_{p\to1} \left( (p-1)\mathcal{V}_{(p)}^{\mu}  \right)= 0$.
\\

These results allow to consider the following non-perturbative four dimensional conformal-critical-regularization of Lovelock-Lanczos gravity : 
\begin{eqnarray}
\begin{split}
I&:= \frac{1}{2 \tilde{\kappa}_0} \int d^4 x \sqrt{-g} \left(  -2 \Lambda_0+ R- \sum_{p=2}^\infty \beta_p \mathfrak{L}_p \right) \,,\\
 \mathfrak{G}^\mu_\nu &:= \Lambda_0 \delta^{\mu}_{\nu} + G^{\mu}_{\nu} -\sum_{p=2}^\infty \beta_p \mathfrak{G}_{\phantom{(p)} \nu}^{(p) \mu}  = \tilde{\kappa_0} T_\nu^\mu \; ,\quad \; \mathscr{E}_\phi :=-\sum_{p=2}^\infty \beta_p  \mathscr{E}^{(p)}_\phi   =0 \,, \label{CCRLLGAction}
 \end{split}
\end{eqnarray}
with $\tilde{\kappa}_0 = 8\pi G_N$. Although we will not derive the spherically symmetric and FLRW sectors of this theory, it is a good candidate to admit a branch of solutions agreeing with the background regularization, as it does for $p=2$, see \cite{Cylindr}. If so, it would automatically admit many regular cosmological and black hole solutions, as we will see in Sec.\ref{Sec.BR}

Remark also that, contrary to the Lovelock theories from which they came from, it is not strictly necessary to consider $p$ as a positive integer (modulo the overall factorials involving it) within these theories, so that one could consider more general combinations of these terms than the infinite sum written above, where $p$ would be an arbitrary real number, keeping in mind that $p>1$ represents high energy modifications to GR while $p<1$ large scale ones. 
\\

Finally, recall that this regularization applied to the Gauss-Bonnet theory was shown to suffer from strong coupling around flat space in \cite{HornGB2} and \cite{Amplitudes}. Although it might be the case that using higher order curvature terms $p>2$ around (A)dS together with the fine-tuning found in the previous section might help in this regard, because it might make the scalar field invisible (at least in a specific dynamical branch) at arbitrary higher than quadratic order, this issue was the reason why an alternative regularization was proposed in \cite{3Dcov1} based on $3$D-covariance. This is the topic of the following section.

\section{~$\,$Dimensionality and regularizations of the Lovelock-Lanczos actions via $3$D-covariant counter-terms}

The class of regularizations of Lovelock-Lanczos gravity that we are now interested in is based on the direct evaluation of the Lagrangian on gauge-fixed but generic metric ansatz. The best one can obtain in this case is for example $3$D-covariant theories propagating two degrees of freedom (due to Lovelock theorem) from an ADM parametrization. A regularization of the Gauss-Bonnet theory following this path has been found recently in \cite{3Dcov1,3Dcov2} using the Hamiltonian formalism and we will now generalize this result to include all the Lovelock invariants.

It is quite clear that the desired dimensional factors cannot come from the usual Lovelock-Lanczos action, because it is not identically vanishing in critical dimension. In order to circumvent this issue, we first need a covariant splitting of the Lovelock-Lanczos scalars of the form $\mathcal{L}= \mathcal{W} - \mathcal{D}\mathcal{V}$, with $\mathcal{W}$ identically vanishing for $d= 2p$ and $\mathcal{D}\mathcal{V}$ a covariant divergence, in such a way that considering instead of $\mathcal{L}$ the theory $\mathcal{L} + \mathcal{D}\mathcal{V}$ would preserve the dynamics and allow to obtain an indeterminate form as $d\to 2p$, once the poles of the dimension are introduced.

In the metric formalism, there are two ways (known to us) to achieve such a covariant splitting, both of which introducing additional structures to the theory. The first one, presented in \cite{Deruelle} by Deruelle, Merino and Olea, requires an ``hybrid" manifold $\bar{\mathcal{M}}$ that allows to construct the covariant divergence in terms of the difference between the Christoffel symbols of the two manifolds, which is a true tensor. If possible, it would be interesting to link this procedure with the necessity to introduce a counter-term in the covariant regularizations of the previous section, where up to boundary terms, the theory to regularize is given in the (critical) conformal case by the difference between the Lovelock scalar $\mathcal{L}_p$ and that of the conformal manifold $\bar{\mathcal{L}}_p$. However, the dimensional dependence of such splitting is not manifest, so that the regularization of the corresponding action might be difficult to achieve in general. Alternatively, a natural way to resolve this issue is to consider the general decomposition of Lovelock-Lanczos gravity found by Horndeski in \cite{HorndeskiVector}.

\subsection{Horndeski-Lovelock-Lanczos Action}

As shown in Appendix \ref{HorndeskiDecomp}, assuming that their exists an arbitrary non-null $d$-dimensional vector field $u$ as an additional structure, the Lovelock-Lanczos Lagrangian \eqref{LagrangianDensity} can be decomposed as  
\begin{eqnarray}
\begin{split}
  \mathcal{L}_{(p)} &= \sum_{s=0}^p \frac{ \sigma_{(p,s)}}{2^p} \;  \delta^{\mu \mu_1 \nu_1 \dots \mu_p \nu_p}_{\nu \sigma_1 \rho_1 \dots \sigma_p \rho_p}  \left( \frac{u_\mu u^{\nu}}{u^{\gamma}u_{\gamma}}   \right) \, \prod_{i=1}^{s} \, R_{\mu_{i} \nu_{i}}^{\sigma_{i} \rho_{i}}  \, \prod_{i=s+1}^{p} \, \frac{\nabla_{\mu_i} u^{\sigma_i} \nabla_{\nu_i} u^{\rho_i} }{u^{\gamma}u_{\gamma}}\\
  &- \sum_{s=0}^{p-1} \frac{ \sigma_{(p,s)}}{2^p} \; \delta^{\mu\nu \mu_1 \nu_1 \dots \mu_{p-1} \nu_{p-1}}_{\sigma\rho \sigma_1 \rho_1 \dots \sigma_{p-1} \rho_{p-1}} \nabla_\mu \left( \left(  \frac{u^{\sigma} \nabla_{\nu} u^{\rho} }{u^{\gamma}u_{\gamma}}  \right) \,  \prod_{i=1}^{s} \,  R_{\mu_{i} \nu_{i}}^{\sigma_{i} \rho_{i}}    \,  \prod_{i=s+1}^{p-1} \, \frac{\nabla_{\mu_i} u^{\sigma_i} \nabla_{\nu_i} u^{\rho_i} }{u^{\gamma}u_{\gamma}} \right)
  \\
  &=: \mathcal{L}_{(p)}^\mathcal{H} -\nabla_\mu V_{(p)}^\mu \,, \label{HorndeskiDecomposition}
\end{split}
\end{eqnarray}
where 
\begin{eqnarray}
\sigma_{(p,s)} := \frac{p!4^{p-s}}{s! \left[ 2 (p-s) -1 \right]!!}   \,.
\end{eqnarray}
From the number of indices appearing in its GKD, it is clear that the bulk term is identically vanishing in critical dimension so that a regularization can be obtained from the Horndeski form of the theory $\mathcal{L}_{(p)}^\mathcal{H}$.  Before seeing this, it is useful to note that this theory corresponds to the covariant form of the ADM Lagrangian of Lovelock-Lanczos gravity, which has a well-defined variational principle, contrary to the usual theory $\mathcal{L}_{(p)}$, so that requiring a regularizable theory automatically provides the correct action.

Indeed, let's define the projector on the hypersurfaces whose normal is given by $u$ as $h_{\mu}^{\nu} := \delta_{\mu}^{\nu} - \frac{u_\mu u^\nu}{u^{\gamma}u_{\gamma}}$, so that we have
\begin{eqnarray}
\delta^{\mu \mu_1 \nu_1 \dots \mu_p \nu_p}_{\nu \sigma_1 \rho_1 \dots \sigma_p \rho_p}  \left( \frac{u_\mu u^{\nu}}{u^{\gamma}u_{\gamma}}   \right)= h_{[\sigma_1}^{\mu_1}  h_{\rho_1}^{\nu_1} \dots h_{\sigma_p}^{\mu_p}  h_{\rho_p]}^{\nu_p} =: h^{\mu_1 \nu_1 \dots \mu_p \nu_p}_{\sigma_1 \rho_1 \dots \sigma_p \rho_p} \,,
\end{eqnarray}
by antisymmetry of the GKD. Therefore, the only quantities appearing in the Horndeski-LL Lagrangian are the extrinsic curvature of these hypersurfaces defined by $K_{\mu\nu} := h_\mu^\sigma h_\nu^\rho \nabla_\sigma u_\rho$, as well as the projection of the $d$-dimensional Riemann tensor 
\begin{eqnarray}
\mathcal{M}_{\mu\nu}^{\sigma\rho} := h_\mu^{\mu'}h_\nu^{\nu'}h_{\sigma'}^{\sigma}h_{\rho'}^{\rho} R_{\mu' \nu'}^{\sigma' \rho'} = \bar{R}_{\mu \nu}^{\sigma \rho} - \frac{K_{[\mu}^\sigma K_{\nu]}^\rho}{u^\gamma u_\gamma}\,,
\end{eqnarray}
with $\bar{R}_{\mu \nu}^{\sigma \rho}$ the components of the Riemann tensor associated with the metric $h_{\mu\nu}$. The equality is simply the Gauss-Codazzi equation. Thus,
\begin{eqnarray}
\mathcal{L}_{(p)}^\mathcal{H} = \sum_{s=0}^p \lambda_{(p,s)} h^{\mu_1 \nu_1 \dots \mu_p \nu_p}_{\sigma_1 \rho_1 \dots \sigma_p \rho_p} \prod_{i=1}^{s} \mathcal{P}_{\mu_i \nu_i}^{\sigma_i \rho_i}  \prod_{i=s+1}^p \mathcal{Q}_{\mu_i \nu_i}^{\sigma_i \rho_i}\,,  \label{HorndeskiLLGLagrangian}
\end{eqnarray}
where we defined $\mathcal{Q}_{\mu\nu}^{\sigma\rho} := \frac{1}{2} \frac{K_{[\mu}^\sigma K_{\nu]}^\rho}{u^\gamma u_\gamma}$ and depending on whether one expands the tensor $\mathcal{M}$ :
\begin{eqnarray}
\left( \lambda_{(p,s)} := \frac{\sigma_{(p,s)}}{2^p} \, ,\;\; \mathcal{P}_{\mu \nu}^{\sigma \rho}=\mathcal{M}_{\mu \nu}^{\sigma \rho} \right) \;\; \text{or} \;\; \left( \lambda_{(p,s)} := \binom{p}{s} \frac{(-1)^{p-s+1} 2^{-s}}{2(p-s)-1} \, ,\;\; \mathcal{P}_{\mu \nu}^{\sigma \rho}=\bar{R}_{\mu \nu}^{\sigma \rho} \right)\,. \label{Joulaka}
\end{eqnarray}
Similarly, considering a non-null boundary of the $d$-dimensional manifold $M$ such that its normal is given by $u$  and using Stokes's theorem directly yields the Myers boundary terms (see \cite{Myers,Olea}) which ensure a well-defined variational principle when added to the usual Lovelock-Lanczos action :
\begin{eqnarray}
\int_M d^d x \sqrt{-g} \nabla_\mu V_{(p)}^\mu = \int_{\partial M}   \left( \frac{u_\mu \sqrt{|h|} d^{d-1}x}{u^\gamma u_\gamma}\right) V_{(p)}^\mu  = \int_{\partial M}   d^{d-1}x  \mathcal{L}_{\text{Myers}}\,,
\end{eqnarray}
where
\begin{eqnarray}
\mathcal{L}_{\text{Myers}} := \sqrt{|h|}  \sum_{s=0}^{p-1} \theta_{(p,s)} \; h^{\nu \mu_1 \nu_1 \dots \mu_{p-1} \nu_{p-1}}_{\rho \sigma_1 \rho_1 \dots \sigma_{p-1} \rho_{p-1}}  \left(  \frac{K_\nu^\rho}{u^{\gamma}u_{\gamma}}  \right) \,  \prod_{i=1}^{s} \,  \mathcal{P}_{\mu_{i} \nu_{i}}^{\sigma_{i} \rho_{i}}    \,  \prod_{i=s+1}^{p-1} \,  \mathcal{Q}_{\mu_i \nu_i}^{\sigma_i \rho_i} \,,
\end{eqnarray}
and
\begin{eqnarray}
\left( \theta_{(p,s)} := \frac{\sigma_{(p,s)}}{2^p} \, ,\;\; \mathcal{P}_{\mu \nu}^{\sigma \rho}=\mathcal{M}_{\mu \nu}^{\sigma \rho} \right) \;\; \text{or} \;\; \left( \theta_{(p,s)} :=2 (p-s) \binom{p}{s} \frac{(-1)^{p-s+1} 2^{-s}}{2(p-s)-1} \, ,\;\; \mathcal{P}_{\mu \nu}^{\sigma \rho}=\bar{R}_{\mu \nu}^{\sigma \rho} \right)\,. \label{Zululu}
\end{eqnarray}
This was also shown in \cite{Deruelle} using Gaussian coordinates. As we will see, for $p>2$, there is a difference when one regularizes directly the Lagrangian \eqref{HorndeskiLLGLagrangian} or the Lovelock-Lanczos Hamiltonian, found in \cite{TeitZan}. It is therefore useful to see how the ADM formalism can be derived from the previous decomposition. 

\subsubsection{Lovelock-Lanczos Hamiltonian} 

Consider, a foliated $d$-dimensional manifold $M=\mathbb{R} \times \Sigma$, where the unit normal to the familly of hypersurfaces $\Sigma_t$ is given by $u_\mu$ with $u^{\gamma}u_{\gamma}= \epsilon = \pm1$, equipped with an ADM metric,
\begin{eqnarray}
ds^2 = g_{\mu\nu} dx^\mu dx^\nu = \epsilon N^2 dt^2 + h_{ab} \left( dx^a + N^a dt \right) \left( dx^b + N^b dt \right)  \,, \label{ADMmetric}
\end{eqnarray}
with the latin indices $a,b = 1, \dots , d-1$, $N$ and $N^a$ are respectively the lapse and the shift while $h_{ab}$ is the induced metric on $\Sigma_t$. We have $\sqrt{-g} =N \sqrt{|h|}$, with $h$ the determinant of the induced metric and $\mathcal{D}$ the covariant derivative compatible with $h_{ab}$. In order to work with $d$-dimensional indices, it will be convenient to consider $N_\mu := \left( s , N_a \right)$ with for example $s=\epsilon N$.

Now let's decompose the Horndeski Lagrangian in the following way : 
\begin{eqnarray}
\mathcal{L}_{(p)}^\mathcal{H} = \lambda_{(p,p)} h^{\mu_1 \nu_1 \dots \mu_p \nu_p}_{\sigma_1 \rho_1 \dots \sigma_p \rho_p} \prod_{i=1}^{p} \mathcal{P}_{\mu_i \nu_i}^{\sigma_i \rho_i} +K_{\mu_p}^{\sigma_p}   \sum_{s=0}^{p-1} \lambda_{(p,s)} h^{\mu_1 \nu_1 \dots \mu_p \nu_p}_{\sigma_1 \rho_1 \dots \sigma_p \rho_p} \left(  \frac{K_{\nu_p}^{\rho_p}  }{\epsilon} \right) \prod_{i=1}^{s} \mathcal{P}_{\mu_i \nu_i}^{\sigma_i \rho_i}  \prod_{i=s+1}^{p-1} \frac{K_{\mu_i}^{\sigma_i}K_{\nu_i}^{\rho_i}  }{\epsilon} \,. \label{DecompLagLLG}
\end{eqnarray}
Using that the extrinsic curvature on $\Sigma_t$ is given by $K_{ab} =\frac{1}{2 N} \left( \partial_0 h_{ab} - \mathcal{D}_{(a} N_{b)} \right)$, and applying this equality to the factorized extrinsic curvature in the second term, it is straightforward to check that :
\begin{eqnarray}
\sqrt{-g} \mathcal{L}_{(p)}^\mathcal{H}  =\pi^{\mu\nu}_{(p)} \partial_0 h_{\mu\nu}  - N \mathscr{H}_{(p)}  - N_\nu \mathscr{H}_{(p)}^\nu -2 \mathcal{D}_\mu\left(N_\nu \pi^{\mu\nu}_{(p)} \right) \,, \label{ADMDECOMP}
\end{eqnarray}
where 
\begin{eqnarray}
\begin{split}
\mathscr{H}_{(p)} &:= - \sqrt{|h|} \; \lambda_{(p,p)} h^{\mu_1 \nu_1 \dots \mu_p \nu_p}_{\sigma_1 \rho_1 \dots \sigma_p \rho_p} \prod_{i=1}^{p} \mathcal{P}_{\mu_i \nu_i}^{\sigma_i \rho_i} \; , \;\;\;\; \mathscr{H}_{(p)\nu} := -2 \mathcal{D}_\mu \pi^{\phantom{(p)}\mu}_{(p)\nu} \\
\pi^{\phantom{(p)}\mu_p}_{(p)\sigma_p} &:=  \frac{1}{2}  \, \sqrt{|h|} \; \sum_{s=0}^{p-1} \lambda_{(p,s)} h^{\mu_1 \nu_1 \dots \mu_p \nu_p}_{\sigma_1 \rho_1 \dots \sigma_p \rho_p} \left(  \frac{K_{\nu_p}^{\rho_p}  }{\epsilon} \right) \prod_{i=1}^{s} \mathcal{P}_{\mu_i \nu_i}^{\sigma_i \rho_i}  \prod_{i=s+1}^{p-1} \frac{K_{\mu_i}^{\sigma_i}K_{\nu_i}^{\rho_i}  }{\epsilon}\,, \label{HamiltonianMomentum}
\end{split}
\end{eqnarray}
which is in agreement with the results of \cite{TeitZan} for $ \lambda_{(p,s)} = \frac{\sigma_{(p,s)}}{2^p}$ and $\mathcal{P}_{\mu \nu}^{\sigma \rho}=\mathcal{M}_{\mu \nu}^{\sigma \rho}$. Therefore, $\pi^{\mu\nu}_{(p)}$ is the momentum conjugate to $h_{\mu\nu}$ while $\mathscr{H}$ and $\mathscr{H}^\mu$ are respectively the Hamiltonian and momentum constraints. Otherwise, in order to prove that this is indeed the named quantities and complete this direct derivation of the Hamiltonian formalism of LLG, one can make use of the relations :
\begin{eqnarray}
\pi^{\gamma \delta}_{(p)} := \frac{\partial \mathcal{L}_{(p)}^\mathcal{H}}{\partial \partial_0 h_{\gamma \delta}} =  \frac{\partial \mathcal{L}_{(p)}^\mathcal{H}}{\partial  \mathcal{Q}_{\mu \nu}^{\sigma \rho} }  \frac{\partial  \mathcal{Q}_{\mu \nu}^{\sigma \rho} }{\partial K_{\alpha\beta}}  \frac{\partial K_{\alpha\beta}}{\partial \partial_0 h_{\gamma \delta}}  \;\;\; , \;\;\; h^{\Bigcdot \mu \nu}_{\Bigcdot \sigma \rho} \frac{\partial  \mathcal{Q}_{\mu \nu}^{\sigma \rho} }{\partial K_{\alpha\beta}}  = h^{\Bigcdot \mu (\alpha}_{\Bigcdot \nu \sigma} h^{\beta)\sigma} K_\mu^\nu\,,
\end{eqnarray}
where the $\Bigcdot$ represents an arbitrary number of indices, so that using the second choice in Eq\eqref{Joulaka} to calculate the first partial derivative yields the result once $\pi_{(p)}$ is re-expressed in terms of $\mathcal{M}_{\mu \nu}^{\sigma \rho}$ rather than $\bar{R}_{\mu \nu}^{\sigma \rho}$ via Eq\eqref{Zululu}. Finally, in terms of these quantities we have 
\begin{eqnarray}
h_{\mu\nu}\pi^{\mu\nu}_{(p)}=\frac{d-2p}{2}\;  \mathcal{L}_{\text{Myers}}\,,
\end{eqnarray}
by using the property\eqref{GdeltaProp} of the GKD associated with the metric $h$. 

\subsubsection{Non-uniqueness \& null boundaries}

Before regularizing the Horndeski Lagrangian \eqref{HorndeskiLLGLagrangian} as well as the Hamiltonian constraint and momentum given in Eq\eqref{HamiltonianMomentum}, it is worth mentioning that the requirement to have an identically vanishing Lagrangian (in order to be regularizable) does not necessarily imply to consider the Horndeski Lagrangian or its associated Hamiltonian, so that the regularizations of Lovelock-Lanczos gravity that will follow in the next section are not unique in this sense.

In order to illustrate this point and show the non-uniqueness of the Horndeski decomposition we will restrict to General Relativity ($p=1$). In that case, given any two vector fields $u$ and $v$ such that $u_\mu v^\mu \neq 0$, 
\begin{eqnarray}
\begin{split}
R \,=& \, \delta_{\nu \nu_1 \nu_2}^{\mu \mu_1 \mu_2} \left( \frac{1}{2} \left( \frac{u_\mu v^\nu}{u_\gamma v^\gamma} \right) R_{\mu_1 \mu_2}^{\nu_1 \nu_2}   + \frac{ 2 u_\mu v^\nu \nabla_{\mu_1} u^{\nu_1} \nabla_{\mu_2} v^{\nu_2} - v_\mu v^\nu \nabla_{\mu_1} u^{\nu_1} \nabla_{\mu_2} u^{\nu_2} }{\left( u_\gamma v^\gamma \right)^2 }  \right) \\
&+ \delta_{\nu_1 \nu_2}^{\mu_1 \mu_2}  \left( \frac{ 2 u_\sigma v^{\nu_2}\nabla_{\mu_1} v^{\nu_1} \nabla_{\mu_2} u^\sigma  - v_\sigma v^\sigma \nabla_{\mu_1} u^{\nu_1} \nabla_{\mu_2} u^{\nu_2} }{\left( u_\gamma v^\gamma \right)^2 }  -2 \nabla_{\mu_1} \left[ \frac{v^{\nu_1} \nabla_{\mu_2} u^{\nu_2}}{u_\gamma v^\gamma } \right]    \right)\,, \label{RicciNull}
\end{split}
\end{eqnarray}
which can be checked by expanding the first GKD, the divergence and using that the antisymmetric commutation of two covariant derivatives yields a Riemann tensor term. Setting $v=u$, we have the following relation, 
\begin{eqnarray*}
\delta_{\nu_1 \nu_2}^{\mu_1 \mu_2}  \left(  2 u_\sigma v^{\nu_2}\nabla_{\mu_1} v^{\nu_1} \nabla_{\mu_2} u^\sigma  - v_\sigma v^\sigma \nabla_{\mu_1} u^{\nu_1} \nabla_{\mu_2} u^{\nu_2} \right) |_{v=u} = \delta_{\nu \nu_1 \nu_2}^{\mu \mu_1 \mu_2}  u_\mu u^\nu \nabla_{\mu_1} u^{\nu_1} \nabla_{\mu_2} u^{\nu_2}\,,
\end{eqnarray*}
which allows to reduce the Eq\eqref{RicciNull} to the Horndeski decomposition of General Relativity Eq\eqref{HorndeskiDecomposition} (i.e. to the covariant form of the ADM action) which is identically vanishing in two and higher dimensions.

Similarly, considering both $u$ and $v$ to be null yields, 
\begin{eqnarray}
\begin{split}
&R  + 2 \delta_{\nu_1 \nu_2}^{\mu_1 \mu_2}  \nabla_{\mu_1} \left( \frac{v^{\nu_1} \nabla_{\mu_2} u^{\nu_2}}{u_\gamma v^\gamma } \right) \\
&= \delta_{\nu \nu_1 \nu_2}^{\mu \mu_1 \mu_2} \left( \frac{1}{2} \left( \frac{u_\mu v^\nu}{u_\gamma v^\gamma} \right) R_{\mu_1 \mu_2}^{\nu_1 \nu_2}   + \frac{ 2 u_\mu v^\nu \nabla_{\mu_1} u^{\nu_1} \nabla_{\mu_2} v^{\nu_2} - v_\mu v^\nu \nabla_{\mu_1} u^{\nu_1} \nabla_{\mu_2} u^{\nu_2} }{\left( u_\gamma v^\gamma \right)^2 } \right)\,, \label{NulldecompGR}
\end{split}
\end{eqnarray}
which is also identically vanishing for $d \geq 2$. Therefore, we see that a two dimensional regularization of GR could in principle be obtained either from the its ADM Lagrangian, or from the RHS of the previous equation, probably leading to different regularized theory. Similarly, we can expect that the regularization of the full Lovelock-Lanczos theory would depend on the specific choices of additional structure required to have an identically vanishing action in critical (and lower) dimensions.
\\

Remark that, just like in the non-null case,  the previous covariant boundary term is precisely what is required in order to have a well-defined variational principle in the presence of a null boundary (see for example \cite{NullBoundariesGR1, NullBoundariesGR2}). Indeed, it was shown in \cite{NullBoundariesGR2} that a unified boundary counter-term, valid for both time(space)-like and null boundaries is given as the integral over the boundary of the term : 
\begin{eqnarray}
\mathcal{C}_0 := 2 \sqrt{-g} \Pi^\mu_\nu \nabla_\mu u^\nu\,,
\end{eqnarray}
where $\Pi^\mu_\nu := \delta^\mu_\nu + v^\mu u_\nu$ is the projector to the boundary and $v^\mu u_\mu=-1$. For a time(space)-like boundary, the choice $v^\mu = - u^\mu / u^\gamma u_\gamma$ yields to the Gibbons–Hawking–York boundary term and thus to a well-defined variational principle. For null boundary $\mathcal{B}$, it is well-known that two null vectors fields are required in order to define the projectors to the boundary. Choosing $u_\mu = l_\mu$ and $v^\mu = k^\mu$ the normals to the boundary with $k^\mu k_\mu= l^\mu l_\mu=0$ and defining $q^\mu_\nu := \delta^\mu_\nu + k^\mu l_\nu + l^\mu k_\nu$, the projector to the boundary orthogonal to both $k^\mu$ and $l^\mu$, yields 
\begin{eqnarray}
\mathcal{C}_0 = 2 \sqrt{-g} \left( \delta^\mu_\nu \nabla_\mu l^\nu -  k^\mu l_\nu \nabla_\mu l^\nu \right) = 2 \sqrt{-g} \left( \delta^\mu_\nu \nabla_\mu l^\nu \right)  =  2 \sqrt{-g} \left( q^\mu_\nu  \nabla_\mu l^\nu  - l^\mu k_\nu   \nabla_\mu l^\nu  \right)\,, \label{UnifBoundarytermNull}
\end{eqnarray}
where the second term ensures a well-defined variational principle, while the first term only contains derivatives of the metric along the null boundary (see Eq(74) of \cite{NullLLG} for more details about that point).

Back to the boundary term appearing in Eq\eqref{RicciNull}, we set $u=l$ and $v=k$. The effect of using Stokes's theorem will be to change the covariant derivative $\nabla_{\mu_1}$ into $l_{\mu_1}$, so that : 
\begin{eqnarray}
2 \delta_{\nu_1 \nu_2}^{\mu_1 \mu_2}  \nabla_{\mu_1} \left( \frac{v^{\nu_1} \nabla_{\mu_2} u^{\nu_2}}{u_\gamma v^\gamma } \right) \longrightarrow - 2 \delta_{\nu_1 \nu_2}^{\mu_1 \mu_2}  l_{\mu_1} k^{\nu_1} \nabla_{\mu_2} l^{\nu_2} \,,
\end{eqnarray}
where we set $\left( l_\mu k^\mu \right) |_{\mathcal{B}}=-1$ to allow a direct comparison with Eq\eqref{UnifBoundarytermNull}. Therefore, 
\begin{eqnarray}
- 2 \delta_{\nu_1 \nu_2}^{\mu_1 \mu_2}  l_{\mu_1} k^{\nu_1} \nabla_{\mu_2} l^{\nu_2}  = -2 l_\alpha k^\alpha \delta^\mu_\nu \nabla_\mu l^\nu + 2 l_\mu k^\nu \nabla_\nu l^\mu = 2  \delta^\mu_\nu \nabla_\mu l^\nu \,,
\end{eqnarray}
which is exactly the unified boundary term given in Eq\eqref{UnifBoundarytermNull} and found in \cite{NullBoundariesGR2}.
\\

Finally, although we will not try here to generalize the regularizable formula \eqref{RicciNull} to the full Lovelock-Lanczos theory for which the null boundary terms are very complicated (see the impressive formulae of \cite{NullLLG} for their variations), it is nonetheless interesting to note that there is a simple algorithm to find the generalizations of the Horndeski decomposition for an arbitrary number of additional structure $u_{(q)}$ of any tensorial structure. To understand this, it suffices to consider a non-minimally coupled action in critical dimension ($d=2p$) :  
\begin{eqnarray*}
I = \int \sqrt{-g} d^{d} x F\left( u_{(q)} \right) \mathcal{L}_{(d/2)}   \,,
\end{eqnarray*} 
where $F\left( u_{(q)} \right)$ is an arbitrary scalar function of the fields $u_{(q)}$.  Such theory yields by construction second order field equations for both the metric and the additional fields, so that it must be possible to decompose it on its corresponding ``Horndeski basis", up to boundary term. In order to do so, it must be possible to write $\mathcal{L}_{(d/2)} =- \nabla_\mu V^\mu\left[ u_{(q)}  ;g\right]$. The corresponding decomposition of Lovelock-Lanczos gravity would then be reached by calculating in generic dimension $\mathcal{L}_{(p)} +\nabla_\mu V^\mu\left[ u_{(q)}  ;g\right]$,
which should by construction be identically vanishing in critical and lower dimensions, i.e. possible to express in terms of a tensorial quantity fully contracted with a GKD containing $2p+1$ antisymmetrized indices.

This kind of argument has been followed through in four dimensions (for the Gauss-Bonnet scalar) in the context of scalar-tensor theories in \cite{Unified} (see Eq(98) and Eq(109) there), where the authors listed all the possible divergences (made of a scalar and a metric field) relating all the possible second order scalar-tensor Lagrangians between themselves. One of these divergences must necessarily correspond to the four dimensional Gauss-Bonnet scalar in the presence of an additional scalar field.

All these alternative formulations of Lovelock-Lanczos gravity could be suited for four-dimensional regularizations, however we leave this for possible future works.

\subsection{Regularizations of the Horndeski-Lovelock-Lanczos action}

From the Horndeski form of LLG and its associated Hamiltonian, we can now proceed towards four dimensional $3$D-covariant regularizations of the theory. Without further assumptions on the form of the ADM metric, the $(d-1)$-dimensional Riemann tensor $\bar{R}_{\mu \nu}^{\sigma \rho}$ cannot be decomposed into a non-trivial form made of Kronecker deltas as it was the case via metric transformations in Eq\eqref{RiemannTransfo}, so that no non-trivial dimensional factors can appear in the present context. Although we will see at the end of the last section of this paper \ref{Sec.BR}\ref{NUniqueReg} that decomposing the $(d-1)$-dimensional spatial metric into $3$-dimensional building blocks enables to regularize the Bianchi I sector of LLG, which is not regularizable otherwise, we are now going to follow the results of \cite{3Dcov1,3Dcov2} on the Gauss-Bonnet theory, in which the spatial metric is not decomposed.

\subsubsection{Choices of regularization}

In that case, the best one can do is to introduce trivial dimensional factors coming from the Weyl decompositions of the quantities  appearing in $\mathcal{L}_{(p)}^\mathcal{H}$, $\mathscr{H}_{(p)}$ and $\pi_{(p)}$. Already at this point, there are different possible choices : either one uses $d$ or $d-1$ dimensional Weyl decompositions , i.e. for quantities with the same symmetries as the Riemann tensor :
\begin{eqnarray}
^{(\text{w})}X_{\mu \nu}^{\sigma \rho} :=X_{\mu \nu}^{\sigma \rho}  - \frac{1}{D-2} \bar{\delta}^{[\sigma}_{[\mu} X_{\nu]}^{\rho]} +  \frac{X}{(D-1)(D-2)} \bar{\delta}^{\sigma}_{[\mu} \bar{\delta}^{\rho}_{\nu]} \,,
\end{eqnarray}
where $X_{\nu}^{\rho} := X_{\mu \nu}^{\mu \rho}$ and $X:=X_{\nu}^{\nu}$, with $D=d, \bar{\delta}=\delta$ or $D=d-1, \bar{\delta}=h$ for respectively $d$ or $d-1$ dimensional  quantities. Using only $d$-dimensional Weyl decomposition from Eq\eqref{HorndeskiDecomposition}, the procedure will end-up yielding $4$D-covariant theories while for $(d-1)$ ones, coming from Eq\eqref{HorndeskiLLGLagrangian}, the invariance will be restricted to $3$D-diffeomorphisms. In both cases, using the property\eqref{GdeltaProp} of the GKD, the Horndeski Lagrangian becomes :
\begin{eqnarray}
\begin{split}
\mathcal{L}_{(p)}^\mathcal{H} =& \sum_{s=0}^p  \sum_{k=0}^s \sum_{l=0}^k \sum_{\bar{k}=0}^{p-s} \sum_{\bar{l}=0}^{\bar{k}} \rho_{(p,s,k,l,\bar{k},\bar{l},D)} \frac{(d-2p-1+k+l+\bar{k}+\bar{l})!}{(d-2p-1)!} \, \mathcal{P}^l \mathcal{Q}^{\bar{l}}  \\
&h^{\mu_1 \nu_1 \dots \mu_{s-k} \nu_{s-k} \alpha_1 \dots \alpha_{k-l} \bar{\mu}_1 \bar{\nu}_1 \dots \bar{\mu}_{p-s-\bar{k}} \bar{\nu}_{p-s-\bar{k}} \bar{\alpha}_1 \dots \bar{\alpha}_{\bar{k}-\bar{l}}}_{\sigma_1 \rho_1 \dots \sigma_{s-k} \rho_{s-k} \beta_1 \dots \beta_{k-l} \bar{\sigma}_1 \bar{\rho}_1 \dots \bar{\sigma}_{p-s-\bar{k}} \bar{\rho}_{p-s-\bar{k}} \bar{\beta}_1 \dots \bar{\beta}_{\bar{k}-\bar{l}}}   \prod_{i=1}^{s-k} {^{(\text{w})}}\mathcal{P}_{\mu_i \nu_i}^{\sigma_i \rho_i}  \prod_{i=1}^{k-l} \mathcal{P}_{\alpha_i}^{\beta_i}  \prod_{i=1}^{p-s-\bar{k}} {^{(\text{w})}}\mathcal{Q}_{\bar{\mu}_i \bar{\nu}_i}^{\bar{\sigma}_i \bar{\rho}_i}  \prod_{i=1}^{\bar{k}-\bar{l}}    \mathcal{Q}_{\bar{\alpha}_i}^{\bar{\beta}_i}
\end{split}
\end{eqnarray}
with
\begin{eqnarray}
\rho_{(p,s,k,l,\bar{k},\bar{l},D)}:= \lambda_{(p,s)} \binom{s}{k}\binom{k}{l}\binom{p-s}{\bar{k}}\binom{\bar{k}}{\bar{l}}  \frac{(-1)^{l+\bar{l}} 2^{2(k+\bar{k})-l-\bar{l}}}{(D-1)^{l+\bar{l}} (D-2)^{k+\bar{k}}} \,.
\end{eqnarray}
For $d$-dimensional Weyl decompositions, $\mathcal{P}_{\mu \nu}^{\sigma \rho} =R_{\mu \nu}^{\sigma \rho} $, $\mathcal{Q}_{\mu\nu}^{\sigma\rho} := \frac{1}{2} \frac{\nabla_{[\mu}u^\sigma \nabla_{\nu]}u^\rho}{u^\gamma u_\gamma}$ and $ \lambda_{(p,s)} := \frac{\sigma_{(p,s)}}{2^p} $, while for $(d-1)$-ones, $\mathcal{P}, \mathcal{Q}$ and $\lambda_{(p,s)}$ satisfy Eq\eqref{Joulaka}.  Remark that one could also have used mixed decompositions, $d$ and $(d-1)$-dimensional ones for either $\mathcal{Q}$ or $\mathcal{P}$, in which case the dependence of $\rho_{(p,s,k,l,\bar{k},\bar{l},D)}$ on $D$ would be modified accordingly.
\\

Among the previous options regarding the Weyl decompositions of $\mathcal{L}_{(p)}^\mathcal{H}$, the first deserving some attention is the one yielding to $4$D-covariant regularization. As an example, consider the Gauss-Bonnet theory ($p=2$), for which the counter-terms reads :
\begin{eqnarray}
\mathcal{L}_{(2)}^{\text{ct}} =h^{\mu_1 \nu_1  \mu_2 \nu_2}_{\sigma_1 \rho_1  \sigma_2 \rho_2} \left(  \frac{8}{3} \, ^{(\text{w})}\mathcal{Q}_{\mu_1 \nu_1}^{\sigma_1 \rho_1} \, ^{(\text{w})}\mathcal{Q}_{\mu_2 \nu_2}^{\sigma_2 \rho_2} + 2  ^{(\text{w})}\mathcal{Q}_{\mu_1 \nu_1}^{\sigma_1 \rho_1} W_{\mu_2 \nu_2}^{\sigma_2 \rho_2}  +\frac{1}{4}  W_{\mu_1 \nu_1}^{\sigma_1 \rho_1} W_{\mu_2 \nu_2}^{\sigma_2 \rho_2}    \right)\,,
\end{eqnarray}
with $W$ the $d$-dimensional Weyl tensor. Now remark the following. As shown above, the Horndeski Lagrangian $\mathcal{L}_{(p)}^\mathcal{H}$ depends on the $(d-1)$-dimensional Riemann tensor and the extrinsic curvature. Moreover, the projection of the $d$-dimensional tensor $^{(\text{w})}\mathcal{Q}$ involves solely the extrinsic curvature. However, the projection of the $d$-dimensional Weyl tensor involves both time derivatives of $K_{\mu\nu}$ and spatial (covariant) derivatives of the acceleration $a_\mu := \epsilon^{-1} u^\nu  \nabla_\nu u_\mu$. Therefore the resulting regularized theory contains higher than two order derivatives and we will not further discuss this possibility, nor its higher curvature generalizations. 
\\

Finally, just like with the metric transformations of the previous section, there is also here the possibility to consider directly the four dimensional regularization of $\mathcal{L}_p$ or the four dimensional restriction of the critical regularization. This choice has no effect on the tensorial structure of the resulting theory, only on the dimensional factors inside $\rho_{(p,s,k,l,\bar{k},\bar{l},D)}$, which would be evaluated at respectively $d=4$ or $d=2p$.  Indeed, it is clear from the previous expression that the terms of the previous sum such that $2p+1-k-l-\bar{k}-\bar{l}>4$ do not contain the required overall factor $(d-4)$. Let's collect these terms into some $\mathcal{L}_{(p)}^{\text{ct}}$ and the remaining ones into some $\mathscr{L}_{(p)}$ such that $\mathcal{L}_{(p)}^\mathcal{H}=\mathscr{L}_{(p)}  +\mathcal{L}_{(p)}^{\text{ct}} $. By construction, only $\mathscr{L}_{(p)}$ is  proportional to $(d-4)$, so that it is this part of the Horndeski-Lovelock-Lanczos action that can be regularized. The general structure of the counter terms is given by 
\begin{eqnarray}
\begin{split}
\mathcal{L}_{(p)}^{\text{ct}}& =  \sum_{s=0}^p \lambda_{(p,s)} h^{\mu_1 \nu_1 \dots \mu_p \nu_p}_{\sigma_1 \rho_1 \dots \sigma_p \rho_p} \prod_{i=1}^{s}  {^{(\text{w})}}\mathcal{P}_{\mu_i \nu_i}^{\sigma_i \rho_i}  \prod_{i=s+1}^p  {^{(\text{w})}}\mathcal{Q}_{\mu_i \nu_i}^{\sigma_i \rho_i}  +(d-2p) \sum_{m=1}^{2p-5} L_m  \prod_{i=1}^{m-1} (d-2p+i) \\
& =: \mathcal{L}_{(p),1}^{\text{ct}} + (d-2p) \mathcal{L}_{(p),2}^{\text{ct}} \,,
\end{split}
\end{eqnarray}
where the first term identically vanishes for $d\leq 2p$, $L_1$ for $d\leq2p-1$, $L_2$ for $d \leq 2p-2$, etc. Thus, considering the critical regularization $\lim_{d\to2p}\left(\mathcal{L}_{(p)}^\mathcal{H}-\mathcal{L}_{(p),1}^{\text{ct}}  \right)  /(d-2p)$ and then using the resulting theory in four dimensions or alternatively regularizing directly in four dimensions, $\lim_{d\to4}\left(\mathcal{L}_{(p)}^\mathcal{H}-\mathcal{L}_{(p)}^{\text{ct}}  \right)  /(d-4)$ simply changes the constant $D$ inside $\rho_{(p,s,k,l,\bar{k},\bar{l},D)}$, so that we will keep that constant free in the following.

\subsubsection{Four dimensional $3$D-covariant regularizations}

Let's focus on the $(d-1)$ Weyl decompositions as it was done in \cite{3Dcov1,3Dcov2}, so that $D=d-1$. In this case, the counter-terms are purely $(d-1)$-dimensional so that the overall theory $\mathscr{L}_{(p)} = \mathcal{L}_{(p)}^\mathcal{H} -\mathcal{L}_{(p)}^{\text{ct}} $ is so as well. We will use the notation $\big\langle  \prod_{i=1}^m X_{(i)} \big\rangle^{\mu_1\dots\mu_n}_{\nu_1\dots\nu_n}  =h^{\beta_1 \dots \beta_m \mu_1\dots\mu_n }_{\alpha_1 \dots \alpha_m \nu_1\dots\nu_n} \prod_{i=1}^m X_{(i)\alpha_i}^{\beta_i} $ for a set of $(1,1)$ tensors $X_{(i)}$.  The regularizable theory becomes :
\begin{eqnarray}
\begin{split}
\mathscr{L}_{(p)} =    \sum_{s=0}^p  \Big(&  a_{p-s} \mathcal{P}^3  \big\langle \mathcal{Q}^3 \big\rangle +a_s \mathcal{Q}^3 \big\langle \mathcal{P}^3 \big\rangle +  c_{p-s}  \mathcal{Q}\mathcal{P}^2 \big\langle \mathcal{Q}^2 \mathcal{P} \big\rangle + c_{s}   \mathcal{Q}^2 \mathcal{P}\big\langle \mathcal{Q} \mathcal{P}^2 \big\rangle     \\
&+ b_{p-s}  \mathcal{Q}\mathcal{P}^3  \big\langle \mathcal{Q}^2 \big\rangle + b_s  \mathcal{Q}^3 \mathcal{P}  \big\langle \mathcal{P}^2 \big\rangle +e_s \mathcal{Q}^2 \mathcal{P}^2  \big\langle \mathcal{Q}\mathcal{P} \big\rangle+f_p \mathcal{Q}^3\mathcal{P}^3 \Big) \mathcal{P}^{s-3} \mathcal{Q}^{p-s-3} 
\\&   \frac{\lambda_{(p,s)} (-2)^p}{3(d-3)^p (d-2)^{p-1}} \frac{(d-4)!}{(d-2p-1)!} \,, \label{RegLag3D}
\end{split}
\end{eqnarray}
with $\mathcal{Q}, \mathcal{P}$ and $\lambda$  satisfying Eq\eqref{Joulaka} and
\begin{eqnarray}
\begin{split}
a_s &:=-4 (d-2)^2 s(s-1)(s-2) \, , \; b_s := 6(d-2)(d-3) s(s-1)\, , \;  c_s:=-12 (d-2)^2 s(s-1)(p-s) \, ,\\
e_s &:= 12 (d-2)(d-3) s(p-s) \, , \; f_p := 3(d-3) \left( d-1-2p(d-2)  \right) \,.
\end{split}
\end{eqnarray}
In order to derive the Hamiltonian of the theory, we first  choose the form  $ \lambda_{(p,s)} = \frac{\sigma_{(p,s)}}{2^p}$ and $\mathcal{P}_{\mu \nu}^{\sigma \rho}=\mathcal{M}_{\mu \nu}^{\sigma \rho}$. From the expression of the Hamiltonian constraint Eq\eqref{HamiltonianMomentum} and the way it is derived from its Lagrangian \eqref{DecompLagLLG}, it is clear that the regularized one is given by the $p^{\text{th}}$ term of the sum : 
\begin{eqnarray}
\mathscr{H}_{(p)} = \sqrt{|h|}\; \left( a_p  \big\langle \mathcal{M}^3 \big\rangle+ b_p \mathcal{M}  \big\langle \mathcal{M}^2 \big\rangle  + f_p \mathcal{M}^3 \right)\mathcal{M}^{p-3} \frac{(-1)^{p+1}}{3(d-2)^{p-1}(d-3)^p} \frac{(d-4)!}{(d-2p-1)!} \,.\label{RegHamCons3D}
\end{eqnarray}
This also corresponds to the regularization of the Lovelock-Lanczos Hamiltonian constraint given by Eq\eqref{HamiltonianMomentum}. By direct calculation we obtain the momentum conjugate :
\begin{eqnarray}
\pi^{\phantom{(p)}\mu}_{(p)\nu} = \frac{\sqrt{|h|}}{12(d-3)^p (d-2)^{p-1}}  \frac{(d-4)!}{(d-2p-1)!}  \sum_{s=0}^{p-1}  \binom{p}{s} \frac{(-1)^{s+1} 2^{p-s}}{2 (p-s)-1}\frac{K^\gamma_\delta}{u^\zeta u_\zeta} h^{[\mu}_{[\nu} \mathcal{S}^{\delta]}_{\gamma]}\mathcal{Q}^{p-s-4} \bar{R}^{s-3}\,, \label{RegPi13D}
\end{eqnarray}
where
\begin{eqnarray}
\begin{split}
\mathcal{S}^{\delta}_{\gamma} :=&   \mathcal{Q} \left( 3 a_{p-s} \bar{R}^3 \big\langle \mathcal{Q}^2 \big\rangle^{\delta}_{\gamma}  + 2 c_{p-s} \bar{R}^2 \mathcal{Q} \big\langle  \mathcal{Q} \bar{R} \big\rangle^{\delta}_{\gamma} + c_s \bar{R} \mathcal{Q}^2   \big\langle \bar{R}^2 \big\rangle^{\delta}_{\gamma} +2 b_{p-s} \bar{R}^3 \mathcal{Q} \big\langle  \mathcal{Q} \big\rangle^{\delta}_{\gamma} + e_s \bar{R}^2 \mathcal{Q}^2 \big\langle   \bar{R} \big\rangle^{\delta}_{\gamma} \right)  
\\
&+h^\delta_\gamma \Big[(p-s-3) a_{p-s} \bar{R}^3  \big\langle \mathcal{Q}^3 \big\rangle +(p-s) a_s \mathcal{Q}^3 \big\langle \bar{R}^3 \big\rangle + (p-s-2) c_{p-s}  \mathcal{Q}\bar{R}^2 \big\langle \mathcal{Q}^2 \bar{R} \big\rangle \\
&\phantom{+\delta^\delta_\gamma \Big[+}+(p-s-1) c_{s}   \mathcal{Q}^2 \bar{R}\big\langle \mathcal{Q} \bar{R}^2 \big\rangle     + (p-s-2) b_{p-s}  \mathcal{Q}\bar{R}^3  \big\langle \mathcal{Q}^2 \big\rangle + (p-s) b_s  \mathcal{Q}^3 \bar{R}  \big\langle \bar{R}^2 \big\rangle\\
&\phantom{+\delta^\delta_\gamma \Big[+} +(p-s-1) e_s \mathcal{Q}^2 \bar{R}^2  \big\langle \mathcal{Q}\bar{R} \big\rangle+ (p-s) f_p \mathcal{Q}^3\bar{R}^3  \Big]\,,
\end{split}
\end{eqnarray}
so that $\sqrt{-g}\mathscr{L}_{(p)}  = - N \mathscr{H}_{(p)} +2  K_\nu^\mu  \pi^{\phantom{(p)}\nu}_{(p)\mu}$, from which follows the usual ADM decomposition Eq\eqref{ADMDECOMP}. Although this expression could be simplified by using identities of the form $h^{[\mu}_{[\nu}  \big\langle  X \big\rangle_{\gamma]}^{\delta]} K^\gamma_\delta = \big\langle X K \big\rangle^\mu_\nu +  X \big\langle K \big\rangle^{\mu}_{\nu}$, $(d-3)h^{[\mu}_{[\nu}h^{\delta]}_{\gamma]} = h^{\mu\delta\sigma}_{\nu\gamma\rho} h_\sigma^\rho$,  in four dimensions $h^{[\mu}_{[\nu}  h^{\delta] \sigma \alpha}_{\gamma] \rho \beta} = h^{[\alpha}_{\nu}  h^{\sigma] \mu \delta}_{\gamma \rho \beta}  + h^{[\delta}_{\gamma}  h^{\mu] \sigma \alpha}_{\nu \rho \beta}$ and hypergeometric functions to evaluate the sum, the result remains quite complicated, so we leave it in the above form.
\\

It is at this point that a difference appears between the regularization of the Horndeski Lagrangian and that of the Hamiltonian : in this last the form of the momentum is more restricted because there is a factorized extrinsic curvature in Eq\eqref{HamiltonianMomentum}, so that one less tensor $\mathcal{Q}_{\mu\nu}^{\sigma\rho}$ can be decomposed. Indeed, regularizing directly the momentum in Eq\eqref{HamiltonianMomentum} yields 
\begin{eqnarray}
\begin{split}
\check{\pi}^{\phantom{(p)}\sigma}_{(p)\mu} =\frac{\left(-1\right)^{p-1} \sqrt{|h|}}{4 \left[(d-3)(d-2)\right]^{p-1}}\frac{(d-4)!}{(d-2p-1)!}   \sum_{s=0}^{p-1}& \sigma_{(p,s)}  \frac{K^\nu_\rho}{u^\zeta u_\zeta} h_{\mu\nu\alpha}^{\sigma\rho\beta} S^\alpha_\beta  \mathcal{Q}^{p-s-2} \mathcal{M}^{s-1} \,, \label{RegPi23D}
\end{split} 
\end{eqnarray}
where 
\begin{eqnarray}
S^\alpha_\beta  :=  \mathcal{Q} \mathcal{M} h^\alpha_\beta -2 (d-2) \left( (p-s-1) \mathcal{M} \mathcal{Q}^\alpha_\beta + s \mathcal{Q} \mathcal{M}^\alpha_\beta \right) \,.
\end{eqnarray}
In both cases, the regularizable terms agree with General Relativity : 
\begin{eqnarray}
\mathscr{L}_{(1)} =2 \bar{R}- \mathcal{M} = \bar{R} + \frac{K^2 - K^{\mu}_\nu K^\nu_\mu }{\epsilon} \; , \;\; \mathscr{H}_{(1)}=- \sqrt{|h|}\;  \mathcal{M}  \; , \;\;  \check{\pi}^{\phantom{(1)}\nu}_{(1)\mu}= \pi^{\phantom{(1)}\nu}_{(1)\mu} =\sqrt{|h|} \left( \frac{K h^\nu_\mu - K^\nu_\mu}{\epsilon} \right)\,,
\end{eqnarray}
and with the result of \cite{3Dcov1} in the Gauss-Bonnet case
\begin{eqnarray}
\begin{split}
&\lim_{d\to4} \frac{\mathscr{L}_{(2)}}{d-4} = \frac{1}{2} \left(8 \bar{R}^2 - 4 \bar{R} \mathcal{M} - \mathcal{M}^2 \right) - \frac{4}{3} \left( 8 \bar{R}^\mu_\nu \bar{R}^\nu_\mu -4 \bar{R}^\mu_\nu \mathcal{M}_\mu^\nu - \mathcal{M}^\mu_\nu \mathcal{M}_\mu^\nu \right) \,,\\
&\lim_{d\to 4} \frac{\mathscr{H}_{(2)}}{d-4}  =  \sqrt{|h|}\; \left(-\frac{3}{2} \mathcal{M}^2+ 4 \mathcal{M}^\mu_\nu  \mathcal{M}_\mu^\nu  \right)\,,\\
&\lim_{d\to 4} \frac{\pi^{\phantom{(2)}\nu}_{(2)\mu}}{d-4}  =\frac{8}{3} \frac{\sqrt{|h|} h_{\mu\beta\rho}^{\nu\alpha\sigma} K^\beta_\alpha}{\epsilon}\; \left( \bar{R}^\rho_\sigma- \frac{1}{4} h^\rho_\sigma  \bar{R} + \frac{1}{2} \left( \mathcal{M}^\rho_\sigma - \frac{1}{4} h^\rho_\sigma \mathcal{M}  \right) \right)\,,
\end{split}
\end{eqnarray}
because $\pi_{(2)} = \check{\pi}_{(2)}$. However, the momenta differ for $p>2$ and thus, also the associated Lagrangians $\mathscr{L}_{(p)} $ and $\check{\mathscr{L}}_{(p)}:=\left(- N \mathscr{H}_{(p)} +2  K_\nu^\mu  \check{\pi}^{\phantom{(p)}\nu}_{(p)\mu} \right) /\sqrt{-g}$. For example, the cubic regularized Lagrangian Eq\eqref{RegLag3D} and Hamiltonian constraint Eq\eqref{RegHamCons3D} are given by  
\begin{eqnarray}
\begin{split}
\frac{\mathscr{L}_{(3)}}{(d-4)(d-5)(d-6)}&= \frac{1}{5(d-3)^3} \left(- \frac{d^2+2d-7}{(d-2)^2} \mathscr{S}_{3,1}-16 \mathscr{S}_{3,2} + \frac{12 (d-1)}{(d-2)} \mathscr{S}_{3,3} \right)\,,\\
\frac{\mathscr{H}_{(3)}}{(d-4)(d-5)(d-6)} &=\frac{\sqrt{|h|}}{(d-3)^3} \left(- \frac{d^2+2d -7}{(d-2)^2} \mathcal{M}^3 -16 \mathcal{M}^\mu_\nu \mathcal{M}_\mu^\gamma \mathcal{M}_\gamma^\nu + \frac{12(d-1)}{(d-2)}\mathcal{M} \mathcal{M}^\mu_\nu \mathcal{M}^\nu_\mu \right) ,
\end{split}
\end{eqnarray}
where 
\begin{eqnarray}
\begin{split}
\mathscr{S}_{3,1} =& \mathcal{M}^3 +2 \mathcal{M} \bar{R}  \left( \mathcal{M} + 4 \bar{R} \right)  - 16 \bar{R}^3\,, \\
\mathscr{S}_{3,2} =& \mathcal{M}^\mu_\nu \mathcal{M}_\mu^\gamma \mathcal{M}_\gamma^\nu + 2 \mathcal{M}^\mu_\nu  \bar{R}_\gamma^\nu \left(\mathcal{M}_\mu^\gamma + 4  \bar{R}_\mu^\gamma\right) - 16\bar{R}^\mu_\nu \bar{R}_\mu^\gamma \bar{R}_\gamma^\nu \,,\\
\mathscr{S}_{3,3} =&  \mathcal{M} \mathcal{M}^\mu_\nu \mathcal{M}^\nu_\mu + \frac{2}{3}\left( \mathcal{M}^\mu_\nu \left(   \mathcal{M}^\nu_\mu \bar{R}+2 \mathcal{M}  \bar{R}^\nu_\mu\right) + 4 \bar{R}^\mu_\nu \left(\mathcal{M} \bar{R}^\nu_\mu + 2 \mathcal{M}^\nu_\mu \bar{R} \right)\right)- 16 \bar{R}^\mu_\nu \bar{R}^\nu_\mu \bar{R}  \,,
\end{split}
\end{eqnarray}
while the difference between the two regularizable momenta (Eqs\eqref{RegPi13D} and \eqref{RegPi23D}) is non-vanishing and given by\footnote{To verify that $\mathscr{L}_{(p)}\neq\check{\mathscr{L}}_{(p)}$, we also checked that $K^\mu_\nu \left(\pi^{\phantom{(p)}\nu}_{(p)\mu}-\check{\pi}^{\phantom{(p)}\nu}_{(p)\mu}\right) \neq 0$.} : 
\begin{eqnarray}
\begin{split}
\frac{\pi^{\phantom{(3)}\nu}_{(3)\mu}-\check{\pi}^{\phantom{(3)}\nu}_{(3)\mu} }{(d-4)...(d-6)}=  \frac{4 \sqrt{|h|}}{5\epsilon(d-3)^3} \bigg(& \frac{\epsilon \pi^{\phantom{(1)}\nu}_{(1)\mu}}{d-2}  \left( 3\left(\mathcal{M}^2 -  \mathcal{M}^\alpha_\beta \mathcal{M}_\alpha^\beta \right)+4 \left( \mathcal{M}\bar{R} -\mathcal{M}^\alpha_\beta\bar{R}_\alpha^\beta \right) + 8 \left( \bar{R}^2 -\bar{R}^\alpha_\beta\bar{R}_\alpha^\beta\right) \right)
\\&-2 K^\sigma_\rho \left( 3 \mathcal{M}^\nu_{[\mu} \mathcal{M}_{\sigma]}^\rho + 2 \mathcal{M}^{[\nu}_{[\mu} \bar{R}_{\sigma]}^{\rho]} + 8  \bar{R}^\nu_{[\mu} \bar{R}_{\sigma]}^\rho \right)   \bigg)+ \mathscr{J}_{(3)\mu}^\nu \,,
\end{split}
\end{eqnarray}
with 
\begin{eqnarray}
\begin{split}
\frac{\check{\pi}^{\phantom{(3)}\nu}_{(3)\mu} }{(d-4)...(d-6)}= \frac{\sqrt{|h|} K^\nu_\rho h_{\mu\nu\alpha}^{\sigma\rho\beta}}{5(d-3)^2(d-2)\epsilon}  \Big(& -4 \left(3 \mathcal{M}^\alpha_\beta \mathcal{M} + 2 \left( \mathcal{M} \bar{R}^\alpha_\beta + \mathcal{M}^\alpha_\beta \bar{R}\right)+ 8 \bar{R}^\alpha_\beta\bar{R} \right) \\& +\frac{1}{d-2} h^\alpha_\beta \left( 3 \mathcal{M}^2 + 4 \mathcal{M} \bar{R} + 8 \bar{R}^2\right)  \Big) \,,
\end{split}
\end{eqnarray}
and in four dimensions,
\begin{eqnarray}
\mathscr{J}_{(3)\mu}^\nu:= \frac{4 \sqrt{|h|}  h_{\mu\beta\rho\delta}^{\nu\alpha\sigma\gamma} K^{\beta}_\alpha}{5\epsilon(d-3)^3}\left( 3 \mathcal{M}^\rho_\sigma \mathcal{M}^\delta_\gamma+4 \mathcal{M}^\rho_\sigma \bar{R}^\delta_\gamma+8\bar{R}^\rho_\sigma \bar{R}^\delta_\gamma \right) =0 \,.
\end{eqnarray}
The quartic case is reported in Appendix\ref{sec.L43Dcov} and the same conclusion applies.

\subsubsection{``Consistent $D\to 4$ Einstein-Lovelock-Lanczos gravity" with two degrees of freedom}

Finally, from these results and following closely \cite{3Dcov1}, it is possible to define two inequivalent families of theories which would seemingly propagate only the two degrees of freedom of the graviton, via the introduction of a gauge fixing condition. To do so, consider the total Hamiltonian given by : 
\begin{eqnarray}
H =\frac{1}{2 \tilde{\kappa}_0} \int d^3x \left( N  \mathscr{H} + N^a \mathscr{H}_a  + \lambda^0 \pi_0 + \lambda^a \pi_a + \lambda_{GF}\mathscr{G}+2 \mathcal{D}_a\left(N_b \pi^{ab}\right) \right) \,, \label{TotalHamilt}
\end{eqnarray}
where $\lambda^0,\lambda^a$ and $\lambda_{GF}$ are Lagrange multipliers, $\pi_0, \pi_a$ and $\pi^{ab}$ are respectively the momenta conjugate to $N,N^a$ and $h_{ab}$, while the Hamiltonian and vector constraints are respectively given by : 
\begin{eqnarray}
\mathscr{H} :=\sqrt{|h|} \left(  2 \Lambda_0 - \mathcal{M} \right) +\sum_{p=2}^\infty \tilde{\alpha}_p \frac{(d-2p-1)!}{(d-1)!}\mathscr{H}_{(p)} \;,\;\;
\mathscr{H}_a := -2 \mathcal{D}_b \pi^b_a \,,
\end{eqnarray}
with $\mathscr{H}_{(p)}$ defined by Eq\eqref{RegHamCons3D} in terms of $K_{ab}$, which is itself defined as a solution of 
\begin{eqnarray}
\pi^a_b =\sqrt{|h|} \left( \frac{K h^a_b - K^a_b}{\epsilon} \right)+ \sum_{p=2}^\infty \tilde{\alpha}_p \frac{(d-2p-1)!}{(d-1)!} \pi^{\phantom{(p)}a}_{(p)b} \,,
\end{eqnarray}
where now $\epsilon=-1$, so that $|h|=h$, and $\pi^{\phantom{(p)}a}_{(p)b}$ is either defined by Eq\eqref{RegPi13D} or Eq\eqref{RegPi23D}, yielding the two different theories.

In order for these to propagate only two degrees of freedom, the constraint $\mathscr{G}$ must be chosen such that $\left\{\mathscr{H}, \mathscr{G} \right\}\not\approx 0$, $\left\{\mathscr{H}_a, \mathscr{G} \right\} \approx \left\{\pi_a, \mathscr{G} \right\} \approx \left\{\pi_0, \mathscr{G} \right\} \approx  0$, see \cite{3Dcov1}. A convenient choice used in that paper was :
\begin{eqnarray}
\mathscr{G} =\sqrt{h} h_{ab}  \mathcal{D}_c \mathcal{D}^c \pi^{ab} \,, \label{Constraint3D}
\end{eqnarray} 
so that performing a Legendre transform from Eq\eqref{TotalHamilt} yields to the Lagrangian density 
\begin{eqnarray}
\mathscr{L}= -2 \Lambda_0 + \left( \bar{R}+  K^a_b K_a^b - K^2 \right) +\sum_{p=2}^\infty \tilde{\alpha}_p \frac{(d-2p-1)!}{(d-1)!}\mathscr{L}_{(p)}\,, \label{action3Dcov}
\end{eqnarray}
where $\mathscr{L}_{(p)}$ is defined by Eq\eqref{RegLag3D} for the choice Eq\eqref{RegPi13D} of the momentum while it would be defined by $\check{\mathscr{L}}_{(p)}:=\left(- N \mathscr{H}_{(p)} +2  K_\nu^\mu  \check{\pi}^{\phantom{(p)}\nu}_{(p)\mu} \right) /\sqrt{-g}$ for the second one given by Eq\eqref{RegPi23D}. However, contrary to the previous sections, the presence of the constraint $\mathscr{G}$ in the Hamiltonian requires to have : 
\begin{eqnarray}
K_{ab} = \frac{1}{2N} \left(\partial_0 h_{ab} - \mathcal{D}_{(a}N_{b)} -h_{ab}  \mathcal{D}_c \mathcal{D}^c  \lambda_{GF}\right) \,,
\end{eqnarray}
when applying the Legendre transform, so that $K_{ab}$ is no more the extrinsic curvature.

Finally, recall that the parameter $d$ is a free constant corresponding to the different regularizations of higher (than two) order Lovelock-Lanczos theories, with the two relevant choices being $d\to2p$, corresponding to the critical regularization or $d\to 4$, corresponding to direct four-dimensional regularization.

\section{Background Regularizations} \label{Sec.BR}

We have now seen three classes of procedures able to make Lovelock-Lanczos gravity (or slight modifications of it) non-trivial in four dimensions. As we already said, one of the interest of these theories is that when evaluated on spherically symmetric spacetimes, they yield the same black hole and cosmological solutions as the more standard background regularization of these sectors of the Lovelock-Lanczos theory, at least in the case of the Gauss-Bonnet one. We will not try to see if this property is preserved for general Lovelock polynomials, but rather investigate and review various results on the background regularizations, valid for any theory that would admit these Lovelock-like background dynamics. 
\\

Following our earlier results, a simple class of background regularization, explaining quite clearly the procedure, is based on conformally flat geometries. Starting with a Weyl decomposition of a Lovelock-like tensor with $q$ free indices on its GKD (denoted as usual by a $\Bigcdot$) :
\begin{eqnarray}
\delta^{\Bigcdot \mu_1 \nu_1 \dots \mu_p \nu_p}_{\Bigcdot \sigma_1 \rho_1 \dots \sigma_p \rho_p} \prod  _{r=1}^p R_{\mu_r \nu_r}^{\sigma_r \rho_r}  = \sum_{k=0}^p \sum_{l=0}^k   \iota_{(d,p,k,l)} R^l \delta^{\Bigcdot \mu_1 \nu_1 \dots \mu_{p-k} \nu_{p-k} \alpha_1 \dots \alpha_{k-l}}_{\Bigcdot \sigma_1 \rho_1 \dots \sigma_{p-k} \rho_{p-k}\beta_1 \dots \beta_{k-l}} \prod  _{r=1}^{p-k} W_{\mu_r \nu_r}^{\sigma_r \rho_r}  \prod  _{r=1}^{k-l} R^{\beta_r}_{\alpha_r}\,, \label{WeylDecompLovelock} 
\end{eqnarray}
where $W$ is the Weyl tensor and
\begin{eqnarray}
\iota_{(d,p,k,l)} := \frac{( d- 2p-q+k+l )!}{( d- 2p-q )!} \binom{p}{k}\binom{k}{l} \frac{2^{2k-l}(-1)^l}{(d-1)^l(d-2)^k} \,,
\end{eqnarray}
we see that for conformally flat background metric $\bar{g}$, the Lovelock tensors reduce to 
\begin{eqnarray}
\mathcal{G}^{(p)\mu}_{\phantom{(p)}\nu}\left[\bar{g}\right] = \sum_{l=0}^p  \frac{( d- p-1+l )!}{( d- 2p-1 )!} \binom{p}{l} \frac{2^{p-l-1}(-1)^{l+1}}{(d-1)^l(d-2)^p} R^l \delta^{\mu \mu_1 \dots \mu_{p-l}}_{\nu \nu_1 \dots \nu_{p-l}}  \prod  _{r=1}^{p-l} R^{\beta_r}_{\alpha_r}  \,,
\end{eqnarray}
so that after being evaluated on such $d$-dimensional ansatz, a factor $(d-4)$ appears in front of all the components of the field equations of Lovelock gravity for $p\leq 3$. Thus, one can multiply these tensors by $\frac{(d-2p-1)!}{(d-1)!}$ and take the limit $d\to 4$, what gives the well-defined regularized Lovelock tensors :
\begin{eqnarray}
\begin{split}
\tilde{\mathcal{G}}^{(2)\mu}_{\phantom{(p)}\nu} \left[\bar{g}\right] &= \frac{1}{72} \left( -R^2 \delta^\mu_\nu +4 R  \delta^{\mu\mu_1}_{\nu\nu_1}  R_{\mu_1}^{\nu_1} -6\, \delta^{\mu \mu_1 \mu_2}_{\nu \nu_1 \nu_2}  R_{\mu_1}^{\nu_1} R_{\mu_2}^{\nu_2} \right) \\
\tilde{\mathcal{G}}^{(3)\mu}_{\phantom{(p)}\nu} \left[\bar{g}\right] &= \frac{1}{432} \left( R^3 \delta^\mu_\nu -6 R^2 \delta^{\mu\mu_1}_{\nu\nu_1}  R_{\mu_1}^{\nu_1} + 18 \,R \delta^{\mu \mu_1 \mu_2}_{\nu \nu_1 \nu_2}  R_{\mu_1}^{\nu_1} R_{\mu_2}^{\nu_2} - 36\, \delta^{\mu \mu_1 \mu_2 \mu_3}_{\nu \nu_1 \nu_2 \nu_3}  R_{\mu_1}^{\nu_1} R_{\mu_2}^{\nu_2} R_{\mu_3}^{\nu_3}  \right)\,,
\end{split}
\end{eqnarray}
where the tilde notation is borrowed from the first section, see Eq\eqref{tilde}.

Similarly, we will now review the background regularization applied to spherically symmetric geometries, which in this case works at arbitrary high curvature order $p$, enabling to consider non-perturbative minisuperspace Lovelock-Lanczos models in four dimensions, admitting, as we will see, regular black hole and cosmological solutions.

\subsection{Dynamical spherically symmetric backgrounds}\label{Sec.DSS}

The main backgrounds for which the regularization has been shown to hold (for all Lovelock-Lanczos scalars \cite{AMS}) are (curved) FLRW and static spherically symmetric spacetimes. They belong more generally to what we will call ``Dynamical Spherically Symmetric backgrounds" (DSS), defined by the warped spacetimes $\mathcal{M}_{(n,k)}=\Sigma\, \times\, \Omega_{(n,k,r)}$, with $n=d-2$, whose interval is given by  
\begin{eqnarray}
ds^2 =d\Sigma^{\,2} +  d\Omega^{\,2}_{n,k,r} \,, \label{DSS}
\end{eqnarray}
 where $r$ is a scalar field on the $2$-dimensional manifold $\Sigma$, defined by the interval $d\Sigma^{\,2} = \gamma_{ab} dx^a dx^b$
with coordinates $x^a$, $a=1,2$ and $\Omega_{(n,k,r)}$ is the topological generalization of the $n$-dimensional sphere of radius $r(x)$, with the unit one defined by :
\begin{eqnarray*}
d\Omega_{n,k,1}^{\,2} = \left\{
  \begin{array}{@{}ll@{}}
 \;\;  d\theta_1^{\,2} + \sum\limits_{i=2}^{n} \prod\limits_{j=1}^{i-1} \sin^2\theta_j d\theta_i^{\,2} \; , \;\;\; \text{for} \;\; k=1 \,,
 
 \\  \;\;  d\theta_1^{\,2} + \sinh^2\theta_1 \left( d\theta_2^{\,2} + \sum\limits_{i=3}^{n} \prod\limits_{j=2}^{i-1} \sin^2\theta_j d\theta_i^{\,2} \right) \; , \;\;\; \text{for}  \;\;  k=-1 \, ,
 
 \\  \;\;  \sum\limits_{i=1}^{n} d\phi_i^{\,2} \; , \;\; \text{for}  \;\;\;  k=0 \, , 
  \end{array}\right.
\end{eqnarray*}
where $k=0,1,-1$ correspond respectively to toroidal, spherical and compact hyperbolic horizon manifolds $\Omega$. We call $R^{(n)}$ the Ricci scalar of $\Omega_{(n,k,r)}$, and $R^{(2)}$ the Ricci scalar of $\Sigma$. It is understood that all the covariant derivatives $\nabla$ are now compatible with the two-dimensional metric $\gamma$. For later use, we define the following quantities,
\begin{eqnarray}
\xi^a := \frac{\partial^a r}{r} \; ,  \;\; Z := \frac{R^{(n)}}{n(n-1)} - \xi^a \xi_a = \frac{k-\partial_c r \partial^c r}{r^2}  \; , \;\;
  \mathcal{Y}^a_b :=\frac{\nabla^a \partial_b r}{r}  \; ,  \;\;  \mathcal{Y} := \mathcal{Y}^a_a   \; ,  \;\;   \mathcal{Y}_2 := \mathcal{Y}^a_b \mathcal{Y}_a^b\;  ,
\end{eqnarray}
so that the non-vanishing components of the Riemann tensor can be written as
 \begin{eqnarray}
R^{ai}_{\;\; bj} = - \mathcal{Y}^a_b \delta_i^j \;, \;\;  R^{ij}_{\;\;kl} = Z  \delta^i_{[k} \delta_{l]}^j  \;, \;\;     R^{ab}_{\;\;ce} = \frac{1}{2} R^{(2)} \delta^a_{[c}\delta_{e]}^b \,.
\label{RiemRic}
\end{eqnarray}

 \subsubsection{Regularized $2$D covariant Actions and field equations}

As shown in \cite{Maeda}, the action of Lovelock-Lanczos gravity Eq\eqref{LLGAction} in this class of spacetime reduces to 
 \begin{eqnarray}
 I\left[ \gamma_{ab} ; r \right] =  \frac{\mathcal{A}_{(n,k,1)}}{2 \kappa_0} \int_\Sigma d^2 x \sqrt{-\gamma} r^{d-2}  \sum _{p=0}^{t}  \alpha_{p}\,  \mathcal{L}_{(p)}\left[  \gamma_{ab} ; r  \right]  \,,
 \end{eqnarray}
 where $\mathcal{A}_{(n,k,1)}$ is the volume of the n-dimensional manifold $\Omega_{(n,k,1)}$ of radius unity, and 
 \begin{eqnarray}
\mathcal{L}_{(p)}   =     \frac{\left(d-2\right)!}{\left(d-2p\right) !} \left( u_p Z^2 + Z \left( v_p \mathcal{Y} + w_p R^{(2)} \right) + x_p \left( \mathcal{Y}^2- \mathcal{Y}_2 \right) \right) Z^{p-2}  \,,  \label{LagrangianDSS}
\end{eqnarray}
where $u_p = (d-2p)(d-2p-1)$, $v_p= -2p(d-2p) $, $w_p = p$, $x_p = 2p(p-1)$. The reduced Lagrangian $\sqrt{-g} \mathcal{L}_{(p)} \left[ g_{\mu\nu} \right] \propto  \sqrt{-\gamma} r^{d-2} \mathcal{L}_{(p)} \left[ \gamma_{ab} \, ; \, r \right]  $ is just a two-dimensional Horndeski theory, because $-2 p \partial_{\partial_c r \partial^c r} \left( Z^{p-1}  \right) = 2p(p-1) Z^{p-2} /r^2$, which is to be expected from LLG, because it yields second order field equations for any backgrounds, so that its dimensional reductions are always belonging to some ``Horndeski class" of scalar-vector-tensor-... theories.

Following the discussion of the previous section, it was to be expected that the dimensional factor in Eq\eqref{LagrangianDSS} is not sufficient to yield a well-defined regularization of the Lovelock-Lanczos Lagrangian in critical dimension. However, using the Horndeski formulation solves this issue : imposing on the arbitrary (non-null) vector field $u$ to follow the isometry of the background and defining $X= \xi_a u^a$, we obtain by direct calculation from Eq\eqref{WHorndeski},
  \begin{eqnarray}
 W_{(s)} =   \frac{(d-2)!}{(d-2p-1)!}   \Big( (d-2p-1) \delta_b^a Z  X^2  + 2  \delta_{bd}^{ac} \left( s Z X \nabla^d u_c -(p-s) X^2 \mathcal{Y}_c^d  \right) \Big)\frac{Z^{p-s-1}  X^{2(s-1)}u_a u^b}{2^s \left(u_e u^e\right)^{s+1}}  
 \end{eqnarray}
 so that for arbitrary order $p$, the Horndeski-Lovelock-Lanczos reduced action (which is linear in $W_{(s)}$) can indeed be regularized in four dimensions.
As we see, all the curvature terms have disappeared, so that the (first) derivative of the two-dimensional metric is present only through the covariant derivatives of the radius scalar field $r$.

Remark however that the form of the regularized minisuperspace Lagrangian is not unique as e.g. :
   \begin{eqnarray}
   \mathcal{L}_{(p)} = \frac{(d-2)!}{(d-2p-1)!}\left( \left( (d-2p-1) Z -2 p  \mathcal{Y}^{ab}  \frac{\xi_a  \xi_b}{\xi_c \xi^c} \right)Z^{p-1}
   +\frac{2  p     r^{2-d}}{d-2p} \, \nabla^a \left(   r^{d-2}Z^{p-1} \frac{\xi_{[b}  \mathcal{Y}_{a]}^b}{\xi_c \xi^c}     \right)  \right) \,,
 \end{eqnarray}
what can be checked by direct computation and using the Horndeski formula for the two-dimensional Ricci scalar, see \cite{Aim}. Although much simpler than the Horndeski decomposition and $2$D-covariant, this splitting probably lacks a background independent generalization.

 Moreover, choosing $u^a = \xi^a$, both are valid only for $\xi_c \xi^c\neq 0$. For instance, the so-called ``exceptional vacuum" as well as Nariai solutions have $\xi_c \xi^c = 0$ because $\xi_c$ is null in the first case and $r(x)$ is constant in the second one (see \cite{Maeda}), so that such Lagrangian cannot be evaluated on these classes of solutions. That's another reason why a generalization of the Horndeski formula for arbitrary null vector fields would be desirable.
\\

As established in \cite{Maeda}, the background Lovelock tensors reduce to : 
\begin{eqnarray}
\mathcal{G}^{(p)\mu}_{\phantom{p}\nu} =  \frac{1}{2} \frac{(d-3)!}{(d-2p-1)!}  \; \text{diag}\left( \Theta^{(p)a}_{\phantom{(p)}b}\, , \; \Phi^{(p)} \delta^i_j \right) Z^{p-2}\,, \label{FieldEqLL}
\end{eqnarray}
where the terms $\Theta$ and $\Phi$ appearing in this block-diagonal decomposition are defined by : 
\begin{eqnarray}
\begin{split}
\Theta^{(p)a}_{\phantom{(p)}b} &:=- (d-2)  \Big( (d-2p-1) Z \delta^a_b -2 p \delta_{b e}^{a c} \mathcal{Y}^e_c \Big) Z \\
\Phi^{(p)} &:=   -p\Big[ R^{(2)}  Z + 2 (p-1) \delta_{ce}^{ab} \mathcal{Y}_a^c \mathcal{Y}_b^e - 2 (d-2p-1) Z \mathcal{Y} \Big] - (d-2p-1)(d-2p-2) Z^2 \,,
\end{split}
\end{eqnarray}
so that this whole sector of the theory is possible to regularize in any dimension $d\geq3$ and for any curvature order $p$, indicating a kind of dimensional universality of this minisuperspace theory. For $d=2$, corresponding to a regularization via ``Kaluza-Klein dimensional reduction", an extra scalar field appears, like in the conformal regularization studied earlier. The regularized Lovelock field equations Eq\eqref{NonPerturbativeFieldEquations} with the suitable normalization of the coupling constants \eqref{Regularization} are given by : 
\begin{eqnarray}
 \sum_{p=0}^t \tilde{\alpha}_p \, \Theta^{(p)}_{ab} Z^{p-2}= 2 \tilde{\kappa_0} T_{ab}  \; , \;\;\;\;  \delta^{i}_j \sum_{p=0}^t \tilde{\alpha}_p\, \Phi^{(p)}Z^{p-2}  =2  \tilde{\kappa_0} T_j^i \,, \label{DSSfullFEq}
\end{eqnarray}
where $\tilde{\kappa}_0 := (d-1)(d-2)\kappa_0=8 \pi G_N$.

However, remark that the components $\mathcal{G}_{a i}$ of the field equations are identically vanishing due to the symmetry assumption, so that it might not be clear whether the four dimensional limit of $\mathcal{G}_{a \,i}^{(p)}/(d-4)$ is well-defined. Moreover, because of Lovelock theorem, we know that no $4$D-covariant regularization can be achieved following this method, implying that it must break down at some order in perturbation theory around these backgrounds. In order to see how close a (background independent) theory admitting these same background field equations can be from Lovelock-Lanczos gravity, we will now reproduce the analysis of the first section, applied to DSS spacetimes. 

\subsubsection{Perturbations \& break down of the regularization} \label{ssSec.PerturbationDSS}

The equations \eqref{FieldEqExp}\eqref{GpnDef}\eqref{Gpn}\eqref{accolades} of the first section are general and valid for any backgrounds. For clarity and because contrary to the case of (A)dS vacua, the indices will matter here, let's rewrite these. Consider the metric expansion $g_{\mu\nu}=\bar{g}_{\mu\nu}  + h_{\mu\nu} + h_{\mu\alpha}h^{\alpha}_\nu + \dots + \left[ h^n \right]_{\mu\nu} + \dots $,
where now $\bar{g}$ is given by the interval Eq\eqref{DSS}. Then the curvature side of the full field equations of LLG \eqref{NonPerturbativeFieldEquations} decomposes as 
\begin{eqnarray}
\mathcal{G}^{\mu}_{\nu} \big[ g_{\mu\nu}  \big] =\sum_{p=0}^t \alpha_p  \mathcal{G}^{(p)\mu}_{(0)\nu}\left[ \bar{g} \right] +  \sum_{n=1}^\infty   \sum_{p=1}^t \alpha_p \mathcal{G}^{(p)\mu}_{(n)\nu}\left[ h^n \right] \;, \;\;  \mathcal{G}^{(p)\mu}_{(n)\nu} :=  -\frac{1}{2^{p+1}} \delta^{\mu \mu_1 \nu_1 \dots \mu_p \nu_p}_{\nu \sigma_1 \rho_1 \dots \sigma_p \rho_p}\left[ \prod  _{r=1}^p   R_{\mu_r \nu_r}^{\sigma_r \rho_r}  \right]^{[n]}   \,,  \label{FullGDSS}
\end{eqnarray}
where $\mathcal{G}^{(p)\mu}_{(0)\nu}\left[ \bar{g} \right]$ is given by Eq\eqref{FieldEqLL}. The argument then proceeds as follows : the dimensional factors needed for the regularization come from the contractions of indices between the GKD and the background Riemann tensor, which can be extracted in the following way,
\begin{eqnarray}
\mathcal{G}_{(n)\nu}^{(p)\mu} = -\frac{1}{2^{p+1}} \delta^{\mu \, \mu_1 \nu_1 \dots \mu_p \nu_p}_{\nu \, \sigma_1 \rho_1 \dots \sigma_p \rho_p} \sum_{k=1}^{p} \binom{p}{k} \prod_{r=1}^{p-k}   R^{[0]\sigma_r \rho_r}_{\phantom{(0)}\mu_r \nu_r}   \left\{ \prod_{r=p-k+1}^{p}   R_{\mu_r \nu_r}^{\sigma_r \rho_r}  \right\}^{[n]}  =: \sum_{k=1}^{p} \binom{p}{k} \mathscr{G}^{(k)\mu}_{(n)\nu}\,, \label{GDSS}
\end{eqnarray}
for $n>0$, where
\begin{eqnarray}
\left\{ \prod_{r=p-k+1}^{p}   R_{\mu_r \nu_r}^{\sigma_r \rho_r}  \right\}^{[n]} := \sum_{\sum_{i=1}^k j_i =n \, ; j_i \neq 0} \prod_{r=p-k+1}^{p}   R^{[j_i]\sigma_r \rho_r}_{\phantom{[j_i]}\mu_r \nu_r}    \,.
\end{eqnarray}
Then, as we saw in the first section, there are two possible ways to regularize LLG : either by imposing a specific form of the coupling constants such that the terms  $\mathscr{G}^{(k)\mu}_{(n)\nu}$ for $k\geq2$ do not appear in the perturbative expansion, or to directly regularize these terms as we did previously with the background sector.

Let's first focus on that last option. For perturbations that are ``GR-like" ($k=1$), we obtain by direct calculation, using the property Eq\eqref{GdeltaProp} of the GKD and the background Riemann tensor Eq\eqref{RiemRic},
\begin{eqnarray}
\begin{split}
  \mathscr{G}^{(1)A}_{(n)B} =   \frac{Z^{p-2} (d-4)!}{(d-2p-1)!} \bigg[ \Big[ (d-2p-1) Z \delta_B^A - 2 (p-1) \delta^{A c}_{B a} \mathcal{Y}_c^a \Big] {}_{(0)}^{(1)}\mathcal{K}^{(1)}_{(n)} + 4 \left(d-3 \right) Z \delta_{B b}^{A a} \left[ {}_{(1)}^{(1)}\mathcal{K}^{(1)b}_{(n)a}  \right]  \bigg],
\end{split}
\end{eqnarray}
where we use capital letters $A,B=0,1$ and $I,J=2,\dots,d$ to emphasize that they are free indices, and where the quantities
\begin{eqnarray}
{}_{(0)}^{(1)}\mathcal{K}^{(1)}_{(n)}:=-\frac{1}{2^{2}} \delta^{i j}_{k l} \left\{  R^{k l}_{i j} \right\}^{[n]} \;\;\; , \;\;\;\;\;\;
{}_{(1)}^{(1)}\mathcal{K}^{(1)b}_{(n)a} :=-\frac{1}{2^{2}} \delta^{i}_{j} \left\{R^{j b}_{i a}  \right\}^{[n]}  \,,
\end{eqnarray}
are the generalizations of the term $\mathcal{K}^{(p)\mu}_{(n)\nu}$ defined in Eq\eqref{KBasis} for the vacuum case and contain the perturbations. These parts of the field equations are thus regularizable at first order in perturbation theory. The remaining components of field equations are 
\begin{eqnarray}
 \mathscr{G}^{(1)A}_{(n)I}  =    \frac{Z^{p-2}(d-4)!}{(d-2p-1)!}    \Bigg[ \Big( 4(p-1) \mathcal{Y}_c^b \delta^{A c}_{a b} - 2 (d-2p-1) Z \delta_a^A \Big)   {}_{(1)}^{(1)}\mathcal{M}^{(1)a}_{(n)I} - 2(d-3) Z \delta_{a b}^{A c}   {}_{(1)}^{(2)}\mathcal{M}^{(1)a b}_{(n)I c}  \Bigg], \label{GaI}
\end{eqnarray}
where the quantities $\mathcal{M}$ are also containing the perturbations,
\begin{eqnarray}
{}_{(1)}^{(1)}\mathcal{M}^{(1)a}_{(n)I} := -\frac{1}{2^{2}}  \delta^{i k}_{j I} \left\{ R^{j a}_{i k}  \right\}^{[n]}  \;\;\; , \;\;\;\;\;\;
{}_{(1)}^{(2)}\mathcal{M}^{(1)a b}_{(n)I c}  :=-\frac{1}{2^{2}}   \delta^{i}_{I} \left\{ R^{a b}_{i c}  \right\}^{[n]}   \,.
\end{eqnarray}
This proves that the four dimensional limit of the background field equations $\mathcal{G}_{a \,i}^{(p)}/(d-4)$ is well-defined, because it is possible to regularize this generalized equation before taking the background limit ($h \to 0$).

However, the last components of the field equations are at first order : 
\begin{eqnarray}
\begin{split}
& \mathscr{G}^{(1)I}_{(n)J}  = Z^{p-3}  \frac{(d-5)!}{(d-2p-1)!}\; \times\\
 & \bigg[ (d-4) Z  \, \bigg( 4 \Big( (d-2p-1) Z  \delta_b^a - 2 (p-1)  \mathcal{Y}_c^e   \delta_{be}^{ac}  \Big)    {}_{(1)}^{(2)}\mathcal{K}^{(1)I b}_{(n)J a} + (d-3) Z \, \delta_{cd}^{ab} \,  _{(1)}\mathcal{P}_{(n)J ab}^{(1)I cd}  \bigg) + \mathcal{B} \,  {}^{(2)}_{(0)}\mathcal{K}^{(1)I}_{(n)J}   \bigg]
\end{split}
\end{eqnarray}
where
\begin{eqnarray}
\mathcal{B} := (p-1)\Big[ R^{(2)}  Z + 2 (p-2) \delta_{ce}^{ab} \mathcal{Y}_a^c \mathcal{Y}_b^e - 2 (d-2p-1) Z \mathcal{Y}_a^a \Big] + (d-2p-1)(d-2p-2) Z^2 
\end{eqnarray}
and
\begin{eqnarray}
^{(2)}_{(0)}\mathcal{K}^{(1)I}_{(n)J}:=-\frac{1}{2^{2}} \delta^{i j \, I}_{k l\, J} \left\{    R^{k l}_{i j} \right\}^{[n]}   \; ,\quad
^{(2)}_{(1)}\mathcal{K}^{(1)I b}_{(n)J a}:=-\frac{1}{2^{2}}  \delta^{i   I}_{j  J} \left\{ R_{i a}^{j b}  \right\}^{[n]}  \; ,\quad
_{(1)}\mathcal{P}_{(n)J ab}^{(1)I cd}:=-\frac{1}{2^{2}}  \delta^{ I}_{J} \left\{  R_{ab}^{cd} \right\}^{[n]} 
\end{eqnarray}
so that there is an issue with the last term $^{(2)}_{(0)}\mathcal{K}^{(1)I}_{(n)J}$, as it does not contain explicitly the factor $(d-4)$, remaining identically vanishing in $d=4$, i.e. for two angular coordinates.

Therefore, the $4$D regularization of the DSS sector of LLG breaks down at first order in perturbation theory, implying that any $4$D theory sharing this whole DSS sector will inevitably depart from LLG at that order.
\\

However, this is not the case for all DSS spacetimes as their exists some sub-sectors for which the scalar $\mathcal{B}$ is proportional to $(d-4)$, so that 
\begin{eqnarray}
\lim_{d\to4} \frac{\mathcal{B}\,  ^{(2)}_{(0)}\mathcal{K}^{(1)I}_{(n)J} }{d-4}  = 0 \,.
\end{eqnarray}
This is the case of curved FLRW metric fields defined by the interval 
\begin{eqnarray}
ds^2 = -dt^2 + a^2(t) \left( \frac{dr^2}{1-q r^2} + r^2 d\Omega^2_{((d-2),1,1)} \right) \,, \label{IntervalFLRW}
\end{eqnarray}
where $q=-1,0,1$ corresponds respectively to open, flat and closed universes, for which we have 
\begin{eqnarray}
Z= H^2 + \frac{q}{a^2} \; , \;\;\; \frac{R^{(2)}}{2} = \frac{\ddot{a}}{a} \; , \;\;\; \mathcal{Y}^a_b = \text{diag}\left( \frac{R^{(2)}}{2},Z\right) \,, \label{FLRWscalars}
\end{eqnarray}
where $H= \dot{a}/a$ is the Hubble parameter and dots represent temporal derivatives, so that 
\begin{eqnarray}
\mathcal{B}= (d-4) Z \left[ (d-2p-1)Z +  (p-1)R^{(2)} \right] \,,
\end{eqnarray}
implying that the full first order perturbations of Lovelock-Lanczos gravity around FLRW spacetimes are regularizable in $4$D, as it was shown in \cite{AleLorenzo}. Higher order perturbations could be investigated in a similar way. The results can be found in Appendix\ref{SecPerturbationLLGDDS}.

On the other hand, for static spherically symmetric background geometries, defined by
\begin{eqnarray}
ds^2 = - f(r) g^2(r) dt + \frac{dr^2}{f(r)}  + r^2 d\Omega^2_{((d-2),k,1)} \,, \label{IntervalSSS}
\end{eqnarray}
we have 
\begin{eqnarray}
Z= \frac{k-f}{r^2}\; , \;\;\; R^{(2)} = -\frac{1}{g}\left( 3 f' g' + g f'' + 2 f g''\right) \; , \;\;\; \mathcal{Y}^a_b =  -\text{diag}\left( \frac{f'}{2 r} +\frac{f g'}{r g}, \frac{f'}{2 r} \right)\,,
\end{eqnarray}
where primes represent radial derivatives, so that even on-shell (where LLG always has a branch with $g(r)=1$), we have :
\begin{eqnarray}
\mathcal{B} = (d-4)(d-3) Z^2 + (p-1)(p-2) r^2 Z'^{\,2} + (p-1) r Z \Big[ 2 (d-3) Z' + r Z'' \Big] \,, \label{QpSSS}
\end{eqnarray}
so that the factor $(d-4)$ is not overall, except for General Relativity $p=1$ or for $Z$ constant, corresponding to the (A)dS backgrounds. Therefore, any theory having this DSS Lovelock background dynamics will admit non-Lovelock (even slowly) rotating black holes. 
\\

Finally, remark that the effective regularization via the fine-tuning of the coupling constants investigated in the first section is not working in the case of DSS backgrounds. To see this explicitly, we derived in Appendix\ref{SecPerturbationLLGDDS} the general perturbative expansion of LLG around DSS. For curved FLRW, it can be put into the form
\begin{eqnarray}
\mathscr{G}^{(k)}_{(n)} = Z^{p-k-1} \frac{(d-2k-2)!}{(d-2p-1)!} \Pi_{(p,n,k)}\,,
\end{eqnarray}
where $\Pi_{(p,n,k)}$ is a well-defined tensor. Although (for these sub-sectors) arbitrary order perturbations $n$ are regularizable for $k=1$, it is not possible to cancel all the $k>1$ terms in this case, because instead of the constant $\Lambda$ enabling to adjust the relative order of the Lovelock terms in the (A)dS case, it is now the spacetime function $Z$, defined in Eq\eqref{FLRWscalars}, which plays this role. Thus, taking specific combinations of the LL terms means in this case to consider spacetime-dependent coupling constants, which is not admissible. Moreover, even it is was, the dependence of $\Pi_{(p,n,k)}$ on the curvature order $p$ prevents to perform the kind of binomial transform that was necessary to cancel the terms $k>1$ up to an arbitrary perturbative order. Only for the (A)dS backgrounds is $\Pi_{(p,n,k)}$ losing its dependence on $p$.

Therefore it is not possible to regularize LLG by fine-tuning in this case, meaning that the regularized theory Eq\eqref{RegularizedPerturbativeLLG} found in the first section seems to describe solely gravitons in (A)dS, but cannot be extended as it is to more general backgrounds.

\subsubsection{General properties of the Lovelock DSS backgrounds}

Having seen the limitations of this approach, we can now turn to the general properties that make any theory sharing this background sector interesting regarding both thermodynamical and singularity aspects. The proof of these results can be found in \cite{Maeda}. First, it was shown in that paper that defining the Generalized Misner-Sharp quasi-local mass (GMS) as,
\begin{eqnarray}
M_{\text{GMS}}\left[ r(x) \right] := \frac{\mathcal{A}_{(n,k,1)}r^{d-1}}{2(d-1)\kappa_0} \sum_{p=0}^{t} \tilde{\alpha}_p  Z^p \,, \label{DSSSolution}
\end{eqnarray}
allows to derive the so-called unified first law of thermodynamics from the background field equations \ref{DSSfullFEq} as : 
\begin{eqnarray}
d M_{\text{GMS}}= \mathcal{A}_{(n,k,r)} \psi_a dx^a + P d V \,,
\end{eqnarray}
where the pressure is given by $P:= - T^a_a /2$, the energy-flux by $\psi_a := T_a^b \nabla_b r + P \nabla_a r$ and the volume $V := r \mathcal{A}_{(n,k,r)} / (d-1)$.

In the context of black hole geometries, the regularized version of the Schwarzschild-Tangherlini black holes are given by the interval \eqref{IntervalSSS} with $g(r)=1$, while $f(r)$ is determined by solving the equation Eq\eqref{DSSSolution} :
\begin{eqnarray}
\frac{2 \bar{M} }{r^{d-1}}= \sum_{p=0}^t \tilde{\alpha}_p \left(\frac{k-f(r)}{r^2} \right)^p \,, \label{WheelerPol}
\end{eqnarray}
with the constant mass parameter defined  as $\bar{M}=  \kappa_0 (d-1) M_{\text{GMS}} / \mathcal{A}_{(n,k,1)}$. As usual, it is possible to associate a temperature to these black holes as $T:=\kappa /2\pi = f'(r_H) / 4 \pi$, where $\kappa$ is the surface gravity and $r_H$ is the horizon radius, so that 
\begin{eqnarray}
T = \frac{1}{4 \pi r} \left(\sum_{p=0}^t \tilde{\alpha}_p (d-2p-1) r_H^{-2p} k^p \right) \left( \sum_{q=0}^t \, q \, \tilde{\alpha}_q  r_H^{-2q} k^{q-1} \right)^{-1} \,.
\end{eqnarray}
Similarly, it is well-known that their Wald entropy is given by :
\begin{eqnarray}
S_W=\frac{2 \pi \mathcal{A}_{(n,k,1)} }{(d-1)\kappa_0} \sum_{p=0}^t      \frac{p \, \tilde{\alpha}_p}{d-2p} \, k^{p-1} r_H^{d-2p} \,,
\end{eqnarray}
As noted in \cite{AMS}, the critical order scalar $d=2p$ yields a constant divergent term in the Wald entropy. However, as the entropy is defined up to a constant, it can be discarded safely, leading to the same result as the dynamical entropy. Moreover, as the precise form of the regularized Wald entropy depends on the critical dimension, it is useful in this case to explicitly impose a four-dimensional regularization. To do so, we set $d=4$ and $\tilde{\alpha}_p = \iota l^{2(p-1)}$ at critical order, with $l$ a length scale and $\iota$ dimensionless, and take the limit $p \to 2$ so that,
\begin{eqnarray}
S_W=\frac{ \pi \mathcal{A}_{(2,k,1)} }{3 \kappa_0} \left( r_H^2 + 4 \iota  k l^2 \log\left(\frac{r_H}{l}\right)+  \sum_{p=3}^t \frac{p \, \tilde{\alpha}_p}{2-p} \, k^{p-1} r_H^{2(2-p)} \right) + cste \,,
\end{eqnarray}
which yields a logarithmic ``one-loop" correction from the regularized Gauss-Bonnet scalar. Using these results yields the first law of black hole thermodynamics $\delta M_{\text{GMS}} = T \delta S_W$.

Finally, let's comment on the asymptotic behaviour of the regularized black holes. For $r\to \infty$, if the vacuum is Minkowski spacetime or (A)dS, we have respectively the usual Schwarzschild or Schwarzschild-(A)dS behaviours, 
\begin{eqnarray}
f(r\to \infty) = k - \frac{2 \bar{M}}{r^{d-3}} \;,\;\;\; f(r\to \infty) = -\Lambda \, r^2  + k  - \frac{2 \bar{M}}{r^{d-3} } \left(\sum\limits_{p=0}^t p\, \tilde{\alpha}_p \Lambda^{p-1} \right)^{-1} \,.
\end{eqnarray}
What is quite special with this class of regularized black holes is that their center behaves as :
\begin{eqnarray}
f(r\to 0) = k - r^2  \left(\frac{2 \bar{M}}{\tilde{\alpha}_t r^{d-1}}\right)^{1/t} \,, \label{Center}
\end{eqnarray}
so that in the non-perturbative limit $t \to \infty$, the black hole core becomes (A)dS, i.e. non-singular and maximally symmetric. Of course their exists some choices of $\tilde{\alpha}_p$ which prevent the singularity resolution, for example by creating new ones at some $r_S>0$, but it is interesting to see that the more generic behaviour seems to be the regular one.

\subsubsection{Non-perturbative toy models admitting exact non-singular black hole and cosmological solutions}

A deeper analysis of black hole and cosmological singularities in the non-perturbative limit $t\to \infty$ can be simplified a lot by considering specific toy models, with definite coupling constants $\tilde{\alpha}_p$. In this part, we set $2 \tilde{\kappa}_0 := 16 \pi G_N = 1$, i.e. $\kappa_0 = 1/ \left[2(d-1)(d-2)\right]$, as well as $d=4$ and $\tilde{\alpha}_1=1$, $\tilde{\alpha}_0=-\Lambda_0<0$. As we will also consider the cosmological case given by the interval \eqref{IntervalFLRW}, we recall that for a barotropic fluid with equation of state $P=\omega \rho$, the equations that enable to determine the scale factor are \cite{AMS} :
\begin{eqnarray}
\sum_{p=0}^\infty \tilde{\alpha}_p J^{2p} = \rho\,,  \label{FLRW}
\end{eqnarray}
where $J^2 =H^2 + \frac{q}{a^2}$ and  
\begin{eqnarray}
\rho\left(a(t)\right) = \rho_0 \left(\frac{a(t)}{a_0}\right)^{-3(\omega+1)} \,, \label{density}
\end{eqnarray}
which comes from integrating the conservation equation $\dot{\rho} +  3 H (\rho + P) = 0$, with $a_0$ the corresponding integration constant.

\paragraph{a. Non-singular Hayward black hole \& associated non-singular universes}
~
\\

Now consider the simplest regularized theory defined for $p\geq1$ by $\tilde{\alpha}_p := l^{2(p-1)}$, where $l$ is a length scale. This model possesses a unique vacuum given by 
\begin{eqnarray}
\Lambda = \frac{\Lambda_0}{1+l^2 \Lambda_0}  \,. \label{HaywardVac}
\end{eqnarray}

In the static spherically symmetric case, it yields to (a cosmological generalization of) the well-known Hayward black hole \cite{Hayward_1} :
\begin{eqnarray}
f(r) = k - \frac{r^2 \left( 2 \bar{M} + r^3 \Lambda_0\right)}{r^3 + l^2 \left(2 \bar{M} + r^3 \Lambda_0 \right)} \,, \label{HaywardBH}
\end{eqnarray}
which has three horizons (inner, event and cosmological) for generic values of the parameters. This solution has first been found in the context of minisuperspace LLG in \cite{Maeda1,Maeda2}. Moreover, it is worth noting that a generalization of this solution has been found from a Renormalization Group improvement of the Schwarzschild solution in \cite{AS1}, generalization which in turns has been recovered from a one-parameter deformation of the DSS sector of LLG in our thesis \cite{Aim}.

This black hole is regular at the center and satisfies the so-called ``Limiting-curvature conjecture" \cite{Mukhanov,Mukhanov_2,Frolov_3,LQGMimetic2,LQGMimetic3}, stating that the curvature should be bounded by a universal constant. Indeed, it can be seen that both limits $r \to 0$ and $\bar{M} \to \infty$ yield the same (A)dS geometry : 
\begin{eqnarray}
f\left(r\to 0 ; \bar{M} \right) =f\left(r ; \bar{M}\to\infty \right) = k - \frac{r^2}{l^2} + \frac{r^5}{2 \bar{M} l^4} + O\left(\frac{r^8}{l^6 \bar{M}^2}\right) \,.
\end{eqnarray}
Its behaviour at infinity is the usual Schwarzschild-de Sitter,
\begin{eqnarray}
f\left(r\to \infty \right) =- \frac{\Lambda_0 \, r^2}{1+l^2 \Lambda_0} + k - \frac{2 \bar{M}}{r \left(1+l^2 \Lambda_0\right)^2} + + O\left( \frac{l^2 \bar{M}^2}{ r^4}\right) \,.
\end{eqnarray}
Due to its horizon structure, the solution has two extremal configurations at
\begin{eqnarray}
r_H^{\pm} = \sqrt{\frac{1+6 l^2 \Lambda_0 \pm \sqrt{1- 24 l^4 \Lambda_0} }{6 \Lambda_0}} \,,
\end{eqnarray}
where $r_H^+$ corresponds to the Nariai limit, where the outer and cosmological horizons merge, while $r_H^-$ corresponds to the merging of the outer and inner horizons, i.e. to a configuration with vanishing temperature,
\begin{eqnarray}
T=  \frac{r_H^2 - 3 l^2 -3 \left(l^2 - r_H^2\right)^2 \Lambda_0}{4 \pi r_H^3} \Bigr\rfloor_{r_H=r_H^-} =0 \,,
\end{eqnarray}
indicating that the Hawking evaporation has stopped, leaving a small regular black hole remnant. Finally, the entropy is given by : 
\begin{eqnarray}
S_W=  16 \pi^2  \left\{ r_H^2 - l^2 \left( \frac{l^2}{l^2 + r_H^2} + 2 \log\left[\frac{l^2 \left(l^2 + r_H^2\right)}{r_H^4}\right]\right)\right\}\,.
\end{eqnarray}
Remark that this solution and more generally any black hole coming from Eq\eqref{FieldEqLL} still have some severe issues. First, they contain an inner horizon, which might make them unstable against the mass inflation instability \cite{MassInflation10}. However, a recent work \cite{Bonbon} on that subject suggests that when the full backreaction is taken into account, the instability disappears. Secondly, they do not possess the proper quantum correction to Newton potential derived in  \cite{Donoghue1, Donoghue2, Donoghue3}.

These undesirable features are absent from many non-singular black hole spacetimes, see e.g. our thesis for more details \cite{Aim}, where it is shown that a class of charged regular black holes that includes the generalizations of the Visser-Hochberg \cite{VisserWormholes,BlackBounce}, D'Ambrosio-Rovelli \cite{DAR1,DAR2} and Modesto semi-polymeric \cite{Modesto} black holes do avoid the mass inflation instability provided that a bound between their mass and charge, of the form $M> l + Q^2 /l$, be satisfied, because, despite being charged, they possess only one horizon in that case. Similarly, one can find in \cite{Aim} many regular black holes with the proper quantum quantum correction to Newton potential.
\\

Turning to the cosmological solutions, the equation \eqref{FLRW} becomes : 
\begin{eqnarray}
\frac{J^2}{1-J^2 l^2} = \rho + \Lambda_0 \,, \label{HaywardCosmo}
\end{eqnarray}
Therefore, in the flat case where $q=0$, i.e. $J^2 = H^2 >0$, as the energy-density $\rho \to \infty$, the Hubble parameter becomes constant $H = 1/l$, meaning that the geometry is de Sitter like in the infinite past, resolving the initial singularity of General Relativity and providing a natural exponential expansion of the early universe, which might be seen as a kind of geometrical inflation, similar to some results found in the context of Quasi-Topological Gravities \cite{geomInfl} or perhaps similar to the so-called Non-Gaussian fixed point driven inflationary era, found in the context of Asymptotic Safety \cite{ASInflation1,ASInflation2,ASInflation3,ASInflation4,ASInflation5}.  This kind of regular cosmological solution have also been found in the context of Mimetic gravity in \cite{ChamMukh}.

More precisely, using the previous equation and Eq\eqref{density}, we obtain
\begin{eqnarray}
\begin{split}
&\sqrt{\frac{1+l^2 \Lambda_0}{\Lambda_0}} \log\left[ \left(\frac{a}{a_0} \right)^{3\left(1+\omega\right)} \left( \Lambda_0 + \Omega \left(1+2l^2 \Lambda_0\right) + 2 \sqrt{\Lambda_0 \Omega \left(1 + l^2 \Lambda_0\right) \left( 1 + l^2 \Omega \right)} \right) \right] \\
&+ l \log\left[ 1 + 2 l \left( l \Omega + \sqrt{\Omega \left(1 + l^2 \Omega \right) }\right) \right]= 3 \left(1+\omega \right) \left( t - t_0 \right)  \,, \label{HaywardCOSMOFLAT}
\end{split}
\end{eqnarray}
where $\Omega = \Lambda_0 + \rho_0 \left(\frac{a}{a_0} \right)^{-3\left(1+\omega\right)}$. We consider $\omega>-1$ to model known physical matter. For $t\to -\infty$ (i.e. $a(t)\to 0$) and redefining $t_0$ to cancel the constant terms arising from this limit gives : 
\begin{eqnarray}
a(t\to -\infty) = a_0 e^{t / l}\,,
\end{eqnarray}
so that the scale factor is asymptotically de Sitter for both $t\to - \infty$ and $t \to \infty$, meaning that the cosmological sector of this theory describes the dynamical transition between high and low energy de Sitter geometries with respective effective cosmological constants given by : $\Lambda_{\text{eff}}= 1 / l^2$ and $\Lambda$, this last corresponding to the unique vacuum of the theory Eq\eqref{HaywardVac}.

Concerning the closed universe solutions ($q=1$), solving Eq\eqref{HaywardCosmo} for the Hubble parameter yields : 
\begin{eqnarray}
\dot{a} = \pm  \sqrt{\frac{-1 + \left( \rho + \Lambda_0 \right)\left( a^2 - l^2 \right)}{1+l^2   \left( \rho + \Lambda_0\right) }} \,. \label{EqCyclic}
\end{eqnarray}
Thus, for positive cosmological constant and energy density, the usual cosmological singularity is avoided because the value of the scale factor is bounded from below by a positive constant, as it must satisfy the inequality $a(t) > l$, for any time $t$. From that observation follows that, depending on the detailed balance between the value of the cosmological constant, current value of the energy-density of the large scale universe and equation of state parameter $\omega$, the solution will either describe a single bounce, replacing the initial singularity, followed (and preceded) by a GR phase and then a de Sitter regime, or it will describe a non-singular cyclic cosmology, by periodically gluing together the positive and negative square-root solutions.

The additional condition allowing to assess the specific nature of the solution is given by 
\begin{eqnarray}
-1 +  \left(\rho_0 \left(\frac{a}{a_0}\right)^{-3(\omega+1)} + \Lambda_0 \right) \left( a^2 - l^2 \right) \geq 0 \,.
\end{eqnarray}
In order to see in more detail the cyclic solutions and as it is just a simple toy model anyway, we set $\Lambda_0=0$. Then for $-1<\omega\leq -1/3$, this inequality yields only a lower bound on the scale factor, indicating that there is a bounce in the early universe, but no future contraction. For instance, setting $w=-1/3$ gives 
\begin{eqnarray}
\rho_0 a_0^2 >1  \;, \;\; a(t) \geq l \sqrt{\frac{a_0^2  \rho_0}{a_0^2 \rho_0 -1}} \,.
\end{eqnarray}
The first inequality is a bound on the possible initial conditions of the associated solution, while the second one indicates the minimal value of the scale factor at which the bounce occurs (i.e. at which $\dot{a}=0$).

On the other hand, for $\omega>-1/3$, this inequality yields both an upper and lower bound on the value of the scale factor. For example, setting $w=1/3$ (i.e. a radiation-dominating universe), we obtain
\begin{eqnarray}
\frac{ \sqrt{a_0^2 \rho_0 - \sqrt{\rho_0\left( a_0^4 \rho_0 - 4 l^2 \right)}}}{\sqrt{2}}  \leq \frac{a(t)}{a_0} \leq \frac{ \sqrt{a_0^2 \rho_0 + \sqrt{\rho_0\left( a_0^4 \rho_0 - 4 l^2 \right)}}}{\sqrt{2}}  \,,
\end{eqnarray}
with $\rho_0 a_0^4 \geq  4 l^2$. Thus, the associated solutions must be cyclic, with the expanding phases given by the positive square-root solutions of Eq\eqref{EqCyclic} and the contracting phases given by the negative ones.

Concerning the open universe solutions, no bound on the scale factor arises when solving Eq\eqref{HaywardCosmo} and when $\rho \to \infty$, the Hubble parameter does not settle to a constant value, so that these solutions remain singular. Thus, $q=-1$ is disfavoured in these models.

\paragraph{b. KMT model and its non-singular solutions}
~
\\

Many other theories admitting regular solutions can be found in this way (see for example the last chapter of our thesis \cite{Aim} and the very recent paper \cite{nonsing}). In particular, G. Kunstatter, H. Maeda and T. Taves have found in \cite{Maeda1} the model given by $\tilde{\alpha}_p =\left(1-(-1)^p\right) l^{2(p-1)}/2$ for $p\geq1$ and still $\tilde{\alpha}_0 = -\Lambda_0<0$, which admits both a dS vacuum $\Lambda_+$, perturbative in $l$, and an AdS one $\Lambda_-$, non-perturbative in $l$ : 
\begin{eqnarray}
\Lambda_\pm = \frac{1}{2 l^4 \Lambda_0} \left( -1 \pm \sqrt{1+4 l^4\Lambda_0^2} \right)\,.
\end{eqnarray} 
In this case, there are two spherically symmetric solutions given by : 
\begin{eqnarray}
f_\pm(r) = k + \frac{r^2}{2 l^4 \left( 2\bar{M}+r^3 \Lambda_0 \right)}  \left( r^3 \mp \sqrt{r^6 +4 l^4 \left( 2\bar{M}+r^3 \Lambda_0 \right)^2} \right)\,. \label{KMTBH}
\end{eqnarray}
Focusing first on spherical horizon topology ($k=1$) for $f_+$, it describes once again a regular dS-core black hole with a Schwarzschild-dS asymptotic behaviour and limited curvature : 
\begin{eqnarray}
f_+(r\to0;M)=f_+(r; \bar{M}\to \infty) = 1- \frac{r^2}{l^2}\;,\;\; f_+(r\to\infty) = -\Lambda_+ r^2 + 1 - \frac{2\bar{M} \Lambda_+}{r \Lambda_0\left( 1+2l^4 \Lambda_0\Lambda_+\right)} \,.
\end{eqnarray}
Similarly, the solution $f_-$ for hyperbolic topology describes a geometry which interpolates between two AdS regimes. Defining first the AdS curvature of the corresponding vacuum as $\Lambda_- = -1/L^2$, we obtain, 
\begin{eqnarray}
f_-(r\to0) = -1 + \frac{r^2}{l^2} \; , \;\; f_-(r\to\infty) = \frac{r^2}{L^2} -1 -\frac{2\bar{M}}{r\Lambda_0\left( 2l^4 \Lambda_0 - L^2\right)}\,,
\end{eqnarray}
where $2l^4 \Lambda_0 - L^2>0$. Given that there exists a unique $r_H>0$ such that $f_-(r_H)=0$, this solution might describe a regular hyperbolic black hole. However, it is also well-known that a metric given by $f=-1+r^2/L^2$ does not describe a black hole, but simply AdS, despite having this same property. For our purpose, it is sufficient to have seen that even such exotic solution turns out regular within this model.

Concerning the cosmological solutions, setting $\Lambda_0=0$ for simplicity, Eq\eqref{FLRW} becomes : 
\begin{eqnarray}
\frac{H^2}{\left(1- H^2 l^2\right)\left(1+ H^2 l^2 \right)} = \rho \,, \label{KMTCOSMO}
\end{eqnarray}
so that once again, $\rho\to \infty$ corresponds to a regular dS geometry with $H =  1/l$.
\\

Finally, a word about the notion of regularity that we used to assess that these solutions do not have singularities. In general the two notions of geodesic incompleteness and divergence of curvature invariants are not equivalent, see \cite{WH1, WH2,WH3,GeodCurvSingWormHole}.  However, these results involve the Palatini (metric-affine) formalism, so that it is not clear if the two notions can be that different in the metric formalism, where the connection is fully determined in terms of the metric (as the Levi-Civita connection). Moreover, as we said earlier, one of the many characterizations of Lovelock-Lanczos gravity is that it is the unique theory for which the Palatini formalism always admits a Levi-Civita branch, so that if this difference between the two notions of singularity is generically absent in the metric formalism, it might also be so in the Palatini one for LLG. Secondly, the specific absence of singularity that we established from the previous examples and from Eq\eqref{Center} is entirely due to the fact that the spacetime approaches a maximally symmetric geometry, for which both notions obviously coincide.

\subsection{Regularizable Bianchi I sectors \& Non-uniqueness} \label{NUniqueReg}

The previous investigation on the regularized black holes and cosmological solutions of RLL gravities has yielded quite contrasted results. At background level, it was shown that regular static black holes as well as regular homogeneous and isotropic universes can be found in the non-perturbative ($t\to \infty$) limit. However, we also saw that, as they are, the perturbations around these solutions are not regularizable beyond first order (and not even at first order for static spherically symmetric ones), preventing to consider more generic four dimensional Lovelock backgrounds as for example axisymmetric ones.

Nonetheless, the background regularization is not unique, meaning that it might still be possible to make sense of concepts such as four dimensional rotating Lovelock black holes. Of course if such regularization is possible to find, it would remain to establish that a well-defined theory admits such solutions, as it has been found in the Gauss-Bonnet case for the black hole \eqref{WheelerPol}, via Horndeski and $3$D-covariant theories. Once again this goes beyond the scope of this paper.
\\

Rather, we will now see some examples of this non-uniqueness by establishing a new method to regularize specific backgrounds of Lovelock-Lanczos gravity that otherwise would not be regularizable.

The case of study will be the Bianchi I sector of this theory, whose interval reads :
\begin{eqnarray}
ds^2 = - dt^2 + \sum_{i=1}^{d-1} a_i^2(t) dx_i^2 \,. \label{BianchiImetric}
\end{eqnarray}
The non-vanishing components of the Riemann tensor are given by $R_{0i}^{0i} = \dot{H}_i + H_i^2 = \frac{\ddot{a}_i}{a_i}$ and  $R_{ij}^{ij} =H_i H_j = \frac{\dot{a}_i \dot{a}_j}{a_i a_j}$. It was shown by Pavluchenko in \cite{Povluchenko2010} that in this case, the Lovelock-Lanczos scalars become:  
 \begin{eqnarray}
\sqrt{-g} \mathcal{L}_{(p)} = 2^p p! (2p-3)!! \Big(Q^{(p)}_{d-1} + (2p-1)\mathcal{R}^{(p)}_{d-1} \Big)\sqrt{-g} \,, \label{BianchiILAG}
\end{eqnarray}
where 
\begin{eqnarray}
\mathcal{R}^{(p)}_{d-1} :=  \sum\limits_{ \substack{ j_1 > \dots > j_{2p}  \\ j_k \in \left\{ 1, \dots, d-1 \right\} }} \prod_{k=1}^{2p} H_{j_k}   \,,\;\;\; 
Q^{(p)}_{d-1} :=  \sum_{i=1}^{d-1} \sum\limits_{ \substack{ \left\{ j_1, \dots , j_{2(p-1)}\right\} \neq i \\ j_1 > \dots > j_{2(p-1)}  \\ j_k \in \left\{ 1, \dots, d-1 \right\} }}   \left(  \dot{H}_i + H_i^2 \right) \prod_{k=1}^{2(p-1)} H_{j_k}  \,.
\end{eqnarray}
Note that the original overall factor in \cite{Povluchenko2010} was $(2 p-3)!!2^{-p}$, involving the double factorial, but it actually does not match the results of LLG in FLRW, given by Eq\eqref{LagrangianDSS}. In order for the FLRW results to match, we propose the factor written above instead.

As we see, the term $Q^{(p)}_{d-1}$ is a total derivative in critical dimension ($d=2p$), where the associated sum collapses to a single term. However it contributes to the field equations otherwise. Therefore, based on the previous considerations, it is from both terms that the factors $(d-2p), (d-2p+1), \dots, (d-1)$ could have come out, but it is not the case : the vanishing of these terms at and beyond the critical dimension is still algebraic.

Indeed, contrary to DSS or perturbation theory around (A)dS vacua, no components of the Riemann curvature can be written as $R_{ij}^{kl} \approx \phi(x) \delta_{ij}^{kl}$ or $R_{ij}^{kl} \approx \psi_i^k(x) \delta_{j}^{l}$, so that the fundamental property of the Generalized Kronecker delta Eq\eqref{GdeltaProp}, from which the analytic dimensional factors come out, cannot be used. Stated in another way, the tensorial and functional structures of the Riemann tensor do not split, because there is a different function $a_i(t)$ associated with each dimension.

Therefore, it is not possible to regularize Lovelock-Lanczos gravities in these cases without additional assumptions. We will now see how to extract the dimensional factors from a ``periodization" of the higher dimensions. However, as we will see by evaluating in different ways the expression $\mathcal{R}^{(p)}_{d-1}$ (the same evaluations for the other term are collected in the Appendix\ref{AppendixBianchi}\ref{AppendixBianchi1}), this procedure is not unique.

\subsubsection{Periodic spatial decompositions of the $d$-dimensional metric}

As we want a four dimensional regularization, let $d=1+3n$, and let us restrict further the Bianchi I metric by imposing the following periodic conditions on the metric functions, for any $k_1  \in \left\{ 1, \dots, n \right\}$ and $k_2 , k_3  \in \left\{ 0, \dots, n-1 \right\}$,
\begin{eqnarray}
a_{3k_1} = a_3 \;\;\; ; \;\; a_{3k_2+1}=a_1 \;\;\; \text{and} \;\; a_{3k_3+2}=a_2 \; .
\end{eqnarray} 
In that case, the spacetime interval reduces to : 
\begin{eqnarray}
ds^2 = - dt^2 + a_1^2(t) \sum_{i=1}^n dx_i^2 + a_2^2(t) \sum_{i=n+1}^{2n} dx_i^2  + a_3^2(t) \sum_{i=2n+1}^{3n} dx_i^2 \,. \label{BianchiI1}
\end{eqnarray}
We can now decompose further the second term in the Lovelock-Lanczos scalars as : 
\begin{eqnarray}
\mathcal{R}^{(p)}_{3n}  = \sum_{i=0}^{2p} \sum_{j=0}^{i}  \binom{n}{2p-i} \binom{n}{i-j} \binom{n}{j} H_1^{2p-i} H_2^{i-j}H_3^j \,.  \label{BianchiIReg1R}
\end{eqnarray}
There are $\binom{3n}{2p}$ terms in the left-hand side sum which are collected in the right-hand-side into symmetric polynomials of $\left( H_1, H_2,H_3 \right)$. Each of these three can be picked $n$ times.  Therefore, by considering a spatial metric made of periodic $3$-dimensional building blocks, the desired factor $(d-4)$ comes out. For instance, for $p=2$, corresponding to the Gauss-Bonnet scalar, we have : 
\begin{eqnarray}
\frac{\mathcal{R}^{(2)}_{3n}}{n-1}  = \frac{n(n-2)(n-3)}{24} \sum H_i^4 +  \frac{n^2 (n-2)}{6} \sum H_i^3 H_j + \frac{n^3}{2} \sum H_i^2 H_j H_k + \frac{n^2 (n-1)}{4}  \sum H_i^2 H_j^2  \label{BianchiIReg2R}
\end{eqnarray}
We will not deal with the field equations here, but note that the scalar constraint is proportional to $\mathcal{R}^{(p)}_{d-1}$, see \cite{Povluchenko2010}, so that it is automatically regularized. 
\\

Of course this ``periodization" is not unique. To see this, let us proceed by setting $d=1+(1+2n)$ and restrict the Bianchi I metric to contain one scale factor $a_1$, $n$ scale factors $a_2$ and $n$ scale factors $a_3$, so that  
\begin{eqnarray}
ds^2 = -dt^2 + a_1^2 dx_1^2 + a_2^2 \sum_{i=2}^{n+1} dx_i^2 + a_3^2 \sum_{i=n+2}^{d-1} dx_i^2 \,.  \label{BianchiI2}
\end{eqnarray}
In this case, we obtain a different, less symmetric, regularization :
\begin{eqnarray}
\mathcal{R}^{(p)}_{1+2n} = H_1 \sum_{i=0}^{2p-1} \binom{n}{2p-i-1}\binom{n}{i} H_2^{2p-i-1} H_3^{i} + \sum_{i=0}^{2p} \binom{n}{2p-i}\binom{n}{i} H_2^{2p-i} H_3^{i}  \,.
\end{eqnarray}
Similarly, setting $d=1+(2+n)$, i.e. only one repeated scale factor, 
\begin{eqnarray}
ds^2 = -dt^2 + a_1^2 dx_1^2 + a_2^2 dx_2^2 +a_3^2 \sum_{i=1}^n dx_{i+2}^2 \,, \label{BianchiI3}
\end{eqnarray}
yields to 
\begin{eqnarray}
\mathcal{R}^{(p)}_{2+n} = H_3^{2(p-1)} \Bigg\{ \binom{n}{2p-1} H_3 \left( H_1+H_2 \right)+ \binom{n}{2p-2} H_1 H_2+ \binom{n}{2p} H_3^2 \Bigg\}\,.\label{BianchiIReg3R}
\end{eqnarray}
Despite this non-uniqueness, it is quite tempting to consider the first regularization more suitable because it preserves more than the others the structure of LLG, which is after all the theory of symmetric polynomials of the curvature.

Moreover, when restricted further to $2$ scale factors (i.e. setting two out of the three Hubble parameters to be equal), this class of metric reduces to the one studied in the previous section, so that it is possible to compare the results together.

\subsubsection{Kantowski-Sachs spacetimes}

In order to do so, consider a general gauge for the DSS metric \eqref{DSS} studied in the previous section :
\begin{eqnarray}
ds^2 =- A(t,x) dt^2 + B(t,x) dx^2 + r(t,x)^2  d\Omega^{\,2}_{n,k,1} \,.
\end{eqnarray}
It can be reduced to a Kantowski-Sachs spacetimes, i.e. a two scale factors anisotropic cosmology, which describes for example the interior dynamical region of a non-rotating black hole, by setting $k=1$, $A=1$, $B(t,x)=a_1^2(t)$ and $r(t,x)=a_2(t)$. In this case, the DSS Lagrangian of LLG Eq\eqref{LagrangianDSS} becomes : 
\begin{eqnarray}
\begin{split}
\mathcal{L}^{(p)}_{\text{DSS}} = \frac{(d-2)!}{(d-2p)!}  \Bigg\{ & 2p \left( H_2 \left[ H_1^2 + \dot{H}_1 \right] + H_1 \left[ (d-2) H_2^2 + 2 (p-1) \dot{H}_2 \right] \right)  \\
&+ (d-2p)  \left( (d-1) H_2^3 + 2p H_2 \dot{H}_2 \right)\Bigg\}H_2^{2p-3} \,. \label{KSDSS}
\end{split}
\end{eqnarray}
Only the last two regularizations of the generic Bianchi I sector are able to give such a result. By direct computation and using the Vandermonde convolution, we checked that it is indeed the case for these lasts in the case $a_3=a_2$.

Therefore, we have shown that the DSS interval \eqref{DSS} overlap the general Bianchi I metric Eq\eqref{BianchiImetric} only for the less symmetric ``periodizations"  given by Eq\eqref{BianchiI2} and \eqref{BianchiI3}, meaning that the four-dimensional Bianchi I sectors that could be associated with the DSS solutions found previously are inherently disymmetric in the three spatial directions.

 This was actually to be expected : the DSS metric \eqref{DSS} is made of a $2$-dimensional spacetime $\Sigma$, warped with an $n$-dimensional sphere $\Omega$, so that the Ricci scalar of $\Sigma$ can contribute only once to the Lovelock scalars, whereas the factor $Z$ (related to the Ricci scalar of $\Omega_{r(x)}$) can be chosen $p-2, p-1$ or $p$ times. Once the resulting theory is regularized on a $(2+2)$-manifold, this produces a strong disymmetry between the curvature of $\Sigma$ and that of the $2$-sphere, as it can be seen in Eq\eqref{LagrangianDSS}.

On the other hand, the FLRW results are all matching together (from the three different regularizations of Bianchi I LLG and from DSS LLG), so that all the results on that sector are uniquely defined. 
\\

Focusing now on the symmetric Bianchi I regularization given by the $d$-dimensional metric Eq\eqref{BianchiI1}, note first as a consistency check that it does reduce to the Bianchi I sector of General Relativity for $p=1$.

Then, reducing it to a Kantowski-Sachs geometry by setting any two scale factors out of the three available to be equal, we can establish the general form of this alternative Regularized LLG Lagrangian describing gravitational system like for example the dynamical interior of black holes. From Eq\eqref{BianchiILAG}\eqref{BianchiIReg1R}\eqref{BianchiIReg1Q}, we obtain that when evaluated on the $(d=1+3n)$-dimensional interval 
\begin{eqnarray}
ds^2 =  - dt^2 + a_1^2(t) \sum_{i=1}^n dx_i^2 + a_2^2(t) \sum_{i=n+1}^{3n} dx_i^2  \,,
\end{eqnarray}
the Lovelock-Lanczos Lagrangian becomes : 
\begin{eqnarray}
\sqrt{-g} \mathcal{L}^{(p)}_{\text{KS}} \left[ a_1(t) \, ; \, a_2(t) \right]= 2^p p! (2p-3)!! \Big(Q^{(p)}_{3n} + (2p-1)\mathcal{R}^{(p)}_{3n} \Big) a_1^{\frac{d-1}{3}} a_2^{\frac{2(d-1)}{3}} \,,\label{KSSym}
\end{eqnarray}
where 
\begin{eqnarray}
\begin{split}
Q&^{(p)}_{3n} = n \left(\dot{H}_1 + H_1^2 \right) \sum_{i=0}^{2(p-1)}  \binom{n-1}{2p-2-i} \binom{2n}{i} H_1^{2p-2-i} H_2^i  \;, \;\;  \mathcal{R}^{(p)}_{3n} = \sum_{i=0}^{2p} \binom{n}{2p-i} \binom{2n}{i} H_1^{2p-i} H_2^i 
\\&+ 2n \left(\dot{H}_2 + H_2^2 \right) \sum_{i=0}^{2(p-1)}  \binom{n}{2p-2-i} \binom{2n-1}{i} H_1^{2p-2-i} H_2^i  \,. \label{RegKS}
\end{split}
\end{eqnarray}
A few examples for the $p=2,3$ cases are given in the Appendix\ref{AppendixBianchi}\ref{AppendixBianchi2}. Note that if the minisuperspace field equations are to be derived from the previous Lagrangian, one must impose by hand the scalar constraint (proportional to the sum of the regularized $\mathcal{R}^{(p)}_{3n}$) because we set the lapse $N(t)=1$ from the beginning of the discussion.
\\

Although it is more symmetric (as a curvature polynomial) than Eq\eqref{KSDSS}, it is also less universal with regard to the dimensions in which it exists : while the regularized version of \eqref{KSDSS} is defined in any dimensions for $d>2$, as we saw in the previous section, the minisuperspace model \eqref{KSSym} is reachable only in dimensions $d=4,7,10,13,...$, i.e. when the number of spatial dimensions is a multiple of three. 

\section{~$\,$Summary and Discussion}

Lovelock-Lanczos gravity being the most natural family of theories that includes General Relativity and provides small scale corrections to its dynamics, which itself generically suffers from singularities, it is tempting to investigate how relevant this class of model could be in four dimensions. As LLG with $p>1$ is trivial or topological in $4$D, one first needs to establish what sectors of this theory could be extended to $4$D by well-defined limiting procedures before attempting to find $4$D background independent theories admitting these (regularized) Lovelock sectors and solutions. Remarkably, such theories have also been found from the background independent regularizations of the full Gauss-Bonnet theory $p=2$, making both types of regularizations related. 
\\

In this paper, we have generalized and clarified many results related to these. Concerning the dynamical spherically symmetric ($2$D-covariant) sector of LLG, we showed that the regularization is well-defined at the background level, from both the minisuperspace action and field equations. However, we explicitly saw that it breaks down at first order in perturbation theory around static spherically symmetric solutions and at second order around (curved) FLRW ones, meaning that any background independent theory admitting these Lovelock sectors will depart from Lovelock gravity at these perturbative orders. To show the non-uniqueness of the background regularization, we found a new method to regularize the Bianchi I sector of Lovelock-Lanczos gravity in inequivalent ways, based on periodic decompositions of the spatial $(d-1)$-dimensional metric \eqref{BianchiIReg1R},\eqref{BianchiIReg2R},\eqref{BianchiIReg3R}. Consequently, a new regularized Kantowski-Sachs model describing alternative black hole interiors was obtained \eqref{KSSym}\eqref{RegKS}.

It is not known whether this method can be extended to more generic backgrounds, like for example axisymmetric ones, nor if there exist some $4$D background independent gravitational theories admitting these new regularized Lovelock sectors. Moreover, we saw that non-singular four dimensional black holes and (past-dS, bouncing and cyclic) universes arise quite generically from the regularized DSS sector of LLG  \eqref{Center},\eqref{HaywardBH},\eqref{HaywardCOSMOFLAT},\eqref{EqCyclic},\eqref{KMTBH},\eqref{KMTCOSMO}, so that it would be interesting to check if this is also the case for the solutions of the Bianchi I minisuperspace models that we found, in order to understand more deeply the singularity resolution within these Lovelock models. 
\\

Although we have not proven whether the regularized DSS sector of LLG for $p>2$ is reachable from background independent theories, we have generalized to arbitrary curvature order $p$ two known methods able to give such theories in the case of Gauss-Bonnet gravity $p=2$ : the regularization of LLG via a conformal transformation \eqref{Lag4DCCRLLG}\eqref{CCRLLGAction}, leading to a shift-invariant Horndeski theory, and its regularization via the breaking of $d$-dimensional covariance down to $d-1$, which in the limit gives $3$D-covariant theories \eqref{RegLag3D}\eqref{RegPi23D}\eqref{action3Dcov}. Both kinds of theory are defined for any curvature order $p$ and it would be interesting to see if they do admit this Lovelock sector for $p>2$, what would automatically make these models quite well-behaved at small distances with respect to singularities.

More generally, we have classified the possible regularizations of Lovelock-Lanczos gravity via metric transformations by finding the general form of the resulting $4$D theories \eqref{RiemannTransfo}\eqref{4DReg}. Whatever the transformation is, there are two relevant classes of regularizations : either considering the first order expansion of the transformed theory $\sqrt{-\bar{g}} \bar{\mathcal{L}}_p$ in powers of $(d-4)$ or $(d-2p)$ and then taking the four dimensional limit of the resulting theory. We mainly focused on the second class of regularization (the ``critical" ones) because in this case the counter-terms needed to take a well-defined limit are boundary terms, so that, in the conformal case, only this class constitutes a regularization of pure Lovelock gravity. However, we saw that non-critical regularized theories also belong to the Horndeski class of scalar-tensor theories, see \eqref{NonCrit1},\eqref{NonCrit2}. As for more general metric transformations like disformal ones, it could be interesting to check if the corresponding regularized theories also admit second order field equations, or perhaps if they belong to the class of DHOST scalar-tensor theories \cite{DHOST1,DHOST2,DHOST3,DHOST4}, which are also avoiding Ostrogradski instabilities. Other metric transformations like those involving a $U(1)$ gauge field might also present some interest, see \cite{U11,U12}. Finally, we noted that for a given curvature order $p$, there are infinitely many regularized theory of the form \eqref{SecondOrderCCRLLG} corresponding to the coefficients of the expansion of the transformed theory in powers of $(d-D)$. These as well might deserve a closer look.

Concerning the regularizations via breaking of $d$-dimensional covariance, although much fewer than the previous ones, some non-uniqueness is nonetheless to be expected from the fact that $3$D-covariant theory \eqref{RegLag3D} is reached by adding counter-terms to the Horndeski-Lovelock-Lanczos (ADM) action, which itself is just a particular decomposition of LLG when a non-null vector field is introduced (i.e. when a particular time-like or space-like foliation is considered). Thus, if a decomposition of LLG based on null vector fields is to be found, it might be possible that the same regularization procedure would yield to a different kind of theory with a different reduced notion of covariance, as exemplified by \eqref{NulldecompGR}. As for the number of degrees of freedom of $3$D-covariant theories with Hamiltonian \eqref{TotalHamilt}, we expect that the same introduction of constraint (see Eq\eqref{Constraint3D}) made in \cite{3Dcov1} is able to make the theory propagates only the two degrees of freedom of the graviton for any curvature order $p$, but this remains to be seen in more details.
\\

Finally, we have found a new class of regularization, which is only working at the perturbative level around a given (A)dS vacuum, based on peculiar combinations of Lovelock invariants \eqref{FineTuning}\eqref{RegularizedPerturbativeLLG}. The effect of this fine-tuning is to cancel all the $1/(d-4)$ poles present in the perturbative expansion of the (normalized \eqref{Regularization}) field equations \eqref{DUDUDUDU} up to an arbitrary perturbative order related to the highest curvature order of such combinations. Thus, if an infinite number of curvature terms related in this way are present, the theory should be possible to regularize non-perturbatively. However, this means that the tensorial structure of such perturbative expansion is indistinguishable from that of General Relativity, the only differences being overall redefinitions of the cosmological and Newton constants, and a different vacuum which is not solely given by the cosmological constant. It is not very clear to us if such regularization is well-defined, but if it is, it could describe gravitons around exotic (A)dS$_4$ vacua, different from the one of GR. We do not know if these specific coupling constants might have an interesting effect on the other regularized theories   \eqref{CCRLLGAction} and \eqref{action3Dcov}, for instance regarding the strong coupling issue of the conformal regularization (see \cite{HornGB2} and \cite{Amplitudes}), but in any case, this could be relevant for higher dimensional Lovelock gravity. Furthermore, we saw that this procedure is conceptually close to effective field theories because the coupling constants have to run with the maximal perturbative order at which the regularization is well-defined. Whether this might have some relevance regarding new counter-terms in perturbative quantum gravity is not known.  We leave these issues for future studies. 

\section*{Acknowledgments}

We thank Karim Noui for very useful discussions and Nathalie Deruelle for interesting and encouraging exchanges in the earlier stages of this work.  The xAct package \cite{xAct} for Mathematica was used to check many formulae found in this paper. 

\section{~$\;$Appendix}

\subsection{Binomial coefficient identities}\label{BCid}

\subsubsection{Proof of Eq\eqref{DecompPerturbationLLG}} \label{BinInversion}

For $p>n$, 
\begin{eqnarray}
\begin{split}
&\sum_{k=1}^{n} \xi_{(p,n,k)} \mathcal{G}^{(k)}_{(n)} 
=\sum_{k=1}^n \Lambda^{p-k} \frac{(d-2k-1)!}{(d-2p-1)!} (-1)^{n+k} \binom{p}{k} \binom{p-k-1}{p-n-1} \mathcal{G}^{(k)}_{(n)} \\
=& \sum_{k=1}^n \sum_{j=1}^k \Lambda^{p-j} \frac{(d-2j-1)!}{(d-2p-1)!} \binom{p}{k}\binom{k}{j} \binom{p-k-1}{p-n-1} (-1)^{n+k} \mathcal{K}^{(j)}_{(n)} \,,
\end{split}
\end{eqnarray}
where in the last equality, we used Eq\eqref{KBasis}. Commuting the sums and using the subset-of-a-subset identity $\binom{p}{k}\binom{k}{j} = \binom{p}{j} \binom{p-j}{k-j}$,
\begin{eqnarray}
\begin{split}
&\sum_{k=1}^{n}\xi_{(p,n,k)} \mathcal{G}^{(k)}_{(n)} 
= \sum_{j=1}^n  \Lambda^{p-j} \frac{(d-2j-1)!}{(d-2p-1)!} \binom{p}{j} \mathcal{K}^{(j)}_{(n)}   \left( \sum_{k=j}^n \binom{p-j}{k-j} \binom{p-k-1}{p-n-1} (-1)^{n+k} \right) \\
=&  \sum_{j=1}^n  \Lambda^{p-j} \frac{(d-2j-1)!}{(d-2p-1)!} \binom{p}{j} \mathcal{K}^{(j)}_{(n)}  =  \mathcal{G}^{(p)}_{(n)} \,,
\end{split}
\end{eqnarray}
where in the last equality we used Eq\eqref{KBasis} again. Thus, we are left to prove that for $p>n$,
\begin{eqnarray}
\sum_{k=j}^n \binom{p-j}{k-j} \binom{p-k-1}{p-n-1} (-1)^{n+k}=1 \,.
\end{eqnarray}
Changing the summation index $k=i+j$ and substituting $j=n-q$, $m=p-n>0$ yields :
\begin{eqnarray}
\begin{split}
&\sum_{k=j}^n \binom{p-j}{k-j} \binom{p-k-1}{p-n-1} (-1)^{n+k}
= \sum_{i=0}^q \binom{m+q}{i}  \binom{m+q-i-1}{m-1} (-1)^{i-q}\\
=&  \sum_{i=0}^q \binom{m+q}{i}   \binom{m+q-i}{m}  \frac{m}{m+q-i} (-1)^{i-q}\\
=&  \sum_{i=0}^q  \binom{m+q}{m}  \binom{q}{i}  \frac{m}{m+q-i} (-1)^{i-q}
= (-1)^{-q} \binom{m+q-1}{m-1}  \sum_{i=0}^q   \binom{q}{i} (-1)^{i} \frac{m+q}{m+q-i} \\
=& (-1)^{-q} \binom{m+q-1}{q}  \sum_{i=0}^q   \binom{q}{i} (-1)^{i} \frac{m+q}{m+q-i}  \,.
\end{split}
\end{eqnarray}
Moreover,  
\begin{eqnarray}
\sum_{l=0}^i \binom{i}{l} \binom{m+q-1}{l}^{-1} =  \binom{m+q-1}{i}^{-1}  \sum_{l=0}^i   \binom{m+q-1-l}{i-l} = \frac{m+q}{m+q-i}  \,,\label{Transform}
\end{eqnarray}
where in the first equality we applied again the subset-of-a-subset identity. Therefore, $\left[  \frac{m+q}{m+q-i} \right]$ is the binomial transform of $\left[ (-1)^l \binom{m+q-1}{l}^{-1} \right]$. The fact that the binomial transform is an involution conclude the proof. 

\subsubsection{Proof of Eq\eqref{EqGamma}} \label{gammarelation}

Now we wish to prove that for $2 \leq k \leq n \leq m < p \leq t$,
\begin{eqnarray}
\gamma_{(p,n,k)}-\gamma_{(p,m,k)}- \sum_{q=n+1}^m \gamma_{(p,m,q)}\gamma_{(q,n,k)} =0 \,.
\end{eqnarray}
This is trivial for $n=m$ by definition of empty product, so we can restrict to the case $n<m$. The previous equation is equivalent to 
\begin{eqnarray}
\begin{split}
\sum_{q=n+1}^m (-1)^{m+n+q} \binom{p-k}{q-k}\binom{p-q-1}{p-m-1}\binom{q-k-1}{q-n-1} 
= (-1)^n \binom{p-k-1}{p-n-1} - (-1)^m \binom{p-k-1}{p-m-1}.
\end{split}
\end{eqnarray}
Using multiple times the subset-of-a-subset identity, the symmetry of the binomial coefficient and its absorption property, we end up with 
\begin{eqnarray}
\begin{split}
&\sum_{q=n+1}^m (-1)^{m+n+q} \binom{p-k}{q-k}\binom{p-q-1}{p-m-1}\binom{q-k-1}{q-n-1} \\
=&  \binom{p-k}{p-m} \binom{m-k}{n-k} \sum_{j=0}^{m-n-1} (-1)^j  \binom{m-n-1}{j} \left[ \frac{(m-n)(p-m)}{(n-k+j+1)(p-n-j-1)} (-1)^{m+1} \right].
\end{split}
\end{eqnarray}
Rewriting 
\begin{eqnarray*}
\frac{(m-n)(p-m)}{(n-k+j+1)(p-n-j-1)}=\frac{(m-n)(p-m)}{p-k} \left( \frac{1}{n-k+j+1} + \frac{1}{p-n-j-1} \right)\,,
\end{eqnarray*}
enables to split the sum, reducing the remaining calculations to two binomial transforms of the same form as Eq\eqref{Transform}. Applying this formula in both cases yields the result. 

\subsection{Boundary terms in the conformal critical regularization} \label{ConfCubic}

Let's define the following scalar densities,
\begin{eqnarray}
\begin{split}
W_{(k,l)} &:= \left( \nabla_\zeta \phi \nabla^\zeta \phi \right)^l \delta^{ \mu_1 \nu_1 \dots \mu_{p-k} \nu_{p-k}  \lambda_1 \dots \lambda_{k-l-1}\alpha}_{ \sigma_1 \rho_1 \dots \sigma_{p-k} \rho_{p-k} \gamma_1 \dots \gamma_{k-l-1}\beta } \nabla_\alpha \phi \nabla^\beta \phi \prod  _{r=1}^{p-k} R_{\mu_r \nu_r}^{\sigma_r \rho_r}  \prod  _{r=1}^{k-l-1} \nabla^{\gamma_r}  \nabla_{\lambda_r} \phi  \\
\Omega_{1(k,l)} &:=  \left( \nabla_\zeta \phi \nabla^\zeta \phi \right)^l \delta^{ \mu_1 \nu_1 \dots \mu_{p-k} \nu_{p-k}  \lambda_1 \dots \lambda_{k-l-1}}_{ \sigma_1 \rho_1 \dots \sigma_{p-k} \rho_{p-k} \gamma_1 \dots \gamma_{k-l-1} } \nabla_\alpha \phi \nabla^\alpha \phi \prod  _{r=1}^{p-k} R_{\mu_r \nu_r}^{\sigma_r \rho_r}  \prod  _{r=1}^{k-l-1} \nabla^{\gamma_r}  \nabla_{\lambda_r} \phi \\
 \Omega_{2(k,l)} &:=  \left( \nabla_\zeta \phi \nabla^\zeta \phi \right)^l \delta^{ \mu_1 \nu_1 \dots \mu_{p-k} \nu_{p-k}  \lambda_1 \dots \lambda_{k-l-1}}_{ \sigma_1 \rho_1 \dots \sigma_{p-k} \rho_{p-k} \gamma_1 \dots \gamma_{k-l-1} }  R^{\sigma_1 \Bigcdot}_{\mu_1 \nu_1}  \nabla_{\Bigcdot} \phi \nabla^{\rho_1} \phi \prod_{r=2}^{p-k} R_{\mu_r \nu_r}^{\sigma_r \rho_r}  \prod_{r=1}^{k-l-1} \nabla^{\gamma_r}  \nabla_{\lambda_r} \phi \\
  \Omega_{3(k,l)} &:=  \left( \nabla_\zeta \phi \nabla^\zeta \phi \right)^l \delta^{ \mu_1 \nu_1 \dots \mu_{p-k} \nu_{p-k}  \lambda_1 \dots \lambda_{k-l-1}}_{ \sigma_1 \rho_1 \dots \sigma_{p-k} \rho_{p-k} \gamma_1 \dots \gamma_{k-l-1} }  \nabla^{\Bigcdot} \nabla_{\lambda_1} \phi  \nabla_{\Bigcdot} \phi \nabla^{\gamma_1} \phi \prod_{r=1}^{p-k} R_{\mu_r \nu_r}^{\sigma_r \rho_r}  \prod_{r=2}^{k-l-1} \nabla^{\gamma_r}  \nabla_{\lambda_r} \phi \label{WOmega}
\end{split}
\end{eqnarray}
where $\Bigcdot$ denotes indices that are contracted together while all others are contracting with the GKD. By expanding the last line of the (determinant form of the) GKD appearing in $W_{(k,l)}$ using the Laplace formula and reorganising the indices, we obtain, 
\begin{eqnarray}
W_{(k,l)}=\Omega_{1(k,l)}-2(p-k)\Omega_{2(k,l)}-(k-l-1)\Omega_{3(k,l)} \,. \label{RecurrenceRelation}
\end{eqnarray}
Moreover, the following divergence decomposes in terms of the previous quantities as :
\begin{eqnarray}
\begin{split}
\mathcal{D}V_{(k,l)} &:=  \delta^{ \mu_1 \nu_1 \dots \mu_{p-k-1} \nu_{p-k-1}  \lambda_1 \dots \lambda_{k-l+1}}_{ \sigma_1 \rho_1 \dots \sigma_{p-k-1} \rho_{p-k-1} \gamma_1 \dots \gamma_{k-l+1} }   \nabla_{\lambda_1} \left( \nabla^{\gamma_1}\phi  \left(\nabla_\alpha \phi \nabla^\alpha \phi\right)^l \prod_{r=1}^{p-k-1} R_{\mu_r \nu_r}^{\sigma_r \rho_r}  \prod_{r=2}^{k-l+1} \nabla^{\gamma_r}  \nabla_{\lambda_r} \phi \right)\\
&=  \Omega_{1(k+1,l-1)} + 2 l  \Omega_{3(k+1,l-1)} - \frac{1}{2} (k-l) \Omega_{2(k,l)} \,. \label{DIVCONF}
\end{split}
\end{eqnarray}
This identity is found by using the differential Bianchi identity and the fact that the commutation of two covariant derivatives yields a Riemann tensor term.

Now let's consider the following quantity, 
\begin{eqnarray}
\begin{split}
Q_p := \sum_{k=0}^p \sum_{l=0}^k &\Xi_{(p,k,l)} (k+l)! \beta_{(k,l)} = \sum_{k=0}^p \sum_{l=0}^k  (k+l)! \beta_{(k,l)} \binom{p}{k}\binom{k}{l} (-1)^{k-1} 2^{k-2l-1} \left(\nabla_\mu \phi \nabla^\mu \phi\right)^l  \\
&\times \left( (k-l) \nabla^\beta \phi \nabla_\alpha \phi -2 \nabla^\beta \nabla_\alpha \phi \right)  \delta^{\Bigcdot \mu_1 \nu_1 \dots \mu_{p-k} \nu_{p-k}  \lambda_1 \dots \lambda_{k-l-1}\alpha}_{\Bigcdot \sigma_1 \rho_1 \dots \sigma_{p-k} \rho_{p-k} \gamma_1 \dots \gamma_{k-l-1}\beta } \prod  _{r=1}^{p-k} R_{\mu_r \nu_r}^{\sigma_r \rho_r}  \prod  _{r=1}^{k-l-1} \nabla^{\gamma_r}  \nabla_{\lambda_r} \phi 
\end{split}
\end{eqnarray}
For $\beta_{(k,l)} =1$, comparing with Eq\eqref{CriticalRegulari}, it corresponds either to the counter-terms needed to regularize the $p^{\text{th}}$ order Lovelock scalar in critical dimension, or to the term in the regularized theory $\mathfrak{L}_p$ multiplied by $\phi$. For $\beta_{(k,l)} = H_{k+l}$, the harmonic numbers, it corresponds to the second term in $\mathfrak{L}_p$.

In terms of the following sequence of constants
\begin{eqnarray}
\alpha_{(k,l)} :=  \binom{p}{k}\binom{k}{l} (-1)^{k-1} 2^{k-2l-1} (k+l)! \,, \label{alphaconf}
\end{eqnarray} 
as well as the scalars defined in Eq\eqref{WOmega}, this quantity can be rewritten as,
\begin{eqnarray}
\begin{split}
Q_p &= \sum_{k=0}^p \sum_{l=0}^k  \alpha_{(k,l)}\beta_{(k,l)}(k-l) W_{(k,l)}  -2 \sum_{k=0}^p \sum_{l=0}^k \alpha_{(k,l)}\beta_{(k,l)} \Omega_{1(k,l-1)}   \\
&=  \sum_{k=1}^p \sum_{l=0}^{k-1}  \alpha_{(k,l)}\beta_{(k,l)}(k-l) W_{(k,l)}  -2 \sum_{k=1}^p \sum_{l=0}^k \alpha_{(k,l)}\beta_{(k,l)} \Omega_{1(k,l-1)}  -2 \alpha_{00}\beta_{00} \Omega_{1(0,-1)} \,.  \label{Qppppp}
\end{split}
\end{eqnarray}
In the second sum we shift the sum over $k$ by setting $k=r+1$.
In the first sum, we use Eq\eqref{RecurrenceRelation} to express $W_{(k,l)}$ in terms of $\Omega_1$, $\Omega_2$ and $\Omega_3$. We break that sum into two parts : one containing  $\Omega_1$, $\Omega_3$ and the other containing $\Omega_2$. In this first, we shift the sums by setting $l=s-1$ and $k=r+1$. Thus, 
\begin{eqnarray}
\begin{split}
 \sum_{k=1}^p\sum_{l=0}^{k-1}&  \alpha_{(k,l)}\beta_{(k,l)}(k-l) W_{(k,l)}  =-2  \sum_{k=1}^p \sum_{l=0}^{k-1}  \alpha_{(k,l)}\beta_{(k,l)}(k-l) (p-k) \Omega_{2(k,l)} 
 \\&+ \sum_{r=0}^{p-1} \sum_{s=1}^{r+1}  \alpha_{(r+1,s-1)}\beta_{(r+1,s-1)}(r-s+2) \left(\Omega_{1(r+1,s-1)}-(r-s+1) \Omega_{3(r+1,s-1)} \right)  \,.
\end{split}
\end{eqnarray} 
In this last term, due to the presence of a binomial coefficient $\binom{r+1}{s-1}$ inside $\alpha_{(r+1,s-1)}$, we can replace $\sum_{s=1}^{r+1}  \longrightarrow \sum_{s=0}^{r+1}$, with the convention that  $\binom{y}{-x}=0$ for $y\geq0$ and $x>0$. In the first term, due to the presence of the factors $(k-l)$ and $(p-k)$ and the binomial coefficient $\binom{k}{l}$ inside $\alpha_{(k,l)}$ (wich vanishes for $l>k$), we can replace $\sum_{k=1}^{p} \sum_{l=0}^{k-1}  \longrightarrow \sum_{k=0}^{p-1} \sum_{l=0}^{k+1} $.

Introducing the divergence Eq\eqref{DIVCONF} from the term containing $\Omega_{2}$, we obtain :
\begin{eqnarray}
\begin{split}
&Q_p= -2 \alpha_{00}\beta_{00} \Omega_{1(0,-1)}  +4  \sum_{k=0}^{p-1} \sum_{l=0}^{k} (p-k) \alpha_{(k,l)}\beta_{(k,l)} \mathcal{D}V_{(k,l)} \\
&+ \sum_{k=0}^{p-1} \sum_{l=0}^{k+1}  \Big(  \alpha_{(k+1,l-1)}\beta_{(k+1,l-1)} (k-l+2) -4  \alpha_{(k,l)}\beta_{(k,l)} (p-k) -2  \alpha_{(k+1,l)}\beta_{(k+1,l)}\Big) \Omega_{1(k+1,l-1)}   \\
&-\sum_{k=0}^{p-1} \sum_{l=0}^{k+1}  \Big( 8 l (p-k) \alpha_{(k,l)}\beta_{(k,l)} + \alpha_{(k+1,l-1)}\beta_{(k+1,l-1)} (k-l+2)(k-l+1) \Big) \Omega_{3(k+1,l-1)} \,.
\end{split}
\end{eqnarray}
Using the definition Eq\eqref{alphaconf}, the last term is vanishing for both $\beta_{(k,l)} = H_{k+l}$ and $\beta_{(k,l)} = 1$, because $\beta_{(k,l)}=\beta_{(k+1,l-1)}$.

For $\beta_{(k,l)} = 1$, the third term vanishes as well, so that
\begin{eqnarray}
Q_p=2^p \mathcal{L}_p +4  \sum_{k=0}^{p-1} \sum_{l=0}^{k} (p-k) \alpha_{(k,l)} \mathcal{D}V_{(k,l)}\,, \label{Qp1}
\end{eqnarray}
where we used the definition \eqref{WOmega} of $\Omega_1$, with $\mathcal{L}_p$ the $p^{\text{th}}$ order Lovelock scalar.

For $\beta_{(k,l)} = H_{k+l}$, we can use the property of the harmonic numbers $\beta_{(k+1,l)} = \beta_{(k,l)} + \frac{1}{k+l+1}$ to find : 
\begin{eqnarray}
Q_p= 4  \sum_{k=0}^{p-1} \sum_{l=0}^{k} (p-k) \alpha_{(k,l)}\beta_{(k,l)} \mathcal{D}V_{(k,l)}  - 2 \sum_{k=0}^{p-1} \sum_{l=0}^{k+1} \frac{\alpha_{(k+1,l)}}{k+l+1} \Omega_{1(k+1,l-1)}\,.  \label{Qp2}
\end{eqnarray} 

Now we can see how the critical conformal regularization can be simplified by these results. For clarity, let's rewrite Eq\eqref{CriticalRegulari},
\begin{eqnarray}
\begin{split}
L^\text{critical}_p :=&\lim_{d\to 2p} \left\{ \sqrt{-\bar{g}} \bar{\mathcal{L}}_p  - \sqrt{-g}  \frac{1}{2^p}   \sum_{k=0}^p \sum_{l=0}^k\, \Xi_{(p,k,l)}   \, (k+l)!  \right\} \frac{1}{d-2p} 
\\
=& \sqrt{-g}   \frac{1}{2^{p+1}}  \sum_{k=0}^p \sum_{l=0}^k  \Xi_{(p,k,l)}  (k+l)! \left(\phi +2 H_{k+l} \right) =:  \sqrt{-g}  \mathfrak{L}_p  \,.
\end{split}
\end{eqnarray}
Therefore, the counter-term needed to take the critical limit of the $d$-dimensional Lovelock scalar of order $p$ is a total derivative, as from \eqref{Qp1} we have 
\begin{eqnarray}
L^\text{critical}_p :=\lim_{d\to 2p} \left\{ \sqrt{-\bar{g}} \bar{\mathcal{L}}_p  - \sqrt{-g} \mathcal{L}_p-  \sqrt{-g} \frac{1}{2^{p-2}}    \sum_{k=0}^{p-1} \sum_{l=0}^{k} (p-k) \alpha_{(k,l)} \mathcal{D}V_{(k,l)}   \right\} \frac{1}{d-2p} \,.
\end{eqnarray}
Moreover, from both Eq\eqref{Qp1}\eqref{Qp2}, we obtain by discarding boundary terms (represented by $\equiv$),
\begin{eqnarray}
\begin{split}
\mathfrak{L}_p &\equiv \frac{1}{2} \phi \mathcal{L}_p + \frac{1}{2^{p-1}}  \sum_{k=0}^{p-1} \sum_{l=0}^{k} (p-k) \alpha_{(k,l)} \phi \mathcal{D}V_{(k,l)} -\frac{1}{2^{p-1}}   \sum_{k=0}^{p-1} \sum_{l=0}^{k+1} \frac{\alpha_{(k+1,l)}}{k+l+1} \Omega_{1(k+1,l-1)} \\
&\equiv \frac{1}{2} \phi \mathcal{L}_p - \frac{1}{2^{p-1}}  \sum_{k=0}^{p-1} \sum_{l=0}^{k} (p-k) \alpha_{(k,l)} W_{(k+1,l)}  -\frac{1}{2^{p-1}}   \sum_{k=0}^{p-1} \sum_{l=0}^{k+1} \frac{\alpha_{(k+1,l)}}{k+l+1} \Omega_{1(k+1,l-1)} \\
&=  \frac{1}{2} \phi \mathcal{L}_p + \frac{1}{2^p} \sum_{k=1}^p \sum_{l=0}^k  \frac{\alpha_{(k,l)}}{k+l}\left( (k-l) W_{(k,l)}  -2 \Omega_{1(k,l-1)} \right)\,,
\end{split}
\end{eqnarray}
where we used the definition of $\alpha_{(k,l)}$ and shifted the sums in the last line. Therefore, we are back to Eq\eqref{Qppppp}, with this time $\beta_{(k,l)} = 1/(k+l)$ and the sum $\sum_{k=0}^p$ replaced by $\sum_{k=1}^p$. Thus, discarding boundary terms and with some care to avoid the terms $\beta_{00}$ that are no more allowed, the same calculation  finally yields 
\begin{eqnarray}
\mathfrak{L}_p \equiv \frac{1}{2} \phi \mathcal{L}_p + \frac{p}{4} \left( \nabla_\zeta \phi \nabla^\zeta \phi \right) \mathcal{L}_{(p-1)} + \frac{1}{2^{p-1}} \sum_{k=1}^{p-1} \sum_{l=0}^{k+1}  \frac{\alpha_{(k+1,l)}}{(k+l)(k+l+1)} \Omega_{1(k+1,l-1)} \,,
\end{eqnarray}
by using the definition of $\Omega_{1(1,0)}$.

\subsection{Horndeski Decomposition of Lovelock-Lanczos gravity} \label{HorndeskiDecomp}

From  Eq\eqref{DIVCONF}, we have 
\begin{eqnarray}
\mathcal{D}V_{(k,-k-1)}=  \Omega_{1(k+1,-k-2)} - 2 (k+1)  \Omega_{3(k+1,-k-2)} - \frac{1}{2} (2 k+1) \Omega_{2(k,-k-1)} \,.
\end{eqnarray} 
For simplicity, let's write $\mathcal{D}V_{(k)}=\mathcal{D}V_{(k,-k-1)}$, $\Omega_{1(k+1)} =\Omega_{1(k+1,-k-2)}$, $ \Omega_{3(k+1)}= \Omega_{3(k+1,-k-2)}$ and $\Omega_{2(k)} =\Omega_{2(k,-k-1)} $. Therefore,
\begin{eqnarray}
\mathcal{D}V := \sum_{k=0}^{p-1} \omega_{(p,k)} \mathcal{D}V_{(k)} = \sum_{r=1}^p \left( \Omega_{1(r)} -2 r \Omega_{3(r)} \right)  \omega_{(p,r-1)} -\frac{1}{2} \sum_{r=0}^{p-1} (2r+1) \omega_{(p,r)} \Omega_{2(r)} \,,
\end{eqnarray}
where we shifted the first sum and introduced a sequence of constants $\omega_{(p,k)}$ that is arbitrary for the moment. Using Eq\eqref{RecurrenceRelation} in the first sum while setting $W_{(k)}=W_{(k,-k-1)}$, we obtain 
\begin{eqnarray}
\begin{split}
\mathcal{D}V &= \sum_{r=1}^p W_{(r)}   \omega_{(p,r-1)} + 2 \sum_{r=1}^p (p-r) \omega_{(p,r-1)} \Omega_{2(r)}    -\frac{1}{2} \sum_{r=0}^{p-1} (2r+1) \omega_{(p,r)} \Omega_{2(r)} \\
&=  \sum_{r=1}^p W_{(r)}   \omega_{(p,r-1)} + \sum_{r=1}^{p-1} \left(2 (p-r) \omega_{(p,r-1)} - \left( r+\frac{1}{2}\right)\omega_{(p,r)} \right) \Omega_{2(r)} - \frac{1}{2} \Omega_{2(0)} \omega_{(p,0)}\,.
\end{split}
\end{eqnarray}
Using Eq\eqref{RecurrenceRelation} for the last term and noticing that $\Omega_{1(0)} = 2^p \mathcal{L}_p$, 
\begin{eqnarray}
\mathcal{D}V =  \sum_{r=1}^p W_{(r)}   \omega_{(p,r-1)}   + \frac{\omega_{(p,0)}}{4p} \left( W_{(0)} - 2^p \mathcal{L}_p \right)
+ \sum_{r=1}^{p-1} \left(2 (p-r) \omega_{(p,r-1)} - \left( r+\frac{1}{2}\right) \omega_{(p,r)} \right) \Omega_{2(r)}
\end{eqnarray}
Therefore, providing that we choose $\omega_{(p,k)}$ such that $\omega_{(p,0)}=4p$ and satisfying
\begin{eqnarray}
2 (p-r) \omega_{(p,r-1)} - \left( r+\frac{1}{2}\right) \omega_{(p,r)}=0\,, \label{SequenceHorndeski}
\end{eqnarray}
we get the Horndeski decomposition of Lovelock-Lanczos gravity  \cite{HorndeskiVector} :
\begin{eqnarray}
\mathcal{L}_{(p)} = \left(  W_{(0)} + \sum_{s=0}^{p-1} \omega_{(p,s)} W_{(s+1)}  \right) - \sum_{s=0}^{p-1} \omega_{(p,s)}  \nabla_{\mu}  V_{(s)}^\mu  \,, \label{HorndeskiFormula}
\end{eqnarray} 
with
\begin{eqnarray}
\begin{split}
W_{(s)} &:=\frac{1}{2^p}  \delta^{\mu \mu_1 \nu_1 \dots \mu_p \nu_p}_{\nu \sigma_1 \rho_1 \dots \sigma_p \rho_p}  \left( \frac{u_\mu u^{\nu}}{u^{\gamma}u_{\gamma}}   \right) \,  \prod_{i=1}^{s} \, \frac{\nabla_{\mu_i} u^{\sigma_i} \nabla_{\nu_i} u^{\rho_i} }{u^{\gamma}u_{\gamma}} \,  \prod_{i=s+1}^{p} \, R_{\mu_{i} \nu_{i}}^{\sigma_{i} \rho_{i}} \,, \\
V_{(s)}^\mu &:=\frac{1}{2^p}  \delta^{\mu\nu \mu_1 \nu_1 \dots \mu_{p-1} \nu_{p-1}}_{\sigma\rho \sigma_1 \rho_1 \dots \sigma_{p-1} \rho_{p-1}}  \left(  \frac{u^{\sigma} \nabla_{\nu} u^{\rho} }{u^{\gamma}u_{\gamma}}  \right) \,  \prod_{i=1}^{s} \, \frac{\nabla_{\mu_i} u^{\sigma_i} \nabla_{\nu_i} u^{\rho_i} }{u^{\gamma}u_{\gamma}} \,  \prod_{i=s+1}^{p-1} \, R_{\mu_{i} \nu_{i}}^{\sigma_{i} \rho_{i}} \,,  \label{WHorndeski}
\end{split}
\end{eqnarray} 
where we substituted $\nabla_\alpha \phi \to u_\alpha$ with respect to the previous Appendix because it does not change  Eq\eqref{DIVCONF} nor Eq\eqref{RecurrenceRelation}. Indeed, the symmetry in $\nabla_\alpha \nabla_\beta \phi$ is never used in these relations as only the quantity $\nabla^\alpha \nabla_\beta \phi$ appears. In that paper, Horndeski made the choice :
\begin{eqnarray}
\omega_{(p,s)}  =  4^{s+1} p \prod_{q=0}^{s-1} \frac{p-q-1}{2q+3}\,,  \label{Horndeskichoice}
\end{eqnarray}
which prevents to incorporate the $W_{(0)}$ term into the sum over $W_{(r)}$, because in this case $\omega_{(p,-1)}=p \neq 1$, by definition of empty product. However, we see from Eq\eqref{SequenceHorndeski} that if $\omega_{(p,0)}=4p$, then $\omega_{(p,-1)}=1$, so that the previous choice of $\omega_{(p,s)}$ made in \cite{HorndeskiVector} is not very fortunate in this regard.

Considering rather Eq\eqref{SequenceHorndeski} as the definition of $\omega_{(p,s)}$, we can write :
\begin{eqnarray}
\mathcal{L}_{(p)} =    \sum_{s=0}^{p} \omega_{(p,s-1)} W_{(s)}  -  \sum_{s=0}^{p-1} \omega_{(p,s)}  \nabla_{\mu}  V_{(s)}^\mu \,.
\end{eqnarray}
Finally, in order to have a meaningful expression for $\omega_{(p,s)}$, we can see from the Appendix A. of the Deruelle-Merino-Olea paper \cite{Deruelle}, establishing the equivalence between the Horndeski divergence and the Myers counter-terms, that changing the sums indices as 
\begin{eqnarray}
 \sum_{s=0}^{p-1} \omega_{(p,s)}  \nabla_{\mu}  V_{(s)}^\mu   =  \sum_{k=0}^{p-1} \omega_{(p,p-k-1)}  \nabla_{\mu}  V_{(p-k-1)}^\mu  \;, \quad  \sum_{s=0}^{p} \omega_{(p,s-1)} W_{(s)} =  \sum_{k=0}^{p} \omega_{(p,p-k-1)}  W_{(p-k)} \,,
\end{eqnarray}
allows to order the products of Riemann tensors and covariant derivatives of the (arbitrary) vector field in the same way as in the Hamiltonian formalism and Myers counter-terms, by simply renaming the indices. Moreover, it was noted in this paper (still \cite{Deruelle}) that 
\begin{eqnarray}
\sigma_{(p,s)} := \frac{p!4^{p-s}}{s! \left[ 2 (p-s) -1 \right]!!}  =  \omega_{(p,p-s-1)}  \,,
\end{eqnarray}
 so that for $s=p$, we do obtain $\omega_{(p,-1)}=1$, while the other cases agree with the choice \eqref{Horndeskichoice}. Thus, the Horndeski decomposition of the Lovelock-Lanczos Lagrangian density can be rewritten as Eq\eqref{HorndeskiDecomposition}.

\subsection{$3$D-covariant regularization of quartic Lovelock-Lanczos theory}\label{sec.L43Dcov}
  
From Eq\eqref{RegLag3D} and Eq\eqref{RegHamCons3D}, we obtain the quartic ($p=4$) Lagrangian and Hamiltonian constraint,
\begin{eqnarray}
\begin{split}
\frac{\mathscr{L}_{(4)}}{(d-4)...(d-8)}&= \frac{1}{35(d-2)(d-3)^4} \left( \frac{(d-1)(15d-29)}{(d-2)^2} \mathscr{S}_{4,1} +64 \mathscr{S}_{4,2} - \frac{24(3d-5)}{(d-2)} \mathscr{S}_{4,3} \right)\\
\frac{(d-2)\mathscr{H}_{(4)}}{(d-4)...(d-8)}&= \frac{\mathcal{M}\sqrt{|h|}}{(d-3)^4} \left( \frac{(d-1)(15d-29)}{(d-2)^2} \mathcal{M}^3 +64 \mathcal{M}^\mu_\nu \mathcal{M}^\nu_\gamma \mathcal{M}^\gamma_\mu  - \frac{24(3d-5)}{(d-2)}  \mathcal{M}^\mu_\nu \mathcal{M}^\nu_\mu \mathcal{M} \right) 
\end{split}
\end{eqnarray}
where 
\begin{eqnarray}
\begin{split}
\mathscr{S}_{4,1} =& 5 \mathcal{M}^4 + 8 \mathcal{M}^3 \bar{R}+ 16 \mathcal{M}^2\bar{R}^2 + 64 \mathcal{M}\bar{R}^3 - 128 \bar{R}^4\,, \\
\mathscr{S}_{4,2} =& 5 \mathcal{M}^\mu_\nu \mathcal{M}^\nu_\gamma \mathcal{M}^\gamma_\mu \mathcal{M}+2 \mathcal{M}^\mu_\nu \mathcal{M}^\nu_\gamma\left( 3 \mathcal{M} \bar{R}^\gamma_\mu + \mathcal{M}^\gamma_\mu \bar{R} \right)+ 8 \mathcal{M}^\mu_\nu \bar{R}^\nu_\gamma \left( \mathcal{M} \bar{R}^\gamma_\mu + \mathcal{M}^\gamma_\mu \bar{R} \right) \\
&+ 16  \bar{R}^\mu_\nu \bar{R}^\nu_\gamma \left( \mathcal{M} \bar{R}^\gamma_\mu  +3 \mathcal{M}^\gamma_\mu  \bar{R}\right) -128   \bar{R}^\mu_\nu  \bar{R}^\nu_\gamma  \bar{R}^\gamma_\mu  \bar{R}\,, \\
\mathscr{S}_{4,3} =& 5 \mathcal{M}^\mu_\nu \mathcal{M}^\nu_\mu \mathcal{M}^2 + 4 \mathcal{M}^\mu_\nu \mathcal{M}\left( \mathcal{M}\bar{R}^\nu_\mu+\mathcal{M}^\nu_\mu\bar{R} \right) + \frac{8}{3} \left( \mathcal{M}^2 \bar{R}^\mu_\nu \bar{R}^\nu_\mu + 4 \mathcal{M}^\mu_\nu \mathcal{M}\bar{R}^\nu_\mu \bar{R}+ \mathcal{M}^\mu_\nu \mathcal{M}^\nu_\mu \bar{R}^2\right) 
\\&+ 32 \bar{R}^\mu_\nu \bar{R}\left( \mathcal{M} \bar{R}^\nu_\mu+\mathcal{M}^\nu_\mu \bar{R}\right) - 128\bar{R}^\mu_\nu \bar{R}^\nu_\mu \bar{R}^2\,.
\end{split}
\end{eqnarray}
Using Eq\eqref{RegPi13D} for the momentum extracted from the previous Lagrangian yields
\begin{eqnarray}
\begin{split}
\frac{\pi^{\phantom{(4)}\nu}_{(4)\mu}-\check{\pi}^{\phantom{(4)}\nu}_{(4)\mu} }{(d-4)...(d-8)}&= \mathscr{J}_{(4)\mu}^\nu+ \frac{16 \sqrt{|h|}}{35\epsilon(d-2)(d-3)^4} \bigg( \epsilon \pi^{\phantom{(1)}\nu}_{(1)\mu}  \left( - \frac{2d-1}{d-2} \mathscr{T}_{4, 1}-4  \mathscr{T}_{4, 2} + \frac{3(2d-3)}{d-2} \mathscr{T}_{4, 3} \right)
\\&-2 K^\sigma_\rho \left( 3\left( 5\mathcal{M} + 2 \bar{R} \right) \mathcal{M}^\nu_{[\sigma} \mathcal{M}_{\mu]}^\rho - 2 \left( 3 \mathcal{M} + 4 \bar{R}\right) \mathcal{M}^{[\nu}_{[\mu} \bar{R}_{\sigma]}^{\rho]} + 8 \left( \mathcal{M} + 6 \bar{R} \right) \bar{R}^\nu_{[\sigma} \bar{R}_{\mu]}^\rho \right)   \bigg)
\end{split}
\end{eqnarray}
with
\begin{eqnarray}
\begin{split}
\mathscr{T}_{4,1} =& 5 \mathcal{M}^3 +6 \mathcal{M}^2 \bar{R} +8 \mathcal{M} \bar{R}^2 + 16 \bar{R}^3 \,, \\
\mathscr{T}_{4,2} =&5 \mathcal{M}^\mu_\nu \mathcal{M}_\mu^\gamma \mathcal{M}_\gamma^\nu + 6 \mathcal{M}^\mu_\nu \mathcal{M}_\mu^\gamma \bar{R}_\gamma^\nu  + 8 \mathcal{M}^\mu_\nu \bar{R}_\mu^\gamma \bar{R}_\gamma^\nu  + 16\bar{R}^\mu_\nu \bar{R}_\mu^\gamma \bar{R}_\gamma^\nu \,, \\
\mathscr{T}_{4,3} =& 5 \mathcal{M} \mathcal{M}^\mu_\nu \mathcal{M}^\nu_\mu + 2 \mathcal{M}^\mu_\nu \left(   \mathcal{M}^\nu_\mu \bar{R}+2 \mathcal{M}  \bar{R}^\nu_\mu\right) + \frac{8}{3} \bar{R}^\mu_\nu \left(\mathcal{M} \bar{R}^\nu_\mu + 2 \mathcal{M}^\nu_\mu \bar{R} \right)+ 16 \bar{R}^\mu_\nu \bar{R}^\nu_\mu \bar{R}  \,,
\end{split}
\end{eqnarray}
and where the momentum $\check{\pi}_4$ is obtained from Eq\eqref{RegPi23D} :  
\begin{eqnarray}
\begin{split}
\frac{\check{\pi}^{\phantom{(4)}\nu}_{(4)\mu} }{(d-4)...(d-8)}&= \frac{4 \sqrt{|h|} K^\nu_\rho h_{\mu\nu\alpha}^{\sigma\rho\beta}}{35(d-3)^3(d-2)^2\epsilon}  \bigg(  -\frac{1}{d-2} h^\alpha_\beta \left( 5 \mathcal{M}^3 + 6 \mathcal{M}^2 \bar{R} +8 \mathcal{M} \bar{R}^2 + 16 \bar{R}^3\right) 
\\& +6 \left(5 \mathcal{M}^\alpha_\beta \mathcal{M}^2 + 2 \mathcal{M} \left( \mathcal{M} \bar{R}^\alpha_\beta + 2 \mathcal{M}^\alpha_\beta \bar{R}\right)+ \frac{8}{3} \bar{R} \left( 2 \mathcal{M} \bar{R}^\alpha_\beta + \mathcal{M}^\alpha_\beta \bar{R} \right) +16 \bar{R}^\alpha_\beta\bar{R}^2 \right)  \bigg)
\end{split}
\end{eqnarray} 
Finally, like in the cubic case we can simplify the above expression for $\pi_4$ by noting that in four dimensions, 
\begin{eqnarray}
\begin{split}
\mathscr{J}_{(4)\mu}^\nu:= -\frac{48 \sqrt{|h|}  h_{\mu\beta\rho\delta}^{\nu\alpha\sigma\gamma} K^{\beta}_\alpha}{35\epsilon(d-2)(d-3)^4}\bigg(& 5 \mathcal{M}^\rho_\sigma \mathcal{M}^\delta_\gamma \mathcal{M}+2 \mathcal{M}^\rho_\sigma \left( 2 \mathcal{M} \bar{R}^\delta_\gamma + \mathcal{M}^\rho_\sigma \bar{R}\right) \\
&+\frac{8}{3} \bar{R}^\rho_\sigma\left( \mathcal{M} \bar{R}^\rho_\sigma+2 \mathcal{M}^\rho_\sigma \bar{R} \right) + 16 \bar{R}^\rho_\sigma \bar{R}^\delta_\gamma \bar{R} \bigg) =0\,.
\end{split}
\end{eqnarray}

\subsection{Perturbations of LLG around DSS backgrounds}\label{SecPerturbationLLGDDS}

Consider the following tensor defined in Eq\eqref{GDSS} in components by,
\begin{eqnarray}
 \mathscr{G}^{(k)\mu}_{(n)\nu} :=-\frac{1}{2^{p+1}} \delta^{\mu\, \mu_1 \nu_1 \dots \mu_p \nu_p}_{\nu\, \sigma_1 \rho_1 \dots \sigma_p \rho_p} \prod_{r=1}^{p-k}   R^{(0)\sigma_r \rho_r}_{\phantom{(0)}\mu_r \nu_r}   \left\{ \prod_{r=p-k+1}^{p}   R_{\mu_r \nu_r}^{\sigma_r \rho_r}  \right\}^{[n]}\,.
\end{eqnarray}
Around a DSS backgrounds, its components in the radial and time-like directions become : 
\begin{eqnarray}
\begin{split}
 & \mathscr{G}^{(k)A}_{(n)B} =  Z^{p-k-1} \frac{(d-2k-2)!}{(d-2p-1)!} \bigg( \Big[ (d-2p-1) Z \delta_B^A - 2 (p-k) \delta^{A c}_{B a} \mathcal{Y}_c^a \Big] {}_{(0)}^{(1)}\mathcal{K}^{(k)}_{(n)} \\
 &\phantom{2^{p-k} Z^{p-k-1} \frac{(d-2k-2)!}{(d-2p-1)!} \bigg(}+ 4k \left(d-2k-1 \right) Z \delta_{B b}^{A a} \left[ {}_{(1)}^{(1)}\mathcal{K}^{(k)b}_{(n)a} + (k-1) \, {}_{(2)}^{(1)}\mathcal{K}^{(k)b}_{(n)a}  \right]  \bigg) \,,
\end{split}
\end{eqnarray}
where
\begin{eqnarray}
\begin{split}
{}_{(0)}^{(1)}\mathcal{K}^{(k)}_{(n)}&:=-\frac{1}{2^{k+1}} \delta^{i_1 j_1 \dots i_k j_k}_{k_1 l_1 \dots k_k l_k} \left\{  \prod_{r=1}^{k}   R^{k_r l_r}_{i_r j_r} \right\}^{[n]} \,, \\
{}_{(1)}^{(1)}\mathcal{K}^{(k)b}_{(n)a} &:=-\frac{1}{2^{k+1}} \delta^{i_1 j_1 \dots i_{k-1} j_{k-1} i_k}_{k_1 l_1 \dots k_{k-1} l_{k-1} j_k} \left\{R^{j_k b}_{i_k a} \; \prod_{r=1}^{k-1}   R^{k_r l_r}_{i_r j_r} \right\}^{[n]} \,,  \\
{}_{(2)}^{(1)}\mathcal{K}^{(k)b}_{(n)a} &:= -\frac{1}{2^{k+1}} \delta^{i_1 j_1 \dots i_{k-2} j_{k-2} i_{k-1} i_k k_k}_{k_1 l_1 \dots k_{k-2} l_{k-2}  j_{k-1} l_{k-1} j_k} \left\{ R^{j_{k-1} l_{k-1}}_{i_{k-1} a} R^{j_{k} b}_{i_k k_k} \,  \prod_{r=1}^{k-2}   R^{k_r l_r}_{i_r j_r} \right\}^{[n]} \,.
\end{split}
\end{eqnarray}
In the time/radial-angular directions,
\begin{eqnarray}
\begin{split}
 & \mathscr{G}^{(k)A}_{(n)I}  =  Z^{p-k-1}  \frac{(d-2k-2)!}{(d-2p-1)!}  \ k  \Bigg[ \Big( 4(p-k) \mathcal{Y}_c^b \delta^{A c}_{a b} - 2 (d-2p-1) Z \delta_a^A \Big)   {}_{(1)}^{(1)}\mathcal{M}^{(k)a}_{(n)I} \\
& +(d-2k-1) Z \delta_{a b}^{A c} \left( -2\,  {}_{(1)}^{(2)}\mathcal{M}^{(k)a b}_{(n)I c} + (k-1)  {}_{(2)}\mathcal{M}^{(k)a b}_{(n)I c} + \frac{4}{3} (k-1)(k-2)\,  {}_{(3)}\mathcal{M}^{(k)a b}_{(n)I c} \right)  \Bigg]\,,
\end{split}
\end{eqnarray}
where the perturbations are contained within the quantities :
\begin{eqnarray}
\begin{split}
{}_{(1)}^{(1)}\mathcal{M}^{(k)a}_{(n)I} &:= -\frac{1}{2^{k+1}}  \delta^{i_1 j_1 \dots i_{k-1} j_{k-1} i_k k_k}_{k_1 l_1 \dots k_{k-1} l_{k-1} j_k I} \left\{ R^{j_k a}_{i_k k_k} \, \prod_{r=1}^{k-1}   R^{k_r l_r}_{i_r j_r} \right\}^{[n]} \,, \\
{}_{(1)}^{(2)}\mathcal{M}^{(k)a b}_{(n)I c}  &:=-\frac{1}{2^{k+1}}   \delta^{i_1 j_1 \dots i_{k-1} j_{k-1} i_k}_{k_1 l_1 \dots k_{k-1} l_{k-1} I} \left\{ R^{a b}_{i_k c} \, \prod_{r=1}^{k-1}   R^{k_r l_r}_{i_r j_r} \right\}^{[n]}  \,, \\
{}_{(2)}\mathcal{M}^{(k)a b}_{(n)I c}  &:=-\frac{1}{2^{k+1}}   \delta^{i_1 j_1 \dots i_{k-2} j_{k-2} i j m}_{k_1 l_1 \dots k_{k-2} l_{k-2}k l I}  \left\{\left(  4 R^{k a}_{i j} R^{l b}_{m c} - R^{a b}_{i j} R^{k l}_{m c}  \right) \, \prod_{r=1}^{k-2}   R^{k_r l_r}_{i_r j_r} \right\}^{[n]}  \,, \\
{}_{(3)}\mathcal{M}^{(k)a b}_{(n)I c} &:=-\frac{1}{2^{k+1}}  \delta^{i_1 j_1 \dots i_{k-3} j_{k-3} i_{k-2} j_{k-2} i_{k-1} j_{k-1} i_k}_{k_1 l_1 \dots k_{k-3} l_{k-3} k_{k-2} k_{k-1} k_k l_k I}   \left\{ R^{k_{k-2} a}_{i_{k-2} j_{k-2}} R^{k_{k-1} b}_{i_{k-1} j_{k-1}} R^{k_k l_k}_{i_k c} \, \prod_{r=1}^{k-3}   R^{k_r l_r}_{i_r j_r} \right\}^{[n]}
\end{split}
\end{eqnarray}
Finally, in the angular-angular directions, it reduces to 
\begin{eqnarray}
\begin{split}
 &\mathscr{G}^{(k)I}_{(n)J}  =  Z^{p-k-2} \frac{(d-2k-3)!}{(d-2p-1)!}\; \Bigg[ \mathcal{B}_1 \;  {}^{(2)}_{(0)}\mathcal{K}^{(k)I}_{(n)J}  +    \mathcal{B}_{2 b}^{\phantom{2}a}  \left(  {}_{(1)}^{(2)}\mathcal{K}^{(k)I b}_{(n)J a} + (k-1){}_{(2)}^{(2)}\mathcal{K}^{(k) I b}_{(n) J a} \right) \\
 &+  \mathcal{B}_3 \, \delta_{cd}^{ab} \, \left( _{(1)}\mathcal{P}_{(n)J ab}^{(k)I cd} + (k-1) _{(2)}\mathcal{P}_{(n)J ab}^{(k) I cd} \right)  + \mathcal{B}_4 \delta^{ab}_{cd}  \left( {}_{(3)}\mathcal{P}_{(n) J ab}^{(k) I cd} + 2 {}_{(4)}\mathcal{P}_{(n)J ab}^{(k) I cd} \right)\Bigg]\,,
\end{split}
\end{eqnarray}
where we have defined the background quantities :
\begin{eqnarray}
\begin{split}
\mathcal{B}_1 :=& (p-k)\Big[ R^{(2)}  Z + 2 (p-k-1) \delta_{ce}^{ab} \mathcal{Y}_a^c \mathcal{Y}_b^e - 2 (d-2p-1) Z \mathcal{Y}_a^a \Big] \\
 &+ (d-2p-1)(d-2p-2) Z^2 \,, \\
 \mathcal{B}_{2 b}^{\phantom{2}a}  :=&4 k  (d-2k-2)   \, Z \Big( (d-2p-1) Z \delta^a_b - 2 (p-k) \mathcal{Y}_c^e \delta^{ac}_{be}  \Big) \,, \\
 \mathcal{B}_3 :=&  k  (d-2k-2) (d-2k-1)  \, Z^2  \,, \\
 \mathcal{B}_4:=& \frac{4}{3} k(k-1)(k-2)(d-2k-2) (d-2k-1)  \, Z^2 \,,
\end{split}
\end{eqnarray}
and the perturbations are given by
\begin{eqnarray}
\begin{split}
^{(2)}_{(0)}\mathcal{K}^{(k)I}_{(n)J}&:=-\frac{1}{2^{k+1}} \delta^{i_1 j_1 \dots i_k j_k \, I}_{k_1 l_1 \dots k_k l_k \, J} \left\{  \prod_{r=1}^{k}   R^{k_r l_r}_{i_r j_r} \right\}^{[n]}\,, \\
^{(2)}_{(1)}\mathcal{K}^{(k)I b}_{(n)J a}&:=-\frac{1}{2^{k+1}}  \delta^{i_1 j_1 \dots i_{k-1} j_{k-1} \, i   I}_{k_1 l_1 \dots k_{k-1} l_{k-1} \, j  J} \left\{ R_{i a}^{j b} \prod_{r=1}^{k-1}   R^{k_r l_r}_{i_r j_r} \right\}^{[n]} \,, \\
^{(2)}_{(2)}\mathcal{K}^{(k)I b}_{(n)J a}&:=-\frac{1}{2^{k+1}} \delta^{i_1 j_1 \dots i_{k-2} j_{k-2} \, i_{k-1} i_k j_k  I}_{k_1 l_1 \dots k_{k-2} l_{k-2} \, k_{k-1} l_{k-1} k_k  J} \left\{ R_{i_{k-1} a}^{k_{k-1} l_{k-1} } R_{i_k j_k}^{k_k b} \prod_{r=1}^{k-2}   R^{k_r l_r}_{i_r j_r} \right\}^{[n]}\,,
\end{split}
\end{eqnarray}
and 
\begin{eqnarray}
\begin{split}
_{(1)}\mathcal{P}_{(n)J ab}^{(k)I cd}&:=-\frac{1}{2^{k+1}}  \delta^{i_1 j_1 \dots i_{k-1} j_{k-1} \,   I}_{k_1 l_1 \dots k_{k-1} l_{k-1} \,  J} \left\{  R_{ab}^{cd} \prod_{r=1}^{k-1}   R^{k_r l_r}_{i_r j_r} \right\}^{[n]}\,, \\
_{(2)}\mathcal{P}_{(n)J ab}^{(k)I cd}&:=-\frac{1}{2^{k+1}} \delta^{i_1 j_1 \dots i_{k-2} j_{k-2} \, i j   I}_{k_1 l_1 \dots k_{k-2} l_{k-2} \, k l J} \left\{  \left( R_{ij}^{cd} R^{kl}_{ab} + 8 R^{k c}_{i a} R_{j b}^{l d} \right) \prod_{r=1}^{k-2}   R^{k_r l_r}_{i_r j_r} \right\}^{[n]} \,, \\
_{(3)}\mathcal{P}_{(n)J ab}^{(k)I cd}&:=-\frac{1}{2^{k+1}}  \delta^{i_1 j_1 \dots i_{k-1} j_{k-1} \,    I}_{k_1 l_1 \dots k_{k-1} l_{k-1} \,  J}   \left\{ \left( 4 R^{\Bigcdot\Bigcdot}_{\Bigcdot a} R^{\Bigcdot d}_{\Bigcdot b} -R_{\Bigcdot\Bigcdot}^{\Bigcdot d} R^{\Bigcdot\Bigcdot}_{ab}  \right) R_{\Bigcdot\Bigcdot}^{\Bigcdot c}  \prod_{r=1}^{k-3}   R^{k_r l_r}_{i_r j_r} \right\}^{[n]} \,, \\
_{(4)}\mathcal{P}_{(n)J ab}^{(k)I cd}&:=-\frac{1}{2^{k+1}} \delta^{i_1 j_1 \dots i_{k-1} j_{k-1} \,    I}_{k_1 l_1 \dots k_{k-1} l_{k-1} \,  J}  \left\{ R^{\Bigcdot\Bigcdot}_{\Bigcdot a} R^{\Bigcdot\Bigcdot}_{\Bigcdot b} R_{\Bigcdot\Bigcdot}^{\Bigcdot c}R_{\Bigcdot\Bigcdot}^{\Bigcdot d} \prod_{r=1}^{k-4}   R^{k_r l_r}_{i_r j_r} \right\}^{[n]}\,,  
\end{split}
\end{eqnarray}
where the $\Bigcdot$ means only here that the indices in the Riemann tensor are written in the same order as in the GKD.

These formulae, although complicated could useful to make predictions about slowly rotating black hole and cosmological solutions in the context of (higher dimensional) LLG.

\subsection{Regularizations of the Bianchi I sector of LLG}\label{AppendixBianchi}

\subsubsection{Scalars $Q^{(p)}_{(d-1)}$}\label{AppendixBianchi1}

We collect here the expressions for the regularized expressions $Q^{(p)}_{(d-1)}$ appearing in the Bianchi I decomposition of LLG. Defining,
\begin{eqnarray}
M_i=\dot{H}_i + H_i^2 \,,
\end{eqnarray}
we obtain in $d=1+3n$ dimensions from metric \eqref{BianchiI1} :
\begin{eqnarray}
\begin{split}
Q^{(p)}_{(3n)}&= n\Bigg[ M_1 \sum_{i=0}^{2(p-1)} \sum_{j=0}^{i} \binom{n-1}{2p-2-i} \binom{n}{i-j} \binom{n}{j} H_1^{2p-2-i} H_2^{i-j} H_3^j \\
&\phantom{= n\Bigg[} +  M_2 \sum_{i=0}^{2(p-1)} \sum_{j=0}^{i} \binom{n}{2p-2-i} \binom{n-1}{i-j} \binom{n}{j} H_1^{2p-2-i} H_2^{i-j} H_3^j \\
&\phantom{= n\Bigg[} + M_3 \sum_{i=0}^{2(p-1)} \sum_{j=0}^{i} \binom{n}{2p-2-i} \binom{n}{i-j} \binom{n-1}{j} H_1^{2p-2-i} H_2^{i-j} H_3^j \Bigg]\,, \label{BianchiIReg1Q}
\end{split}
\end{eqnarray}
while in $d=1+(1+2n)$, for the metric \eqref{BianchiI2} :
\begin{eqnarray}
\begin{split}
&Q^{(p)}_{(1+2n)} = n \Bigg[ M_2 \Bigg\{ H_1 \sum_{i=0}^{2p-3}  \binom{n-1}{2p-3-i} \binom{n}{i} H_2^{2p-3-i} H_3^i+ \sum_{i=0}^{2p-2}  \binom{n-1}{2p-2-i} \binom{n}{i} H_2^{2p-2-i} H_3^i \Bigg\}\\
&\phantom{n \Bigg[x}  +M_3 \Bigg\{ H_1 \sum_{i=0}^{2p-3}  \binom{n}{2p-3-i} \binom{n-1}{i} H_2^{2p-3-i} H_3^i+ \sum_{i=0}^{2(p-1)}  \binom{n}{2p-2-i} \binom{n-1}{i} H_2^{2p-2-i} H_3^i \Bigg\} \Bigg] \\
&\phantom{n \Bigg[x}  +M_1\sum_{i=0}^{2(p-1)} \binom{n}{2p-2-i} \binom{n}{i} H_2^{2p-2-i} H_3^i \,,
\end{split}
\end{eqnarray}
and finally in $d=1+(2+n)$ from the third ``periodization" \eqref{BianchiI3} :
\begin{eqnarray}
\begin{split}
Q^{(p)}_{(2+n)} =& M_1 \left[ \binom{n}{2p-2} H_3^2 + \binom{n}{2p-3} H_3 H_2 \right] H_3^{2(p-2)} \\ 
+ &M_2 \left[ \binom{n}{2p-2} H_3^2 + \binom{n}{2p-3} H_3 H_1 \right] H_3^{2(p-2)}  \\
+ n &M_3  \left[ \binom{n-1}{2p-2}  H_3^2 + \binom{n-1}{2p-3} H_3 \left( H_1 + H_2 \right) + \binom{n-1}{2p-4} H_1 H_2 \right] H_3^{2(p-2)}\,.
\end{split}
\end{eqnarray}

\subsubsection{Regularizations of the Kantowski-Sachs  sector of the Gauss-Bonnet and cubic LLG}\label{AppendixBianchi2}

After evaluating Eq\eqref{RegKS} in the Gauss-Bonnet case $p=2$, setting $n=(d-1)/3$, dividing by $(d-4)$ and finally setting $d=4$, we obtain :
\begin{eqnarray}
\begin{split}
\mathcal{R}^{(2)}_{3} &= \frac{1}{36} \left( H_1^4 -4 H_1^3 H_2 + 6 H_1^2 H_2^2 + 8 H_1 H_2^3 - 2 H_2^4 \right) \;, \\
Q^{(2)}_{3} &= \frac{1}{6} \left( - H_1^4 + 4 H_1^3 H_2 + 4 H_1 H_2 \dot{H}_1 - H_1^2 \left(\dot{H}_1-2 \dot{H}_2 \right) + 4 H_2^2 \left(\dot{H}_2+ H_2^2 \right) \right) + b.t.
\end{split}
\end{eqnarray}
where the boundary term in $Q^{(2)}_{d-1}$,
\begin{eqnarray}
b.t. = \frac{1}{27}(d-1)^2 H_2 \left( 3(d-3) H_1^2 H_2 + (2 d-5) \left( \dot{H}_1 H_2 + 2 H_1 \left( \dot{H}_2 + H_2^2 \right) \right) \right)\,,
\end{eqnarray}
has been discarded to avoid a divergence when $d\to4$. For $p=3$,
\begin{eqnarray}
\begin{split}
\mathcal{R}^{(3)}_{3} &= \frac{1}{180} \left( 2 H_1^6 - 6 H_1^5 H_2 + 5 H_1^4 H_2^2 + 4 H_1 H_2^5 - 2 H_2^6 \right) \;, \\
Q^{(3)}_{3} &= \frac{1}{36} \Big( -\left(\dot{H}_1+ H_1^2\right) \left(3 H_1^4 - 8 H_1^3 H_2 +6 H_1^2 H_2^2 + 2 H_2^4 \right) \\
&\phantom{= \frac{1}{36} \Big(}+2 \left(\dot{H}_2+ H_2^2\right) \left( H_1^4 - 2 H_1^3 H_2 - 4 H_1 H_2^4 + 2 H_2^4 \right)\Big)\,.
\end{split}
\end{eqnarray}

\newpage{\pagestyle{empty}\cleardoublepage}

\end{document}